\documentclass[
floats,floatfix,amssymb,amsmath,prd,superscriptaddress,nofootinbib,aps,
twocolumn
]{revtex4-2}

\usepackage{graphicx}
\usepackage{dcolumn}
\usepackage{txfonts}
\usepackage{bm}
\usepackage{hyperref}
\usepackage[usenames, dvipsnames]{xcolor}
\hypersetup{colorlinks,bookmarksnumbered,
citecolor=cyan, urlcolor=magenta}
\usepackage{natbib}
\usepackage{multirow}
\usepackage[caption=false]{subfig}

\newcommand{\chisq}{$\chi^2$}
\providecommand{\adsurl}[1]{\href{#1}{ADS}}
\usepackage{aas_macros}

\begin{document}

\preprint{APS/123-QED}
\title{Identifying type II strongly lensed Gravitational-Wave Images in Third-Generation Gravitational-Wave Detectors}
\author{Yijun Wang}
\email{yijunw@caltech.edu}
\affiliation{California Institute of Technology, Pasadena, CA 91125, USA}
\author{Rico K.L.\ Lo}
\affiliation{California Institute of Technology, Pasadena, CA 91125, USA}
\author{Alvin K.Y.\ Li}
\affiliation{California Institute of Technology, Pasadena, CA 91125, USA}
\author{Yanbei Chen}
\affiliation{California Institute of Technology, Pasadena, CA 91125, USA}

\date{\today}

\begin{abstract}
Strong gravitational lensing is a gravitational wave (GW) propagation effect that influences the inferred GW source parameters and the cosmological environment. Identifying strongly lensed GW images is challenging as waveform amplitude magnification is degenerate with a shift in the source intrinsic mass and redshift. However, even in the geometric-optics limit, type II strongly lensed images cannot be fully matched by type I (or unlensed) waveform templates, especially with large binary mass ratios and orbital inclination angles. We propose to use this mismatch to distinguish individual type II images. Using planned noise spectra of Cosmic Explorer, Einstein Telescope and LIGO Voyager, we show that a significant fraction of type II images can be distinguished from unlensed sources, given sufficient SNR ($\sim 30$).  Incorporating models on GW source population and lens population, we predict that the yearly detection rate of lensed GW sources with detectable type II images is 172.2, 118.2 and 27.4 for CE, ET and LIGO Voyager, respectively. Among these detectable events, 33.1\%, 7.3\% and 0.22\% will be distinguishable via their type II images with a log Bayes factor larger than 10. We conclude that such distinguishable events are likely to appear in the third-generation detector catalog; our strategy will significantly supplement existing strong lensing search strategies.
\end{abstract}

\maketitle

\section{Introduction}
\label{section:intro}
Successful detection of gravitational wave (GW) signals from compact binary mergers by the Advanced Laser Interferometer Gravitational-Wave Observatory (aLIGO) and Virgo collaboration has greatly enriched our understanding of gravity and many aspects of astrophysics \citep[see,  e.g.][]{GW1908142020,GW1905212020}. To extract physical information from detector data, proper signal interpretation is crucial. For this purpose, it is important to study changes in the waveform as it propagates through the universe, since, if unaccounted for, propagation effects can be confused with intrinsic GW features and introduce bias in subsequent analysis. On the other hand, results of propagation effects depend on properties both of the GW and the objects along its path that it interacts with \citep[see, e.g.,][]{Lai2018,Meena,Pardo,GWTC1TestGR,Perkins,Vijaykumar}. Therefore, identifying such signatures also maximizes the scientific output of GW detection. 

One GW propagation effect is strong gravitational lensing, in which the rays of a GW are bent strongly enough by a gravitational potential and form multiple images with different magnifications. Gravitational lensing of gravitational waves has attracted enormous interest.  It has been estimated that third-generation detectors can detect up to hundreds of strongly lensed events~\cite{Biesiada2014,Li2018,Oguri2018,Yang}; such events can then be used to study cosmological structures~\cite{Ding2015,liao2017precision,Dai2017,Li2018,Ng2018} and fundamental physics~\cite{fan2017speed,Collett,Mukherjee_MM,Mukherjee_Grav}. 

However, identifying strongly lensed images is challenging since the predominant effect of strong lensing, namely the amplitude magnification by $\sqrt{\mu}$, is degenerate with scaling down the luminosity distance, $D_L$, by $\sqrt{\mu}$ and keeping the redshifted mass, $M_\bullet(1+z_s)$, constant \citep[see, e.g.,][]{DaiPop2017}, where $M_\bullet$ is the total mass of the binary and $z_s$ is the redshift of this GW source. This degeneracy stems from the fact that General Relativity is a scale-free geometric theory, and that GW frequency evolution is unaffected by strong lensing \cite{DaiPop2017}.

Current search strategies typically look for multiple events in a catalog that are consistent in intrinsic properties and sky locations, and have orbital phase related in characteristic ways \citep[see, e.g.,][]{Hannuksela2019,Dai2020}. For example, this has been used to study the series of events GW170104, GW170814 and a sub-threshold trigger, GWC170620, as potential candidates for lensed images~\cite{Dai2020}. In Ref.~\cite{Broadhurst}, the event GW170814 and GW170809 are analyzed as potential strongly lensed companion images using a similar consistency test. 

It is also proposed that a sharp transition in the inferred source intrinsic mass distribution at high mass values could single out strongly lensed images \cite{Ng2018,Broadhurst_pop}. This mass distribution anomaly argument, however, must be made in reference to an expected GW source distribution. Currently, such source population models are subject to considerable uncertainties.

The above strategies share two other drawbacks: (1) without prior knowledge of the lensed source parameters, all pairs of cataloged events must be searched over to find strongly lensed candidates. As detector sensitivity improves and next-generation detectors start observing, the computational cost of such analysis will surge with the increased number of detected events; (2) it is also required that more than one lensed images are detected. If all but one of the images are missed, the methods described above cannot ascertain if a GW image is strongly lensed. 

For the above reasons, an intrinsic waveform distortion in a lensed image can be both a more definitive and efficient indicator of strong lensing. If such a lensed image is found, its estimated parameters help narrowing down the search space in the more general pair-wise search method mentioned above. An example is the frequency-specific GW diffraction in weak lensing \cite{Nakamura1998,Takahashi2003}. Diffraction signature was searched for in current detected events, but it has yet to be found \cite{Hannuksela2019}. It is also predicted that, for GWs within the frequency range of LIGO, diffraction becomes important when the lens mass ranges from $1\sim100~M_\odot$ \cite{Diego,Meena,Cheung}. Strong lensing by such small lenses requires a small impact parameter, which places stringent requirement on the alignment of the GW source, the lens and the observer \citep[see, e.g.,][]{Schneider1992}. Consequently, we expect such events to be rare.

Though diffraction is negligible for strong lensing (within a similar frequency range as LIGO),  waveform distortion does occur in the geometric-optics limit when an image originates from a saddle-point solution to the lens equation \citep[see, e.g.,][]{Schneider1992}. Such images are called type II images, their waveforms are the Hilbert transforms of the corresponding unlensed waveforms.  By contrast, waveforms of type I and type III images are identical to the unlensed waveform, up to a rescaling --- and, for type III images, a sign flip.  In \cite{Dai2017}, it is pointed out that type II images are degenerate with type I images with an azimuthal angle shift of $\pi/4$, if only the dominant (2,$\pm$2) modes are considered. For highly eccentric orbits, this degeneracy is partially lifted. For quasi-circular binaries, the degeneracy is also broken if higher multipoles are considered. Recently, Ref.~\cite{Esquiaga2020} has systematically examined the type II images of a wide range of GW sources, including the effects of spin precession and orbital eccentricity. It was found that the type I/II waveform difference is still small upon tuning the azimuthal angle, the polarization angle and relative phases between GW modes. In this paper, we build upon the work in Ref.~\cite{Esquiaga2020} by exploring whether such type II images can be distinguished from regular images despite the small waveform mismatch.

For third-generation detectors, such as the LIGO Voyager\cite{Voyager2020}, the Einstein Telescope\footnote{\texttt{http://www.et-gw.eu/}} \cite{ET2011} (ET) and LIGO Cosmic Explorer\footnote{\texttt{https://cosmicexplorer.org/}} \cite{CE2019} (CE) with current models, we expect to be able to detect a non-trivial number of such distinguishable type II events thanks to the expected high Signal-to-Noise Ratio (SNR).
 
This paper is organized as follows. In Section~\ref{section:theory} we review the geometric optics theory for GW lensing. In Section~\ref{section:mismatch}, we calculate the best-match overlap between type II and type I waveforms over a range of detector-frame binary mass, mass ratio and orbital inclinations. We briefly discuss the implication of waveform mismatch for detection triggering in the current LIGO pipeline framework. In Section~\ref{section:distinguish} we discuss the distinguishability of type II images in the high-SNR regime by comparing the log likelihoods under type I and type II image hypothesis. Based on this, we compute the fraction of distinguishable type II images. In Section~\ref{section:pop}, we incorporate population models on GW sources and lensing galaxies, and predict the expected number of events with distinguishable type II images for LIGO Voyager, ET and CE. We then discuss the results and draw the conclusion.

Throughout this work, we assume a $\Lambda$CDM universe with $(\Omega_M,\Omega_\Lambda)=(0.3,0.7)$ and a Hubble Constant of $H_0=70~\rm{km}~s^{-1}Mpc^{-1}$.

\section{Lens Theory and Image Type}
\label{section:theory}
The geometric optics treatment of gravitational lensing is thoroughly investigated and well established by many authors \citep[see, e.g.,][]{Schneider1992,Nakamura1998,NakamuraDeguchi1999}. In this section, we summarize and discuss scenarios where type II images are distinctive from type I counterparts. We closely follow the discussion in \cite{NakamuraDeguchi1999} and keep mostly consistent notations.

\subsection{Thin gravitational lens: geometric-optics limit}

We adopt the thin lens model, in which the line-of-sight lens dimension is much smaller than separations between the GW source, the lens and the observer. The source plane and the lens plane are defined by the GW source and the lens center, and both planes are perpendicular to the optical axis connecting the lens center and the observer. All the lens mass is projected onto the lens plane. Lensing deflection to GW paths occurs only on the lens plane. 

On each plane, the origin is established as its intersection with the optical axis. The source position has the dimensionless coordinate $\vec{y}= \vec{\eta}D_d/(r_*D_s)$ and the GW path intersects the lens plane at $\vec{x}=\vec{\xi}/r_*$. $\vec{\eta},\vec{\xi}$ are coordinates with physical units of length, $r_*$ is the lens' Einstein radius, while $D_d,D_s$ are the observer's angular diameter distance to the lens and the source. 

The amplitude of the observed image is then expressed as a Kirchhoff integral over the lens plane \citep[see also, e.g.,][]{Schneider1992,Takahashi2003},
\begin{equation}
\label{eqn:kirchhoffamp}
    F(\omega,\vec{y}) = \frac{\omega}{2\pi i}\int d^2\vec{x} e^{i\omega t\left(\vec{x},\vec{y}\right)}\;,
\end{equation}
where $\omega$ is the source-frame GW frequency and $t\left(\vec{x},\vec{y}\right)$ is the GW travel time difference between lensed paths and the unlensed path,
\begin{equation}
    t\left(\vec{x},\vec{y}\right)\approx \frac{1}{2}|\vec{x}-\vec{y}|^2+t_\Phi\;,
\end{equation}
where the first term accounts for the geometrical extra path length in the small deflection limit and the second term, $t_\Phi$, is the Shapiro time delay inside the lens' gravitational potential. In the geometric optics limit, only paths very close to the stationary points of $t$ contribute to the integral, and we may Taylor-expand the time delay around the $j$-th stationary point, 
\begin{equation}
    t\left(\vec{x},\vec{y}\right) = t\left(\vec{x}_j,\vec{y}\right)+\frac{1}{2}dx^adx^bT_{,ab}\left(\vec{x}_j,\vec{y}\right)+\mathcal{O}(|d\vec{x}|^3)\;,
\end{equation}
where $dx^a$ is a component of the two-dimensional vector $d\vec{x}\equiv \vec{x}-\vec{x}_j$ on the lens plane, and $|,|$ denotes partial derivatives and repeated upper and lower indices imply summation. The integral in Eq.\eqref{eqn:kirchhoffamp} then reduces to two Gaussian integrals after diagonalizing the time delay Jacobian, $T_{,ab}$. 

When ${\rm{det}}(T_{,ab})>0$, phase shifts from both the $\omega/i$ prefactor and the two Gaussian integrals depend on the sign of $\omega$. When ${\rm{Tr}}(T_{,ab})>0$, the phase factor is 1, giving type I images. When ${\rm{Tr}}(T_{,ab})<0$, the phase shift is $-{\rm{sgn}}(\omega)\pi$, where the function sgn returns the sign of its argument. This phase shift gives type III images, which differ from type I by an overall phase of $\pi$. (Note that $\pm \pi$ phases are equivalent.)

When ${\rm{det}}(T_{,ab})<0$, the two Gaussian integrals give opposite phase shifts regardless the sign of $\omega$, and no longer contribute to the overall phase of $F(\omega,\vec{y})$. The overall phase shift is then $-{\rm{sgn}}(\omega)\pi/2$, giving type II images which are equivalent to a Hilbert transform of type I images. 

\subsection{Gravitational waves from circular, non-spinning binaries}

For compact binaries, the complex GW strain at infinity can be  written as 
\begin{equation}
\label{hf}
    h= h_+ -ih_{\times} = \sum_{l,m}{}_{-2}Y_{lm}(\iota,\phi)h_{lm}\;,
\end{equation}
where the subscripts $+,\times$ denote plus and cross polarizations, and ${}_{-2}Y_{lm}(\iota,\phi)$ is the $s=-2$ spin-weighted spherical harmonics.  For non-spinning binaries with quasi-circular orbits, we choose the coordinate system such that the orbital angular momentum is along the $z$ axis. In this way, arguments $\iota$ and $\phi$ of the spin-weighted spherical harmonic also corresponds to the orbital inclination angle and the azimuthal angle, respectively.

Let us start out by considering $m \neq 0$ modes. The contribution from modes with $m=\pm m_0$, where $m_0$ is a positive integer, is
\begin{equation}
    \tilde{h}_{{\rm I},m_0} = \sum_{l}\sum_{m=\pm m_0}{}_{-2}Y_{lm}(\iota,\phi)\tilde{h}_{{\rm I},lm}\;,
\end{equation}
where the subscript I denotes the regular type I waveforms. The quantity $\tilde{h}_{{\rm I},lm}$ is the Fourier transform of $h_{lm}$ in Eq.\eqref{hf} via
\begin{equation}
\label{hlmf}
    \tilde{h}_{{\rm I},lm}(f)=\int^{\infty}_{-\infty}h_{lm}(t)e^{-2\pi i ft}dt\;.
\end{equation}
We note that $\phi$ appears only in the factor of $\exp(im\phi)$ in ${}_{-2}Y_{lm}(\iota,\phi)$. Furthermore, for non-spinning, circular binaries, with orbital angular momentum along the $z$ axis,  in frequency domain, $m>0$ modes only have negative frequency components and the inverse is true for $m<0$ modes. Therefore, the Hilbert transform of $\tilde{h}_{{\rm I},m_0}$, $\tilde{h}_{{\rm II},m_0}$ is written as
\begin{equation}
\begin{split}
    \tilde{h}_{{\rm II},m_0}(\iota,\phi)&=-i~{\rm sgn}(f)~\tilde{h}_{{\rm I},m_0}(\iota,\phi)\\
    & = \tilde{h}_{{\rm I},m_0}\left(\iota,\phi+\frac{\pi}{2m_0}\right)\;.
\end{split}
\end{equation}
Therefore, for each subset of GW modes with $m=\pm m_0$, the Hilbert transform is degenerate with an additional orbital azimuthal angle $\Delta \phi = \pi/(2m_0)$. For example, the required angle change is $\Delta\phi=\pi/4$, provided that only the $(l,\pm2)$ GW modes are considered. Modes with different $|m|$ require different angle changes to compensate for the Hilbert transform (e.g., the $(l,\pm3)$ modes require $\Delta\phi=\pi/6$). This difference in the compensation requirements breaks the degeneracy between Hilbert-transformed signals and orbital azimuthal angle change. 

\begin{figure*}[!htb]
    \centering
    \includegraphics[width=0.9\textwidth]{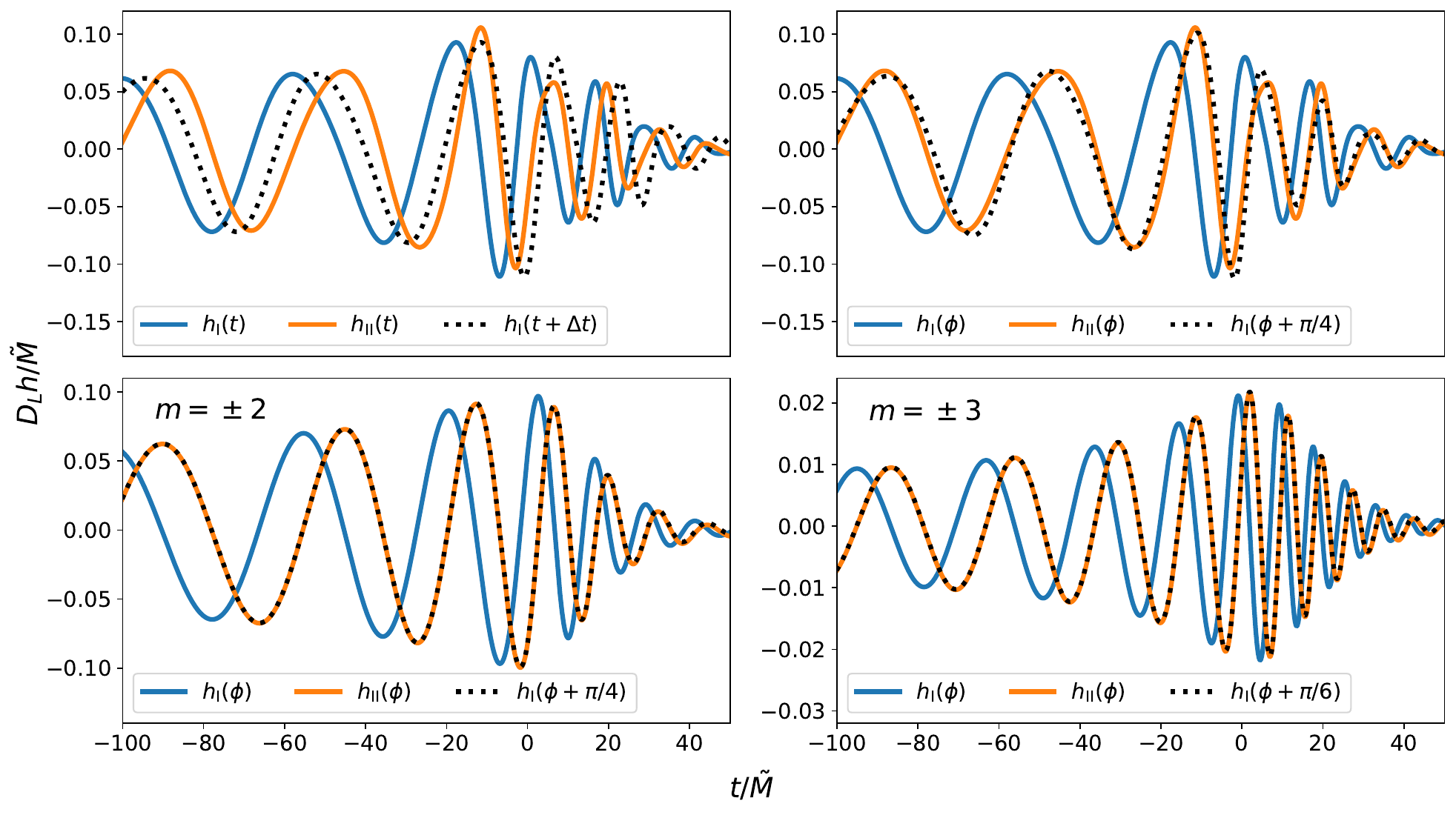}
    \caption{\small{type I/II \texttt{NRSur7dq4} surrogate model waveforms from a binary with $\tilde{M}=150~M_\odot,q=2.2,\iota=80~\deg$. The binary is non-spinning in a quasi-circular orbit. The black dotted line shows the type I waveform with a $\pi/4$ shift in the orbital azimuthal angle, and the shifted waveform is completely degenerate with the type II image. The orange dotted line shows the type II waveform, such that its peak overlaps with that of the type I waveform. We observe that the type I/II waveform offset cannot be compensated by a time shift.}}
    \label{fig:waveform}
\end{figure*}

Physically, $|m|\neq 2$ modes can be significant when the orbit is significantly eccentric \cite{Dai2017}. For binaries with significant mass ratios and inclination angles, the $(3,\pm3)$ modes become significant, breaking degeneracy. Figure~\ref{fig:waveform} is analogous to Figure 2 in \cite{Dai2017} and plots example type I/II waveforms from a binary with a detector-frame mass $\tilde{M}=150~M_\odot$, a mass ratio $q=2.2$ and an orbital inclination angle $\iota=80\deg$. The binary is non-spinning in a quasi-circular orbit, and all multipoles with $l\leq4$ are included. The top two panels show that the type II image is not degenerate with the type I image with an additional time shift. The bottom panels show that, when we include only the $m=\pm m_0$ modes, the Hilbert transform is degenerate with the original waveform with $\Delta\phi=\pi/(2m_0)$.

For $m=0$ modes, $h$ is independent from $\phi$, and one cannot recover its Hilbert transform via shifting $\phi$.  This in principle further breaks the degeneracy, although $m=0$ modes are generally weak for non-spinning binaries in circular orbits.  However, note that these are where the GW memory effects take place \cite{Favata,Favata2009,Hubner}. 

In this paper, we systematically explore GW sources which are non-spinning binary black holes in quasi-circular orbits. The distinguishable signature of type II images will be due to higher order GW modes, which is related to binary mass ratio, $q$, and orbital inclination, $\iota$. 

\section{Waveform Mismatch}
\label{section:mismatch}
In this section, we quantify the mismatch between type I/II waveforms for non-spinning binaries, in preparation for discussion on their distinguishability in the next section. We also discuss the implication of this mismatch for the GW signal veto process, namely, whether the mismatch leads to type II signal rejection in the current LIGO data analysis pipeline.

\subsection{Best-match Overlap}

In this section, we describe the procedure to compute the type I/II waveform difference over a large parameter space. We model only non-spinning binaries in quasi-circular orbits. Highly spinning binaries or those with highly eccentric orbits are expected to be fewer than the population we consider \citep[see, e.g.,][]{Zaldarriaga,search_ecc}. Since, the optical depth for type II images is also small, on the order of $10^{-3}\sim10^{-4}$ \cite{Oguri2018,DaiPop2017,Li2018}, we exclude these less frequent sources  from our analysis. 

For this source population, frequency-domain GW strain is given by the Fourier transform of Eq.~\eqref{hf},
\begin{equation}
\label{eqn:hHf}
     \tilde{h}_{\rm I}(f) = \sum_{l,m}{}_{-2}Y_{lm}(\iota,\phi)\frac{\tilde{H}_{{\rm I},lm}(\tilde{M},q,f)}{D_L}e^{-2\pi ift_0- i\Phi}\;,
\end{equation}
where $\tilde{H}_{{\rm I},lm}(\tilde{M},q,f)/D_L$ is equal to $\tilde{h}_{{\rm I},lm}(f)$ in Eq.\eqref{hlmf}, with the dependence on $D_L$ explicitly shown. The waveform is a function of the detector-frame mass (or equivalently, the redshifted mass), $\tilde{M} = (1+z) M$ (where $M$ is the intrinsic mass), the mass ratio, $q\equiv \tilde{M}_1/\tilde{M}_2\geq 1 ~(\tilde{M}_1+\tilde{M}_2=\tilde{M}$), and the luminosity distance, $D_L$. The polarization angle, $\Phi$, and signal time-of-arrival, $t_0$, add additional phase shifts to the signal. 

For any two waveforms, $\tilde{h}_1, \tilde{h}_2$, we define the overlap by
\begin{equation}
\label{overlap}
    {\rm overlap}= \frac{\mathfrak{Re}(\langle \tilde{h}_1|\tilde{h}_2\rangle)}{\sqrt{\langle \tilde{h}_1|\tilde{h}_1\rangle\langle \tilde{h}_2|\tilde{h}_2\rangle}}=1-\epsilon \;,
\end{equation}
where $\epsilon$ is the mismatch and $\langle\cdot\rangle$ denotes inner product given by
\begin{equation}
    \langle a |b\rangle =  \int ^{\infty}_{-\infty}  \frac{a^*(f)b(f)}{S_{n}(f)}df\;,
\label{innerproduct}
\end{equation}
where $S_n(f)$ is the two-sided noise power spectral density. By applying the optimal matched filter, the SNR of $\tilde{h}$, $\rho$, is given by $\sqrt{\langle \tilde{h}|\tilde{h}\rangle}$. 

Throughout this paper, we use Roman numeral subscripts to denote the image types and Arabic numeral subscripts to represent any individual waveform. We also adopt the simplifying assumption that both GW polarizations can be independently detected, i.e., the time-domain waveform is taken to be complex, as in Eq.~\ref{hf}. In Section~\ref{section:conclusion}, we discuss in more detail the validity of this assumption. 

To obtain highly accurate models for $\tilde{h}$, we adopt the time-domain Numerical Relativity surrogate waveform model, \texttt{NRSur7dq4}, \cite{sur2019} extracted through the Python package, \texttt{gwsurrogate} \cite{gwsur2018}. This surrogate model provides all $l\leq 4$ mode waveforms, $h_{{\rm I},lm}(t)$, through the inspiral, merger and ringdown phases.

To avoid spurious edge effects due to the finite-length of surrogate waveforms, we apply a time-domain kaiser window function from \texttt{numpy.kaiser} \cite{numpy} with $\beta=4$. The window is centered at the waveform amplitude peak to maximally preserve waveform features. The signal is zero-padded prior to the Fourier transform to ensure sufficiently smooth transformed waveform. 

To maximize the overlap, Ref.~\cite{Esquiaga2020} separately tunes the azimuthal angle, polarization angle, and the relative phases between each of the GW multipoles. We adopt a different approach by tuning the intrinsic parameters of the GW sources. We adopt a nested search method. We first generate a type II signal template, $\tilde{h}_{\rm{II},0}$, with $(\tilde{M}_0,q_0,\iota_0,\phi_0)$. Since the waveform amplitude scaling does not contribute to the overlap, we fix $D_L=3$ Gpc for all waveforms. We make a $(\tilde{M},q)$ grid, with the mass range centered on $\tilde{M}_0$ and mass ratio between 1 and 4 (the range of $q$ used to train the surrogate model). At each grid point, we construct the type I template and use the Python module \texttt{scipy.optimize.dual\_annealing} \cite{scipy} to find the $(\iota,\phi)$ that maximize the overlap between the type I template and the type II target. The spin-weighted spherical harmonics are computed using the Python package \texttt{spherical\_functions}\footnote{\texttt{https://github.com/moble/spherical\_functions}} and \texttt{quaternion}\footnote{\texttt{https://github.com/moble/quaternion}}. To implicitly maximize over $t_0$ and $\Phi$, we take the Fourier transform of the integrand in Eq. \eqref{overlap} and pick the element with the largest absolute value \citep[see, e.g.,][]{Moore2018}:
\begin{equation}
\label{eqn:max_eps}
\begin{split}
    (1-\epsilon)_{\rm{max}}&= {\rm{max }_{t_0}}\left|\frac{1}{a}\int_{-\infty}^\infty df \frac{\tilde{h}^*_{\rm I}(f)\tilde{h}_{\rm II}(f)}{S_n(f)}e^{-2\pi ift_0}\right|,\\
    a &= \sqrt{\langle \tilde{h}_{\rm I}|\tilde{h}_{\rm I}\rangle\langle \tilde{h}_{\rm II}|\tilde{h}_{\rm II}\rangle}\;.
\end{split}
\end{equation}

\begin{figure}
    \centering
    \includegraphics[width=.45\textwidth]{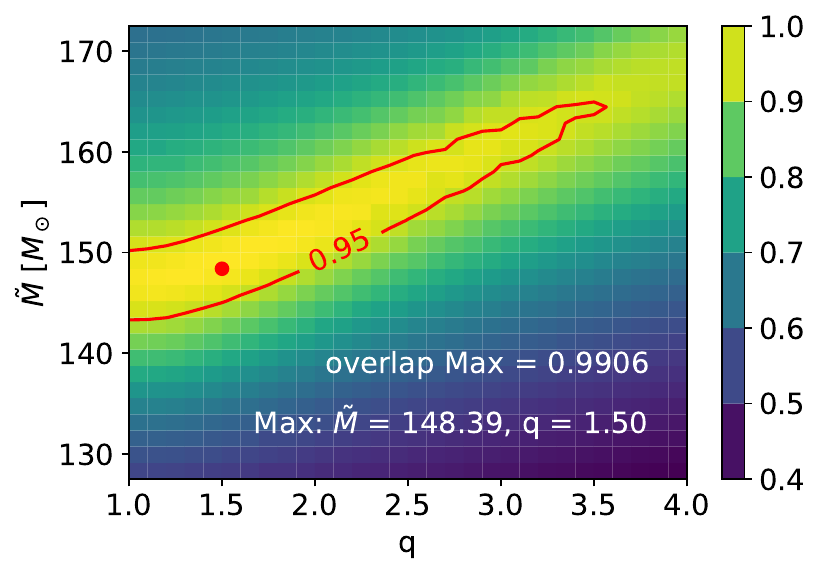}
    \caption{\small{Contour plot for maximized overlap for a type II waveform with $\tilde{M}=150~M_\odot,~q=1.7$ and $\iota=70~\deg$. Grid point with the maximum overlap is shown with the red dot at $\tilde{M}=148.39~M_\odot,~q=1.50$. 95\% overlap contour is shown in red.}}
    \label{fig:cf}
\end{figure}

Figure~\ref{fig:cf} shows an example maximization result contour plot for a type II signal with $\tilde{M}=150~M_\odot$, $q=1.7$ and $\iota = 70~\deg$ with CE noise curve. Due to the waveform mismatch, the best-match template has different parameter values from those of the true signal, with a maximal overlap of 99.06\%.

We calculate the overlap for sample points on the grid $X(\tilde{M})\bigotimes Y(q)\bigotimes Z(\iota)$, with
\begin{equation*}
\begin{split}
    X(\tilde{M})=\{&60,80,100,150,200,230,260,\\&300,400,500,600,700,800\}[M_\odot]\;,
\end{split}
\end{equation*}
$$Y(q)=\{1.2,1.7,2.2,2.7,3.2\}\;,$$ 
$$Z(\iota)=\{15,30,40,50,60,70,80\}[\deg]\;.$$
We then interpolate between the samples using the \texttt{scipy.interpolate} module \cite{scipy} to construct a function $\epsilon(\tilde{M},q,\iota)$. For non-spinning binaries, the interpolated function ensures that $90~\deg<\iota<180~\deg$ is symmetric to $0~\deg<\iota<90~\deg$. We perform the same analyses for CE, ET and LIGO Voyager with their respective noise power spectral density (PSD) \cite{VoyagerPSD,Voyager2020,ET2011,CE2019}.

\begin{figure}
    \centering
    \includegraphics[width=.9\columnwidth]{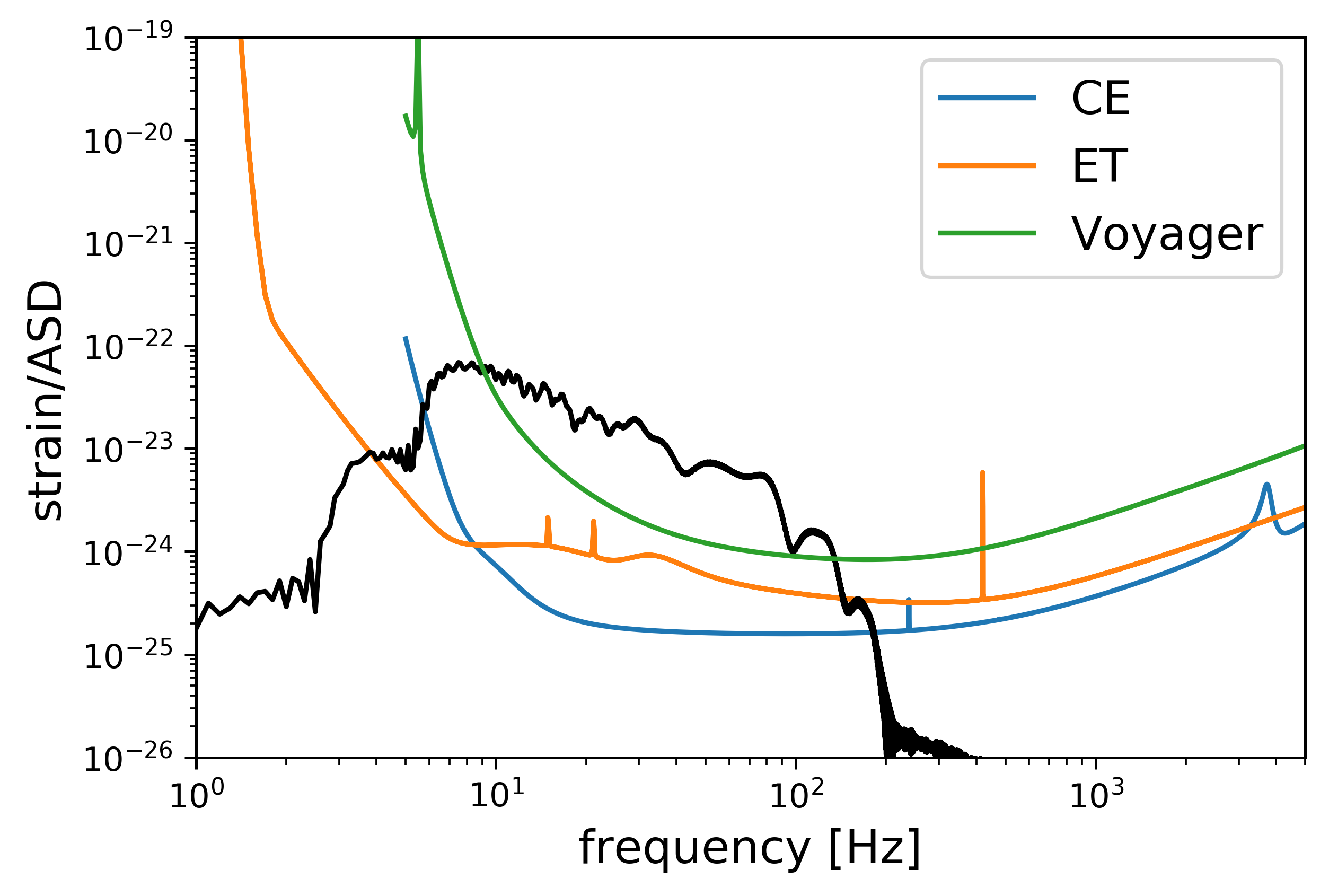}
    \caption{Positive frequency band waveform for a binary with $\tilde{M}=200~M_\odot$, $q = 2.2$, $\iota = 80~\deg$ and $D_L$ = 1 Gpc, plotted in black. The amplitude spectral densities (ASDs) for CE, ET and LIGO Voyager are plotted with colored traces. Note that ASDs for CE and LIGO Voyager are available starting from 3 Hz.}
    \label{fig:hf}
\end{figure}

Figure~\ref{fig:hf} shows the amplitude spectral density of CE, ET and LIGO Voyager \cite{VoyagerPSD,Voyager2020,ET2011,CE2019}, as well as the waveform of a binary with $\tilde{M}=200~M_\odot,q=2.2,\iota=80~\deg$ at $D_L=1$ Gpc as an example. The low-frequency amplitude loss of the surrogate waveform is due to the finite length of the \texttt{NRSur7dq4} waveforms. For less massive binaries, this effect results in significant loss of $\rho$, especially in the case of CE, where the low-frequency sensitivity degrades slower. 

To estimate how this $\rho$ loss affects the overlap values, we compute the maximum overlap for a $(\tilde{M}_0=60~M_\odot,q=3,\iota=80~\deg)$ binary with the CE PSD, filtering all frequency components below 30 Hz, where the loss of $\rho$ becomes significant. Compared with the unfiltered case ($(1-\epsilon)_{\rm{max}}=0.981)$, the overlap decreases only by $3.3\times10^{-3}$. Since $\tilde{M}=60~M_\odot$, $q=3$ and $\iota=80~\deg$ are roughly the smallest redshifted mass, largest mass ratio and inclination we consider, other binaries within our parameter space should have a smaller loss of the overlap. Considering the small size of the difference, we do not filter signals in subsequent analysis.

\begin{figure*}[!htb]
\centering
\includegraphics[width=\linewidth]{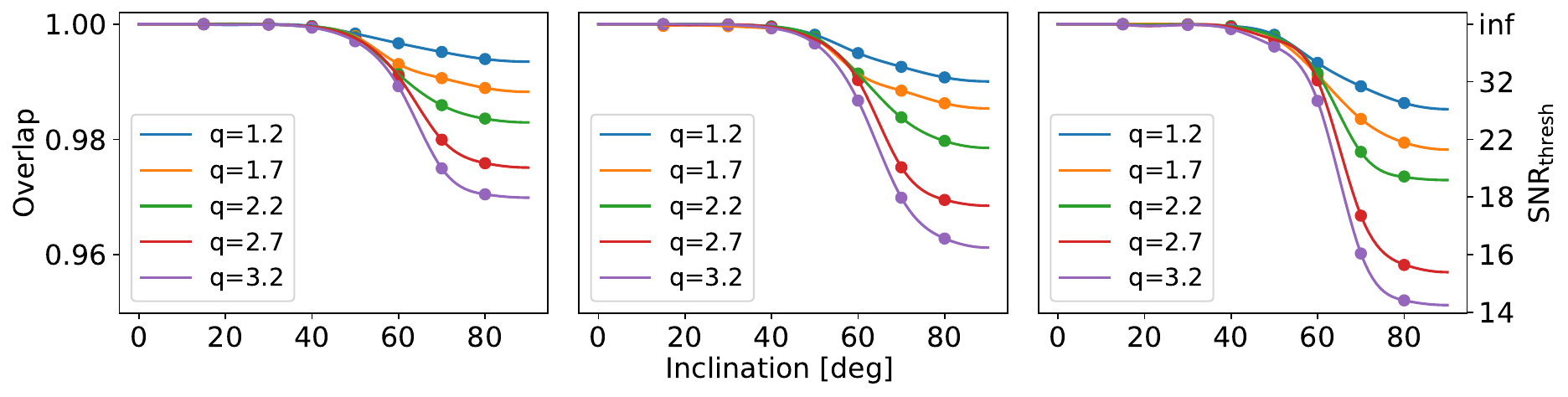}
\caption{\small{Overlap between type I and type II waveforms for $\tilde{M}=150~M_\odot$ at selected mass ratio values. The axis on the right shows the threshold $\rho$ to distinguish such type II images from type I counterparts by a log Bayes factor of 10, for the corresponding waveform overlap value. See Section~\ref{section:distinguish} for details. Panels from left to right are overlaps for CE, ET and LIGO Voyager respectively. In all panels, data points are shown with dots, and the interpolated overlap functions are shown in smooth curves.}}
\label{fig:overlap}
\end{figure*}

Figure~\ref{fig:overlap} shows the best-match overlap for GW waveforms with a redshifted mass of $150~M_\odot$ for the three GW detectors at selected mass ratio values. Maximization data points are shown with solid dots, and the interpolation functions are shown as smooth curves. The right axis shows the required $\rho$ to distinguish type I/II waveforms with a log Bayes factor of 10 at the corresponding overlap values on the left axis. See discussion in Section~\ref{section:distinguish}. Consistent with intuition, the best-match overlap is the lowest for high mass-ratio signals at large inclinations. Over our parameter space, the mismatch value for such signals is typically on the order of 2\%. We note that the same type II waveforms have the largest mismatch with type I waveforms in LIGO Voyager, as the LIGO Voyager PSD emphasizes high-frequency waveform components, where the Hilbert transform effect is more pronounced. 

\subsection{Signal Veto}

An ensuing concern from the mismatch is whether the difference in waveforms could lead to type II signals vetoed or assigned a lower significance value during observing runs. For the current GW data analysis pipelines, once a threshold $\rho$ is reached, the data typically go through a \chisq~veto test to screen out spurious signals. In this section, we calculate the non-central parameter in the \chisq~statistic distribution from using type I templates to match type II signals. 

The \chisq~veto was described in detail in \cite{Allen2005}. This test characterizes the distribution of $\rho$ over frequency bins and vetoes detector ``glitches'', or loud bursts of non-Gaussian noise that might have a high $\rho$, but have a frequency distribution very different from that of a genuine GW signal. 

Suppose the best-match template to the signal, $\tilde{n}+\tilde{h}_0$ is $\tilde{h}_T$, where $\tilde{n}$ is noise and $\tilde{h}_0$ is the embedded waveform. We divide the detector sensitive frequency range into $p$ disjoint sub-bands, $\Delta f_j$, such that the template $\rho$ in each bin is $1/p$ of its total $\rho$,
\begin{equation}
    \rho_{T,j} = \int_{-\Delta f_j,\Delta f_j}\frac{|\tilde{h}_T|^2}{S_n(f)}df = \frac{1}{p}\int_{_\infty}^\infty\frac{|\tilde{h}_T|^2}{S_n(f)}df \;.
\end{equation}

We then calculate the signal $\rho$ in each frequency bin as:
\begin{equation}
    s_j\equiv \frac{1}{\sqrt{\langle \tilde{h}_T|\tilde{h}_T\rangle}}\int_{-\Delta f_j,\Delta f_j}\frac{\tilde{h}_T^*(\tilde{n}+\tilde{h}_0)}{S_n(f)}df\;.
\end{equation}
We then define the \chisq statistic as
\begin{equation}
    \chi^2 \equiv p\sum_{i=1}^p (s_i- s/p)^2,~s\equiv \sum_{j=1}^p s_j\;.
\end{equation}

In the case where the best-match template in the template bank does not exactly match the embedded waveform, the distribution of \chisq ~over many Gaussian noise realizations is a classical \chisq ~distribution with a non-central parameter,
\begin{equation}
    \langle\chi^2\rangle = p-1+\kappa \langle s\rangle^2\;,
\end{equation}
where $\langle\cdot\rangle$ denotes the average over noise realizations. The factor, $\kappa$, in the non-central parameter is bound by 
\begin{equation}
    0<\kappa<\frac{1}{(1-\epsilon)^2}-1\approx 2\epsilon\;,
\end{equation}
where $\epsilon$ is the minimized mismatch between the template and the underlying waveform, as is defined in Eq.~\ref{eqn:max_eps}. The approximate equality is satisfied when $\epsilon\ll1$. This bound is agnostic of the specific waveform of the signal and templates. Consequently, the non-central parameter introduced by using type I templates on type II signals is smaller than $0.12\langle s\rangle^2$ in most cases, if we take the largest mismatch to be 6\%. If such a non-central parameter lies within the \chisq~threshold during detection, type II images are unlikely to be vetoed.

\subsection{type II Signal Recovery} 

There have been ongoing efforts to look for possible weaker (sub-threshold) strongly lensed counterparts of confirmed GW detections, assuming the latter being strongly lensed signals themselves \cite{Li:2019osa,McIsaac:2019use}. One method is to simulate lensed injections of a super-threshold GW event, then use a generic template bank to search for these injections through an injection run, and produce a targeted template bank for searching possible lensed counterparts of the target event by retaining only templates that can find the injections.

However, only type I lensed images have been considered for current searches. The question we would like to investigate is: Should type II lensed images be present in the data, would a type I template bank be able to find them? The answer to this question may be a crucial step for us to identify possible lensed GWs that we might have already detected but still not being discovered.

As a preliminary test to this question, we apply the search method detailed in \cite{Li:2019osa} to the high-mass-ratio compact binary coalescence event GW190814 \cite{GW190814}. Using the waveform approximant \texttt{IMRPhenomXPHM} \cite{Pratten2020}, we generate a set of simulated lensed injections for GW190814. They are then injected into real LIGO-Virgo data in two ways: (1) by treating them as type I images, and (2) by treating them as type II images, i.e. applying Hilbert transform to the waveform in the frequency domain as discussed previously. Through the GW CBC search pipeline GstLAL \cite{gstlal2017}, we apply the previously generated type I image target bank to search for these injections in both tests, and finally we compare the number of missed injections to roughly estimate the effectiveness of a type I image bank to look for type II images.

As discussed in \cite{gstlal2017, Li:2019osa}, each GW candidate found in the GstLAL search will be assigned a log likelihood ratio statistic $\ln \mathcal{L}$ to measure its significance. The False-Alarm-Rate (FAR) can be calculated accordingly, which corresponds to how often noise will produce a trigger with ranking statistic $\ln\mathcal{L}$ larger or equal to the ranking statistic $\ln\mathcal{L}^{*}$ of the trigger we are considering. In the search, an injection is said to be found if its FAR passes the usual threshold $1$ in $30$ days, as usual for a generic gravitational-wave search \cite{GWTC1:2018mvr}.

In both tests, we have injected a total of $8036$ simulated lensed injections. We assume that the injected events are registered by both detectors in the aLIGO network and the Virgo detector. In test {\bf A}, we apply a type I image bank to look for injected type II images. For test {\bf B}, 
we use the same image bank and look for the type I counterpart of the injections in test {\bf A}. In test {\bf A}, $638$ injections are missed, whereas in test {\bf B} the missed count is $536$. We observe that the number of missed injections increases when the injections were treated as type II images, indicating that the current search method for sub-threshold lensed GWs may be missing possible type II lensed signals.

However, it is important to remark that our current results are inconclusive since: (1) we have only been testing on one particular GW event, and (2) the exact reason for the extra number of injections to be missed are yet to be investigated. Nevertheless, our results indicate there could be improvements to the current search method for sub-threshold lensed GW signals, and further investigation will be done as future work.

\section{Distinguishing type II Events}
\label{section:distinguish}
While we have systematically examined the type I/II waveform mismatch, whether it enables us to distinguish type II images in actual GW experiments deserves further discussion. In this section, we use the waveform overlap and quantify the fraction of strongly lensed GW sources that have distinguishable type II images.

\newcommand{\data}{\vec{d}}
\newcommand{\para}{\vec{\theta}}
\subsection{Bayes Factor}
Using a Bayesian model (or equivalently hypothesis) selection framework, we quantify the distinguishability between a type I image and a type II image by computing the Bayes factor $\mathcal{B}$, which is the ratio of the probability  of observing the data $\data$ under the hypothesis that the signal is of type II over that under the hypothesis that the signal is of type I, namely
\begin{equation}
\label{Eq:Bayes Factor}
\begin{aligned}
	\mathcal{B} & = \frac{p(\data{}|\;\mbox{type II image})}{p(\data{}|\;\mbox{type I image})} \\
	& = \frac{\int d\para{}\; \mathcal{L}(\para{}|\;\mbox{type II image}) \pi(\para{}|\;\mbox{type II image})}{\int d\para{}\; \mathcal{L}(\para{}|\;\mbox{type I image}) \pi(\para{}|\;\mbox{type I image})},
\end{aligned}
\end{equation}
where $\mathcal{L}(\para{})$ is the (Whittle) likelihood as a function of the waveform parameters $\para{}$, and $\pi(\para{})$ is the prior distribution, which is different under the two hypotheses. The log likelihood function, up to a normalization constant, is given by
\begin{equation}
\begin{aligned}
	\ln \mathcal{L}_{i} (\para{}) & \propto - \frac{1}{2} \langle d - h_{i}(\para) | d - h_{i}(\para) \rangle \\
	& \propto - \frac{1}{2} \langle d | d \rangle + \langle d | h_{i} \rangle - \frac{1}{2} \langle h_{i} | h_{i} \rangle,
\end{aligned}
\end{equation}
where the subscript $i = {\rm I},{\rm II}$ denotes the assumed image type.
In an actual inference analysis, we do not know a priori the `true' waveform parameters. Therefore, we usually evaluate the integrals in Eq.\eqref{Eq:Bayes Factor} using a sampling algorithm that explores the parameter space spanned by $\para$ stochastically.

\begin{figure}
\centering
\includegraphics[width=\columnwidth]{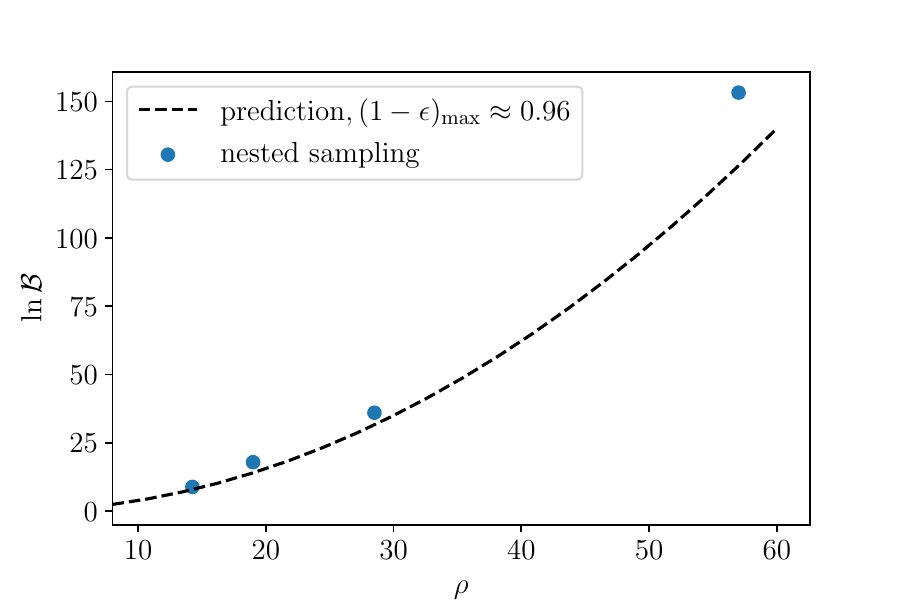}
\caption{\label{Fig:Log Bayes Factor Scaling}The log Bayes factor $\ln \mathcal{B}$ as a function of the SNR $\rho$ of the injections with different luminosity distances and fixed mismatch $\epsilon$, computed using Eq. \eqref{Eq:Bayes Factor} with nested sampling and Eq. \eqref{Eq:Bayes Factor Approximation}. We see that the simulation results roughly follow the expected quadratic scaling with the SNR.}
\end{figure}

Still, we can give an analytical approximate of the Bayes factor for distinguishing a type II image from a type I image using only the SNR $\rho$ and the mismatch $\epsilon$ we calculated in Sec.~\ref{section:mismatch}.
Following the treatment in Refs. \cite{Cornish:2011ys, Vallisneri:2012qq, DelPozzo:2014cla}, with the Laplace approximation we can write the log Bayes factor as
\begin{equation}
\label{Eq:Laplace approximation}
\begin{aligned}
	\ln \mathcal{B} \approx	& \ln \left[ \frac{\mathcal{L}_{\rm II}(\para_{\rm MLE})}{\mathcal{L}_{\rm I}(\para_{\rm MLE})} \right] + \ln \left( \frac{\sigma^{\rm posterior}_{\rm II}}{\sigma^{\rm posterior}_{\rm I}} \right),
	\end{aligned}
\end{equation}
where $\sigma^{\rm posterior}_{i}$ is the posterior (uncertainty) volume assuming that the lensed GW is of type-$i$.
The log likelihood ratio in Eq.\eqref{Eq:Laplace approximation} can be shown \cite{DelPozzo:2014cla}, in the high SNR limit, to scale as 
\begin{equation}
	\ln \left[ \frac{\mathcal{L}_{\rm II}(\para_{\rm MLE})}{\mathcal{L}_{\rm I}(\para_{\rm MLE})} \right] \approx  \epsilon \; \rho^2,
\end{equation}
when the (minimized) mismatch $\epsilon \ll 1$.
If we ignore the correlation between the parameters, we can estimate the posterior volume $\sigma^{\rm posterior}_{i}$  roughly as
\begin{equation}
		\sigma^{\rm posterior}_{i} \approx \prod_{j=1}^{N} \sqrt{2\pi} \Delta \theta^{j, \;{\rm posterior}}_{i},
\end{equation}
with $j$ loops over the $N$-dimensional vector $\para$ and $\Delta \theta^{j, \;{\rm posterior}}_{i}$ is the uncertainty of the 1D marginal posterior distribution for $\theta_{j}$ assuming that the image is of type-$i$. Note that here we assumed that identical prior was used when calculating the Bayes factor, except for the image type. The posterior volume ratio also scales with the the mismatch, actually. Since
$\langle h_{\rm I} | h_{\rm I} \rangle \approx (1 - \epsilon)^2 \; \langle h_{\rm II} | h_{\rm II} \rangle$
and that $\Delta \theta^{j, \;{\rm posterior}}_{i} \propto 1/\sqrt{\langle h_{i} | h_{i} \rangle}$,
therefore we have
\begin{equation}
	\ln \left( \frac{\sigma^{\rm posterior}_{\rm II}}{\sigma^{\rm posterior}_{\rm I}} \right) \approx - N\epsilon.
\end{equation}
Indeed, in the high SNR limit, the first term in Eq.~\eqref{Eq:Laplace approximation} is much larger than the second term as $N \sim 10$ and $\epsilon \ll 1$. Hence, we will ignore the contribution from the log posterior volume ratio in this paper.  
Therefore, we can estimate the log Bayes factor simply as~\footnote{Note that posterior volume also depends on dependences of $h_{\rm I}$ and $h_{\rm II}$ on $\theta_j$}
\begin{equation}
\label{Eq:Bayes Factor Approximation}
	\ln \mathcal{B} \approx \epsilon \; \rho^2.
\end{equation}
Figure \ref{Fig:Log Bayes Factor Scaling} shows the log Bayes factor as a function of the SNR $\rho$ using nested sampling with the help of the library \texttt{bilby} \cite{Ashton:2018jfp} and \texttt{dynesty} \cite{2020MNRAS.493.3132S} as in Eq. \eqref{Eq:Bayes Factor}, as well as its approximate using only the optimal SNR and the mismatch using Eq.\eqref{Eq:Bayes Factor Approximation}. Here we use the \texttt{IMRPhenomXHM} waveform model \cite{Garcia-Quiros:2020qpx} for both the simulated signals and the inference. All simulated signals have a redshifted total mass of $\tilde{M} = 150 M_{\odot}, q = 3.2$ viewing at an inclination angle of $\iota = 80 \deg$ with different luminosity distances to adjust the optimal SNR. We see that the simulation results roughly follow the expected quadratic scaling with the optimal SNR. Indeed, by performing a least-squares fit we found that the exponent is $2.02 \pm 0.07$.

Since $\ln\mathcal{B}$ scales as SNR$^2$, even a small type I/II mismatch could lead to significant $\ln\mathcal{B}$ in the high-SNR regime. For instance, for a mismatch of 3\%, an SNR of 20 would yield a log Bayes factor larger than 10, favoring the type II waveform hypothesis, thereby identifying this event as a strongly lensed image regardless whether other images are detected. The right axis in Figure~\ref{fig:overlap} shows the required SNR to produce $\ln\mathcal{B}=10$ for the corresponding type I/II overlap values. While such SNR is high for the current aLIGO, for third-generation detectors, it occurs frequently. For example, an equal-mass binary with a detector frame total mass of $100~M_\odot$ at $D_L=8$ Gpc has an $\rho=30$ for LIGO Voyager. The same source with $D_L=17$ Gpc has $\rho=131$ for CE.

\subsection{Threshold Inclination}

In this section, we find the range of parameters, $(\tilde{M},q,z_s$), where type II images can be distinguished via the log Bayes factor test. We choose $\ln \mathcal{B}_{\rm{thresh}}=10$ as the criterion for distinguishability.  

We begin by computing the distinguishable threshold inclination, $\iota$, for sources with certain redshifted mass, mass ratio and redshift. Since both $\rho$ and $\epsilon$ in Eq.\eqref{Eq:Bayes Factor Approximation} depend on $\iota$, it is more straightforward to first fix $\tilde{M},q$ and $\iota$ to obtain $\epsilon$, and then scale $\rho$ via $D_L$ to achieve the $\ln\mathcal{B}_{\rm thresh}$ condition. Inverting $D_{L,\rm{thresh}}(\tilde{M},q,\iota)$ yields $\iota_{\rm{thresh}}(\tilde{M},q,D_L(z_s))$, where $z_s$ is the GW source redshift.

To calculate $\rho$, we assume both GW polarizations can be detected, and the total amplitude is $\sqrt{h_+^2+h_{\times}^2}$. In the Discussion Section, we further discuss the justifications for this assumption in the context of third-generation GW detectors. However, the finite length of the surrogate model waveform can lead to significant loss in $\rho$, even though the effect on waveform overlap is negligible, as demonstrated in Section~\ref{section:mismatch}. For a binary with $\tilde{M}=60~M_\odot,q=3$, approximately 15\% of $\rho$ is lost in the case of CE. For LIGO Voyager, the noise increase starts earlier and steeper towards lower frequencies; consequently, the $\rho$ loss for the same binary is only $\sim5\%$. To accurately estimate $\rho$, we supplement the surrogate model waveform with analytical inspiral stage waveform, whose amplitude scales as $f^{-7/6}$ \cite{Cutler1994}. The inspiral amplitude is matched to the surrogate waveform amplitude at $0.5~f_{\rm{ISCO}}$, where $f_{\rm{ISCO}}$ is the Innermost Stable Circular Orbit frequency, approximated as \citep[see, e.g.,][]{Sesana2004}
\begin{equation}
    f_{\rm{ISCO}}=\frac{c^3}{6^{3/2}\pi G \tilde{M}}\;.
\end{equation}
We note that, by compensating for the lost $\rho$, our result is optimistic in estimating the distinguishability; while the early inspiral phase contributes significantly to $\rho$, the type I/II waveform mismatch is less pronounced. For the same $\rho$, the compensated inspiral waveform does not offer as much information as the higher frequency GW phases for distinguishing type II images. Nonetheless, this overestimate is significant only for systems towards the low mass limit, where the expected detectable number of events is low due to the small $\rho$. 

The mismatch $\epsilon$ is available from the interpolation function in Section~\ref{section:mismatch}. We do not consider binaries with best-match overlap larger than 0.999, i.e., we consider such mismatch a result of systematic errors and does not reflect actual waveform difference. As discussed in Section~\ref{section:mismatch}, the truncated surrogate waveform leads to errors in the best-match overlap, though for high mass systems, the error will be much smaller than $3.3\times10^{-3}$ for the $\tilde{M}=60~M_\odot,q=3,\iota=80~\deg$ example binary. For computational cost concerns, we also limit the grid density in the nested maximization process. If the actual best-match binary is not on the grid points, the maximization result will deviate from the true value, and the size of the deviation depends on the distance between the true best-match and its closest grid point. Aside from systematic errors in the waveform and overlap optimization process, interpolation for $\epsilon$ also introduces errors. In particular, the cubic spline fit may introduce spurious trace curves to guarantee smoothness when connecting the limited number of samples. Especially in the case of CE, $\rho$ can be very large, thus exaggerating the physical significance of such a small mismatch. The exact value of this threshold is tuned to exclude spurious interpolation function results. In the next subsection, we discuss our choice of the upper limit value for the best-match overlap and assess the impact of this mismatch resolution in the next subsection. 

\begin{figure*}[!htb]
\includegraphics[width=0.8\textwidth]{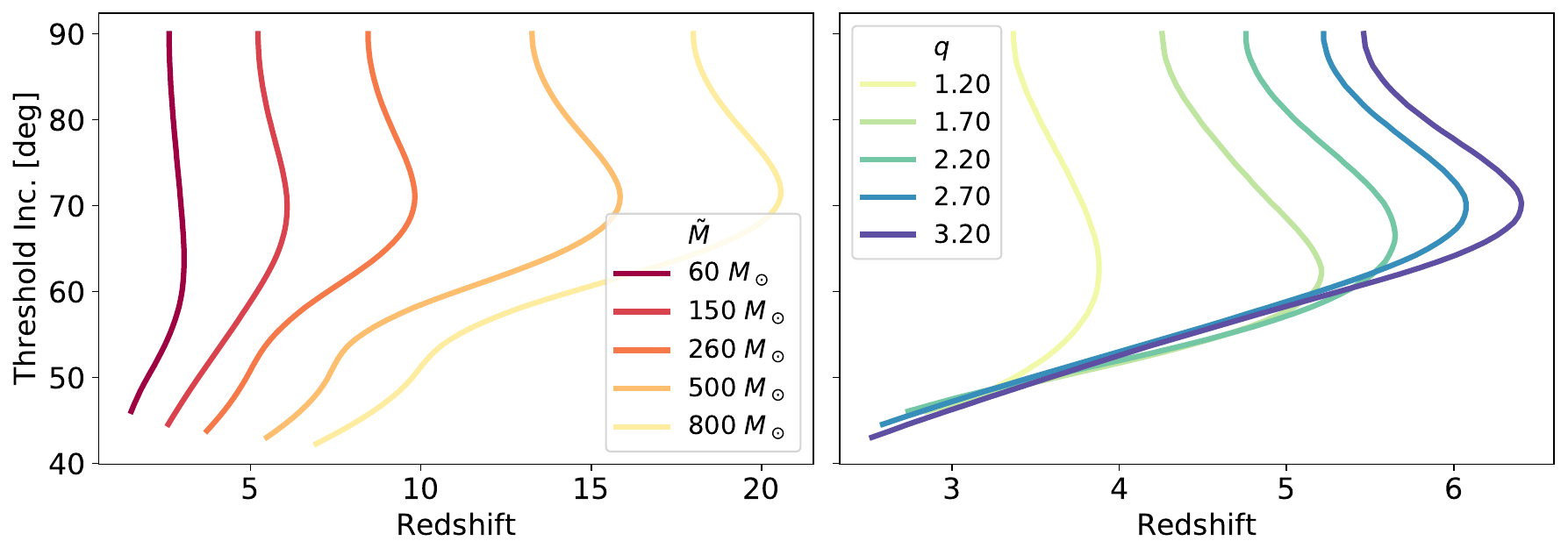}
\caption{\small{Inclination threshold curves for distinguishable type II sources as a function of redshift assuming CE sensitivity. \textit{Left:} inclination threshold curves for fixed mass ratio $q=2.67$ at selected redshifted mass values. \textit{Right:} inclination threshold curves for binaries with redshifted mass $\tilde{M}=150~M_\odot$ with selected mass ratio values. The curve-crossing at low inclination values are due to systematic errors; see text for discussion. }}
\label{fig:inc_thresh}
\end{figure*}

Figure~\ref{fig:inc_thresh} shows the threshold inclination as a function of source redshift assuming CE sensitivity. The left panel shows threshold inclination with fixed redshifted mass $\tilde{M}=150~M_\odot$. We observe that the mass ratio becomes an increasingly important factor at high inclinations. At low redshift, the threshold inclination is constrained primarily by the mismatch $\epsilon$; at higher redshift (e.g., $z_s\sim4.2$ for $q=1.73$), the high inclination regions start to be excluded despite the large mismatch value, as $\rho$ becomes too small. Beyond a certain redshift (e.g., $z_s\sim5.2$ for $q=1.73$), no combination of $\rho$ and $\epsilon$ meets the $\ln\mathcal{B}_{\rm{threshold}}$ condition, and no more type II images can be distinguishable. The right panel shows similar threshold cures fixing the mass ratio to be 2.67. We observe a similar curve shape, although lighter binaries have smaller $\rho$ and consequently a larger threshold inclination. 

\subsection{Distinguishable Image Fraction}

\begin{figure*}[!htb]
\includegraphics[width=0.8\textwidth]{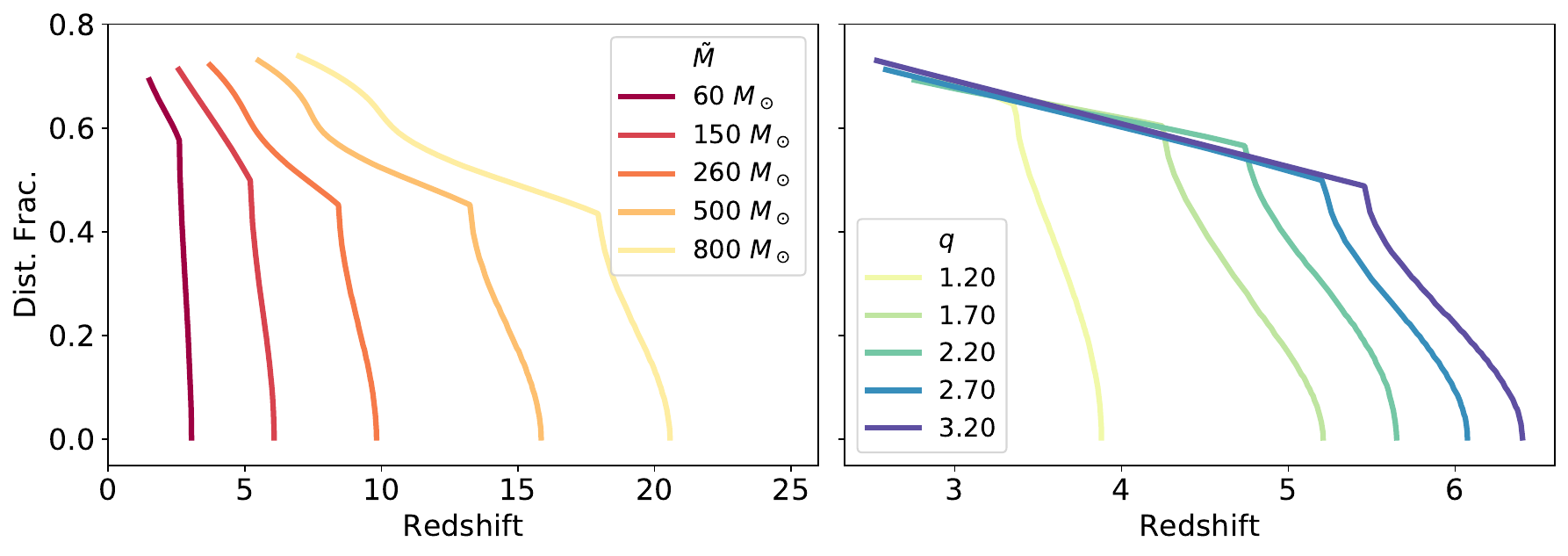}
\caption{\small{The fraction of distinguishable type II images as a function of redshift for CE sensitivity. \textit{Right:} distinguishable fraction, ${\rm{fr}}(\tilde{M},q,z_s)$, for constant $\tilde{M}=150~M_\odot$. \textit{Left:} distinguishable fraction for constant $q=2.67$. The cusps in the fraction correspond to the exclusion of high-inclination binaries with sub-threshold $\rho$. The fraction curves directly correspond to the threshold inclination curves in Figure~\ref{fig:inc_thresh}.}}
\label{fig:frac_z}
\end{figure*}

\begin{figure*}[!htb]
\includegraphics[width=\linewidth]{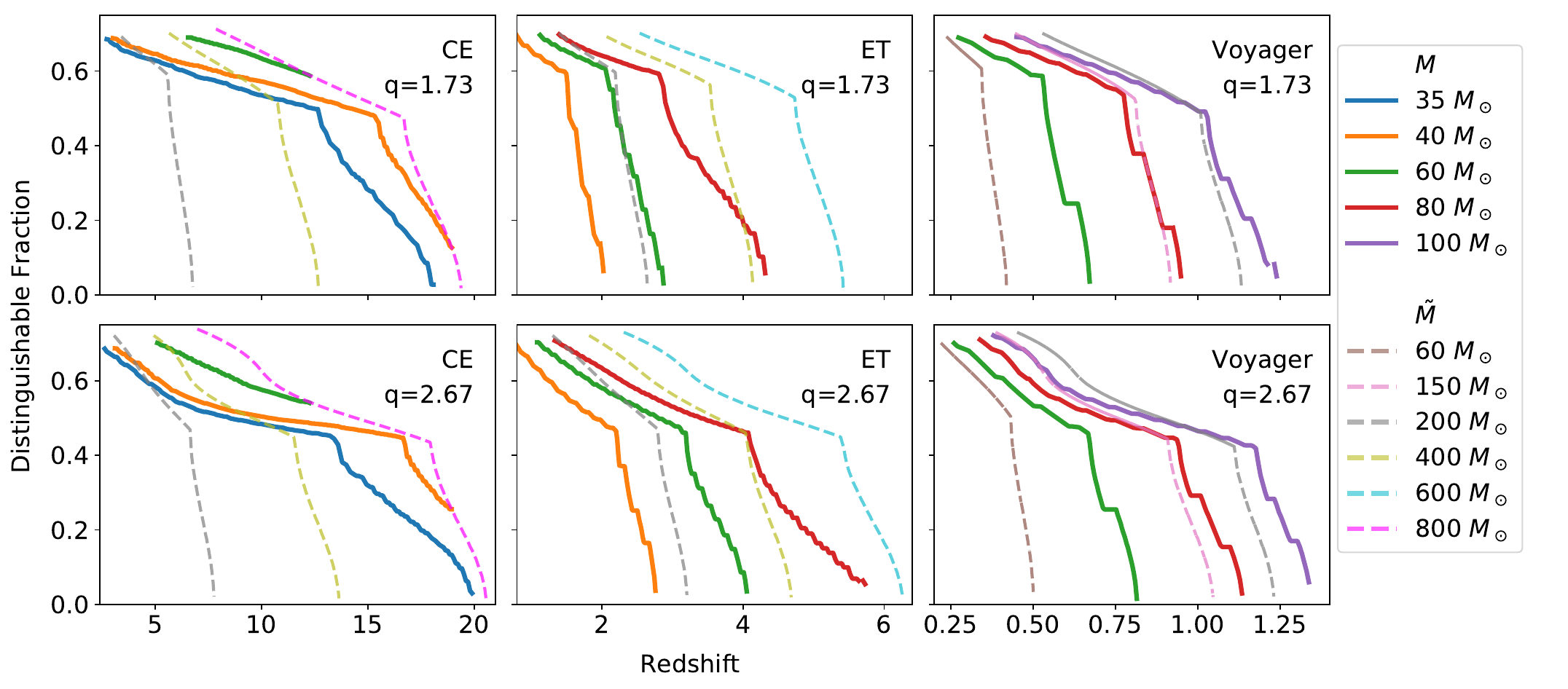}
\caption{\small{The fraction of distinguishable type II images as a function of redshift. \textit{Top Row:}} GW sources have constant mass ratio of 1.73. \textit{Bottom Row:} GW sources have constant mass ratio of 2.67. The \textit{Left, Middle,} and \textit{Right} columns show distinguishable fractions assuming CE, ET and LIGO Voyager sensitivity, respectively. The fractions for fixed redshifted mass, ${\rm{fr}}(\tilde{M},q,z_s)$, are shown in dashed lines, and those for fixed apparent mass, ${\rm{fr}}_{\rm{app}}(M,q,z_s)$, are shown in solid lines. The mass values are shown in the legend. Note that the fraction curves are jagged due to interpolation errors and limited data density.}
\label{fig:M_int}
\end{figure*}

From the threshold inclination, we can further calculate the fraction of GW sources with distinguishable type II images, ${\rm{fr}}(\tilde{M},q,z_s)$. For simplicity, we assume that GW sources and type II images are isotropically distributed, therefore the fraction of distinguishable type II images scales as the area of the celestial sphere within the $\iota$ threshold limits. The differential fraction is then proportional to $\sin\iota$. Figure~\ref{fig:frac_z} shows the distinguishable fraction of type II sources for the same binaries as in Figure~\ref{fig:inc_thresh}. The cusps mark the redshift when high inclination regions start to be excluded due to smaller $\rho$. We observe that, for CE, large fractions of sources with type II images can be distinguished via the log Bayes factor test out to high redshift. Similar plots for ET and LIGO Voyager are shown in Figure~\ref{fig:M_int} as dashed lines. 

We have so far considered only the redshifted mass (detector-frame mass), $\tilde{M}$, as it is the direct input to the surrogate model, which assumes an asymptotically flat and stationary universe. The ``apparent'' total mass of the binary, $M$, is related to the redshifted mass by $M=\tilde{M}/(1+z_s)$. Due to lensing magnification, this inferred ``apparent'' total mass could be larger or smaller than the actual GW source total mass. We discuss magnification effects in Section~\ref{section:pop}. Therefore, the fraction of distinguishable type II sources with apparent mass $M$, mass ratio $q$ at redshift $z_s$ is given by,
\begin{equation}
    {\rm{fr}}_{\rm{app}}(M,q,z_s) = {\rm{fr}}(M(1+z_s),q,z_s)\;.
\end{equation}

Figure~\ref{fig:M_int} shows ${\rm{fr}}_{\rm{app}}(M,q,z_s)$ for selected apparent mass values in solid traces. The top row shows the fractions with a fixed mass ratio of 1.73, and the bottom row shows similar plots with mass ratio fixed at 2.67. The left, middle and right columns show results for CE, ET and LIGO Voyager, respectively. The distinguishable fractions for fixed redshifted mass, ${\rm{fr}}(\tilde{M},q,z_s)$, are plotted for reference in dashed lines. The exact mass values are specified in the legend. 

As expected, the resulting traces show similar trends and features as in Figure~\ref{fig:frac_z}: at lower redshifts, the distinguishable fraction decreases with $\rho$. It then undergoes a cusp where the high inclination regions start to be excluded before continuing to decrease. The trace is jagged due to the finite spacing of the interpolation data points, rather than any physical jumps in the fraction. 

In most cases, there is a significant fraction of GW sources with distinguishable type II images via the log Bayes factor test. As Figure~\ref{fig:overlap} suggests, the mismatch value is not drastically different across the three detectors with different noise curve shapes. The redshift reach is rather primarily determined by $\rho$, related to the overall sensitivity level of different detectors. For example, for type II images with apparent mass $\tilde{M}=60~M_\odot$ and mass ratio $q=1.76$, 60\% can be distinguished in CE out to $z_s\sim12.5$. For ET, 60\% of the same population can be distinguished out to $z_s\sim2$. Due to the lower sensitivity of LIGO Voyager, a similar fraction of such type II images can be identified only out to $z_s\sim0.5$. However, for type II images with a higher apparent mass of $100~M_\odot$, 50\% can still be registered out to $z_s\sim1$.

\begin{figure}
\centering
\includegraphics[width=\linewidth]{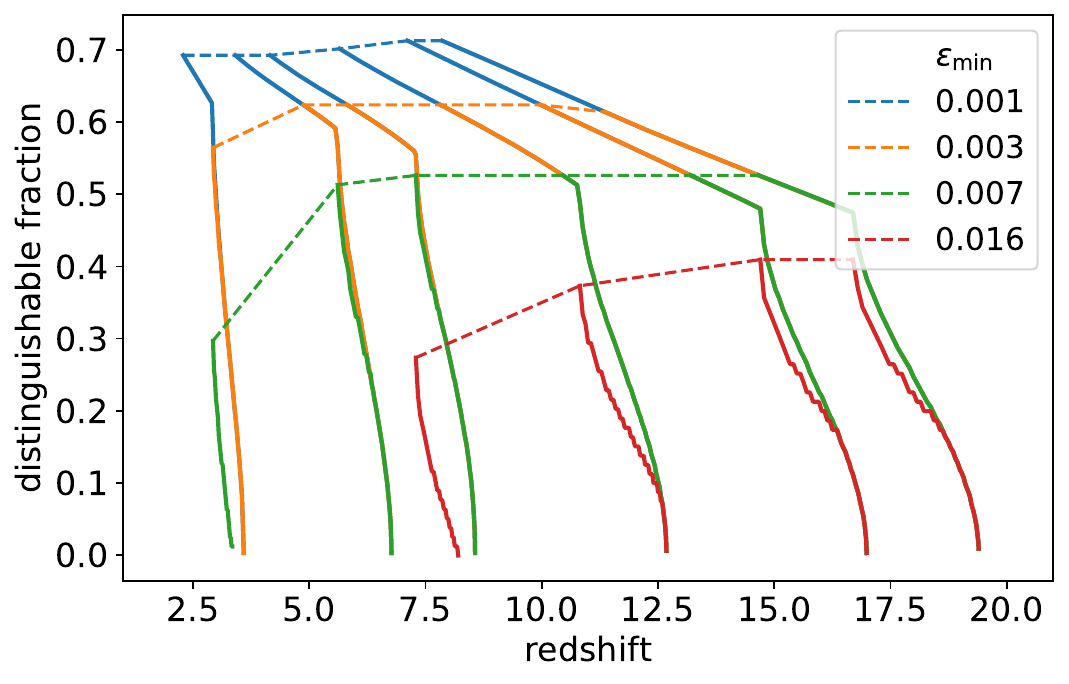}
\caption{\small{Distinguishable fractions of sources with type II images assuming a minimum resolvable mismatch of 0.001, 0.003, 0.007 and 0.016 with CE sensitivity. The solid lines show the distinguishable fractions for selected redshifted mass. The mass ratio is fixed to be 1.73. From left to right, the traces correspond to a redshifted mass of $100~M_\odot,~200~M_\odot,~260~M_\odot,~400~M_\odot,~600~M_\odot,$ and $800~M_\odot$. Dashed lines show the maximum fraction for this range of redshifted mass. For a mismatch resolution of 0.016 (or 0.984 overlap), fraction traces for type II images with redshifted mass $\tilde{M}=100,200~M_\odot$ (the leftmost two traces) are absent, since the resolution is larger than the maximum possible waveform mismatch for such GW sources.}}
\label{fig:eps_thresh}
\end{figure}

Finally, we assess the impact of mismatch resolution. Throughout this paper, we adopt a minimum mismatch value of $\epsilon=0.001$. Figure~\ref{fig:eps_thresh} shows the changes in the distinguishable fraction of GW sources with type II images for CE. The mass ratio is fixed to be $q=1.73$, and the solid lines from left to right represent $\tilde{M}=100~M_\odot,~200~M_\odot,~260~M_\odot,~400~M_\odot,~600~M_\odot,$ and $800~M_\odot$. The dashed horizontal traces show the largest distinguishable fraction as a function of the redshifted mass. 

As the waveform mismatch resolution becomes coarser, the distinguishable fraction decreases significantly. For instance, 70\% of all sources with type II images with redshifted mass $m=100~M_\odot$ and $q=1.73$ have distinguishable type II images out to redshift $z_s\sim2.5$ if a mismatch of 0.001 is resolvable, but the fraction drops to 30\% if the mismatch resolution is 0.007. With a mismatch resolution of 0.016, no such type II images are distinguishable. This critical role of the minimum resolvable mismatch suggests that the distinguishability of type II images does not solely depend on the SNR. In the era of third-generation GW detectors, the possible scientific output from GW detection events is not solely determined by the noise level. As is discussed, the waveform template bank density in the matched filtering search limits the waveform difference resolution. In addition, the detector calibration must also be sufficiently accurate such that we can be confident that the mismatch from the data reflects a real signal difference, rather than an instrument systematic error. Otherwise, we cannot take full advantage of the large SNR offered by exquisite detector sensitivity. Since these factors during the third-generation GW detector era are still subject to much uncertainty, to our knowledge, we have chosen $\epsilon=0.001$ as the fiducial value. The analysis should be refined as such information becomes available.

\section{Detectable Population}
\label{section:pop}
The distinguishable fraction calculations depend only on the waveform mismatch, and do not assume astrophysical estimates on GW source and lens distributions. In this section, we describe how results of image distinguishability can be combined with astrophysical models to give a more detailed prediction of the detectable GW events with type II images and those with distinguishable type II images. In the following subsections, we first offer an overview of the procedures for calculating the lensed population, followed by more detailed discussions on each ingredient.

\subsection{Overview}
We define a GW event as a particular binary black hole (BBH) merger with possibly multiple images due to strong lensing. For our calculation, a GW image is detectable if its single-detector $\rho\geq 8$; we defer the detector network scenario to future studies. A GW event has a distinguishable type II image if this image satisfy the log Bayes factor threshold. The differential detectable and distinguishable merger rate is given by 
\begin{widetext}
\begin{equation}
    \frac{\partial^3 \dot{N}_{\rm{II,det}}}{\partial M_\bullet\partial q\partial z_s} = \tau_{\rm II}(z_s)\frac{\partial^3 \dot{N}}{\partial M_\bullet\partial q\partial z_s} \int d\log_{10}\mu~\left[\frac{\partial P_{\rm II}(\mu,z_s)}{\partial \log_{10}\mu}\int_0^{\pi/2} d\iota~\sin\iota~\Theta \left(\sqrt{\mu}~\rho(M_\bullet,q,z_s,\iota)-8\right)\right] \;,
\label{eqn:NdetII}
\end{equation}

\begin{equation}
    \frac{\partial^3 \dot{N}_{\rm{II,dis}}}{\partial M_\bullet\partial q\partial z_s}=\tau_{\rm{II}}(z_s)\frac{\partial^3 \dot{N}}{\partial M_\bullet\partial q\partial z_s}\int d\log_{10} \mu~\frac{\partial P_{\rm{II}}(\mu,z_s)}{\partial\log_{10}\mu}{\rm{fr}}\left(\tilde{M},q,\tilde{z}_s\right)
    \;,
\label{eqn:Ns}
\end{equation}
\end{widetext}
where $\partial^3 \dot{N}/\partial M_\bullet \partial q\partial z_s$ is GW event rate per intrinsic binary mass, $M_\bullet$, mass ratio, $q$, and GW source redshift, $z_s$, measured in the observer frame. The weighting factor $\sin \iota$ comes from the assumption that BBH mergers are distributed evenly on the sky. $\Theta$ is the Heaviside function. Multiplying with the optical depth $\tau_{\rm{II}}(z_s)$, we obtain the rate of events with at least one type II images. The quantity $\partial P_{\rm{II}}(\mu,z_s)/\partial \log_{10} \mu$ describes the distribution of magnification $\mu$ for type II images for sources at $z_s$, normalized such that 
\begin{equation}
    \int d\log_{10}\mu~\frac{\partial P_{\rm{II}}(\mu,z_s)}{\partial \log_{10} \mu}  = 1\;.
\end{equation}
Due to magnification, the source appears to have the same redshifted mass, but the inferred luminosity distance is different. Therefore, 
\begin{equation}
\begin{split}
    \tilde{M}&=M_\bullet(1+z_s)\\
    D_L(\tilde{z}_s) &= D_L(z_s)/\sqrt{\mu}\;.
\end{split}
\end{equation}
The differential merger number per observer time is calculated as \citep[see also][]{DaiPop2017}
\begin{equation}
    \frac{\partial^3 \dot{N}}{\partial M_\bullet\partial q\partial z_s} = R_{\rm{mrg}}(M_\bullet,q,z_s)\frac{1}{1+z_s}\frac{dV_c}{dz_s}\;,
\end{equation}
where $dV_c/dz_s$ is the differential comoving volume. The $1/(1+z_s)$ factor accounts for the cosmological redshift and converts the source-frame merger rates into detector-frame merger rates. In Figure~\ref{Rmrg}, we plot this ``modified'' differential comoving volume and the total merger rates for reference. Since we have adopted the same population models, Figure~\ref{Rmrg} replicates Figure 1 in \cite{Li2018}.  

For fast calculation of $\rho(M_\bullet,q,z_s,\iota)$, we use the phenomenological model \texttt{IMRPhenomHM} \cite{PhenomHM}, called from the Python package \texttt{pycbc.waveform} \cite{pycbc}. 

We note that for a type I/II waveform mismatch of 6\%, the required $\rho$ to be distinguishable is approximately 13, larger than the threshold SNR of 8, and none of the GW sources we consider have a larger waveform mismatch. Therefore, we may assume that the \textit{distinguishable} images are all \textit{detectable}, leading to the omission of the Heaviside function in Eq.\eqref{eqn:Ns}. In addition, only 0.2\% of all sample lens systems have a brighter type II image than the type I image. Considering errors from the lens-equation solution algorithm and the small number of events with distinguishable type II images, we may assume that the events with detectable or distinguishable type II images will most certainly have a detectable type I companion image. 

In the following subsections, we compute the type II image optical depth and the magnification distribution. We then summarize procedures to calculate the total BBH merger rates. Detailed steps and adopted parameter values are presented in Appendix~\ref{app:rmerg}. We then make concrete detection population for CE, ET and LIGO Voyager and discuss results. 

\subsection{Optical Depth and Magnification}

To obtain $\tau_{\rm II}(z_s)$ and $\partial P_{\rm{II}}(\mu,z_s)/\partial \log_{10} \mu$, we perform a Monte Carlo simulation. We consider elliptical galaxies as lenses, as they are expected to be the predominant lensing objects \cite{Oguri2018}. While the lens geometry and properties are expected to be more complex and varied in nature, studies showed that a simple lens model, such as the singular isothermal sphere model, is sufficient to capture most of the results from more sophisticated hydrodynamic simulations of the Universe \cite{Robertson}. In this study, we adopt the slightly more generalized singular isothermal ellipsoid model following the examples of Refs.~\cite{Oguri2018,Li2018}. Refs.~\cite{Biesiada2014,Ding2015} adopt the singular isothermal sphere lens model and predict the detectable strongly lensed events for ET. For one of the BBH evolutionary scenarios they investigate, it is predicted that $57.2$ strongly lensed events can be detected out of the $2.08\times10^5$ total detectable BBH events per year, roughly a factor of three smaller than our prediction (see Table~\ref{tab:sum}). In the future, we can adapt our analysis using different lens models and systematically study the uncertainty in the strong-lensing population predictions.

We restrict the GW source redshift to $0.05\leq z_s \leq 7$. In the low-redshift limit, GW sources in our local universe ($z_s\ll 0.05$) is unlikely to be strongly lensed, since lensing rates are expected to be low, and there are not sufficiently many massive galaxies in between to compensate. We set the upper limit of the galaxy redshift to $z_s=7$, since such galaxies are faint and robust observational data is relatively scarce for developing a reliable phenomenological model of the mass function \cite{Finkelstein}. 

At each redshift, we generate samples of lenses, parameterized by surface velocity dispersion, $\sigma_v$, ellipticity, $e$, lens redshift, $z_l$, and the lens-plane angular coordinates of the lens, $\vec{\theta}=(x, y)$. For the number of lenses per unit $\sigma_v$ per comoving volume, $\Psi(\sigma_v,z_l)$, we first adopt the modified Schechter function \cite{Choi}, which is calibrated to observation on galaxies in the solar neighborhood,

\begin{equation}
    \Psi(\sigma_v,0) = \phi_*\left(\frac{\sigma_v}{\sigma_*}\right)^\alpha {\rm{exp}}\left[-\left(\frac{\sigma_v}{\sigma_*}\right)^\beta\right]\frac{\beta}{\sigma_v\Gamma\left(\alpha/\beta\right)}\;,
\label{eqn:psiz0}
\end{equation}
where $\phi_* = 8.0\times10^{-3}h^3~\rm{Mpc}^{-3}$, $\sigma_*=161~\rm{km/s}$, $\alpha=2.32$ and $\beta=2.67$. $h$ is the Hubble parameter. 

\begin{figure}
    \centering
    \includegraphics[width=0.9\columnwidth]{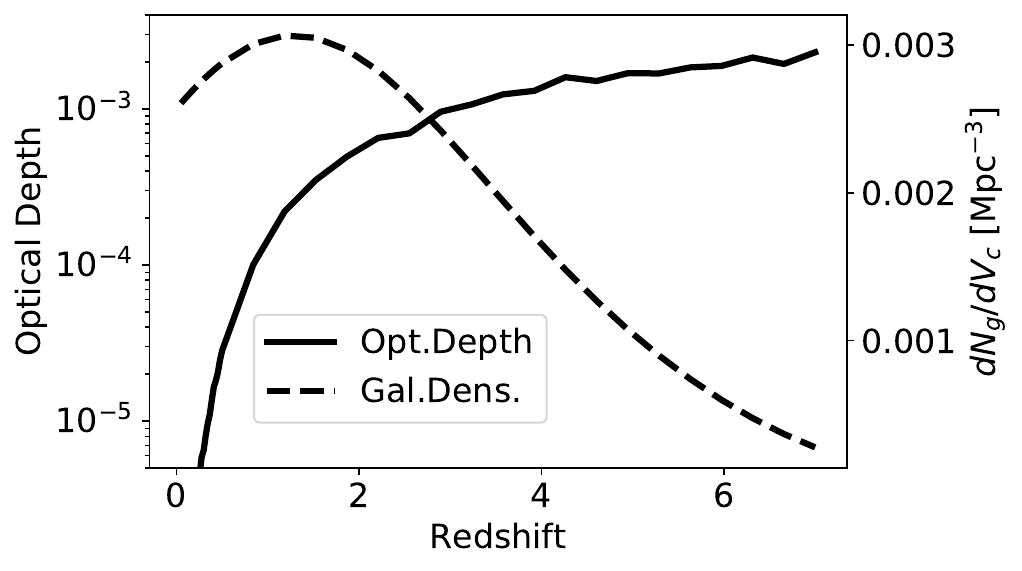}
    \caption{\small{\textit{Left} axis: optical depths, $\tau_{\rm II}(z_s)$, for GW sources with at least one type II image as a function of source redshift. Optical depths lower than $\sim10^{-5}$ are omitted, as they are too low to predict an observable GW source at such redshifts with type II images in future detectors. \textit{Right} axis: comoving number density of all galaxies modeled as lenses.}}
    \label{optdepth}
\end{figure}

To account for the redshift dependence, we follow the prescription in \cite{Oguri2018}, in which
\begin{equation}
    \Psi(\sigma_v,z_l) = \Psi(\sigma_v,0)\frac{\Psi_{\rm{hyd}}(\sigma_v,z_l)}{\Psi_{\rm{hyd}}(\sigma_v,0)}\;,
\label{eqn:psi_vz}
\end{equation}
where $\Psi_{\rm{hyd}}(\sigma_v,z_l)$ is the velocity dispersion function derived from hydrodynamical simulation in \cite{Torrey}. The redshift depedence of the galaxy comoving number density is shown on the right axis of Figure~\ref{optdepth}. We truncate $\sigma_v$ at $50$ km/s and 400 km/s to include the major part of the distribution in Eqn.~\eqref{eqn:psiz0}. In general, the galaxy number density peaks around $z\sim 1,2$ and decreases towards higher redshift as they have less time to form.

For galaxy ellipticity, we adopt the same Gaussian distribution as in \cite{Li2018}, where the mean and standard deviation are 0.7 and 0.16, truncated at $e=0.2$ and $e=1$. The lens redshift is uniformly sampled from $[0,z_s]$. Since strong lensing occurs only when the angular separation between the lens and the source is small, we uniformly sample the lens positions within a square region centered at the source with the side length equal to four times the Einstein radius of the lens, given by \citep[see][]{Li2018}

\begin{equation}
    \theta_E = 4\pi \left(\frac{\sigma_v}{c}\right)^2\frac{D_{ls}}{D_s}\;,
\label{eqn:thetaE}
\end{equation}
where $D_{ls}$ and $D_s$ are the lens-source and observer-source separations, respectively. 

For each sampled lens parameter set $(\sigma_v,e,z_l,\vec{\theta})$, we solve the lens equation with the Python package \texttt{lenstronomy}\footnote{\texttt{https://github.com/sibirrer/lenstronomy}}\cite{lenstronomy} and obtain the number of images, image types and magnifications. Since our interest in distinguishable type II images is to identify strongly lensed GW sources, we compute the ``source-based'' optical depth, the fraction of GW sources with type II images, rather than the fraction of all images that are type II. Each sample with at least one type II image contributes to $\tau_{\rm II}$, while depending on the solution for $\mu$, it contributes to $\partial P_{\rm{II}}/\partial \log_{10}\mu$ accordingly. To account for the lens population, each sample is weighted by the expected count of such a galaxy within the defined lens position range.

Figure~\ref{optdepth} shows the type II image optical depth at various source redshifts on the left axis. Optical depths smaller than $\sim10^{-5}$ are truncated, as they are too low to produce a possible lensed source. We observe that the optical depth is on the order of $10^{-3}\sim10^{-4}$, consistent with results from ray-tracing studies using N-body simulations \citep[see, e.g.,][]{Hilbert2007,Hilbert2008}. 

In the generated sample, the probability of a strongly lensed GW event (i.e., with multiple images) to have no type II images is smaller than 0.01\% and therefore negligible. We conclude that the type II image optical depth is effectively identical to the strong lensing optical depth. We calculate that roughly 91.5\% of all sources with multiple images have a type II image as the second ``brightest'' image, which suggests that if multiple images were to be detected, it is likely that at least one of the images may be a candidate for type II image distinction via the log Bayes factor test. 

For larger redshifts, we do not extrapolate optical depth due to the lack of information on extremely high redshift galaxy velocity dispersion function from hydrodynamical simulations. Instead, we take the conservative limit and assume the optical depth to be constant beyond $z_s=7$.

\begin{figure*}[!htb]
\includegraphics[width=0.8\linewidth]{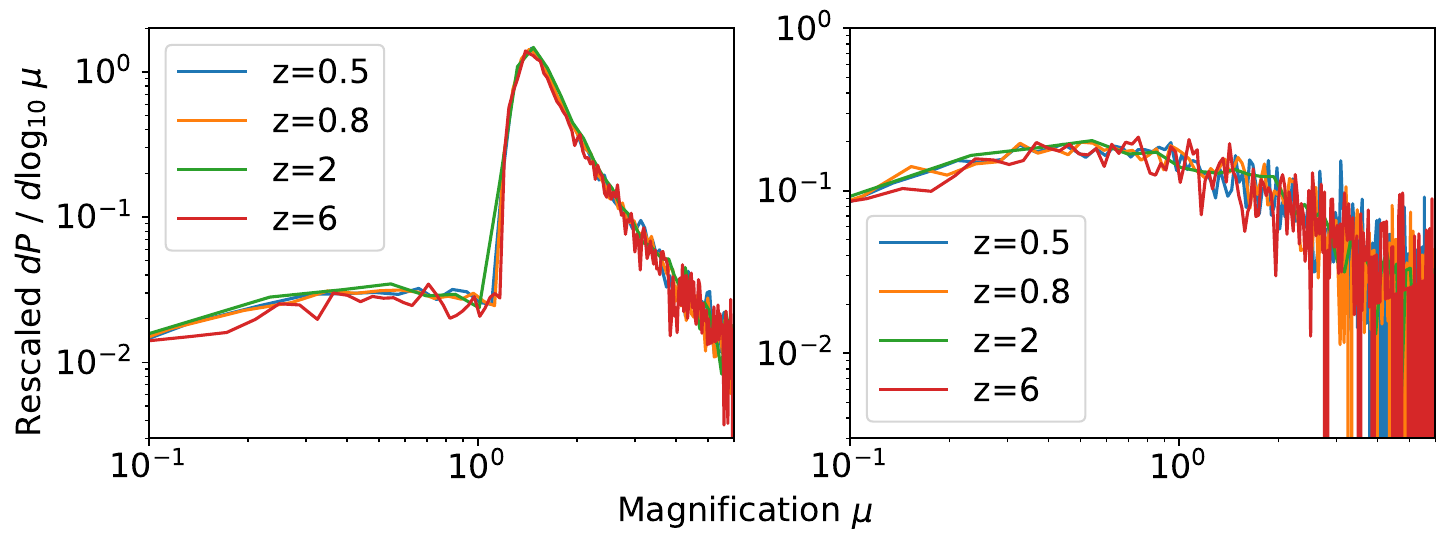}
\caption{\small{Rescaled magnification distribution at redshift $z=0.5,0.8,2,6$. The \textit{left} panel shows $\partial P(\mu,z_s)/\partial \log_{10} \mu$, including all images from the Monte Carlo samples. The \textit{right} panel shows the rescaled distribution of only type II images, $\partial P_{\rm II}(\mu,z_s)/\partial \log_{10} \mu$ . The traces are rescaled such that the largest image count is normalized to 1.}}
\label{P_dist}
\end{figure*}

For the magnification distribution, we extract the type II images from the Monte Carlo simulation samples for each redshift. Figure~\ref{P_dist} shows the rescaled image magnification distribution per $\log_{10}\mu$ at selected redshifts $z_s=0.5,0.8,2,6$. The \textit{left} panel includes all images with the peak dominated by the slightly magnified type I images. The \textit{right} panel contains only type II images, which constitute the demagnified image population. The rescaling normalizes the highest image count in each case to 1. Since the magnification for all images peaks around 1, we ignore it when calculating the detectable strongly lensed GW events; instead, the detection rate of all strongly lensed events can be estimated by multiplying the detectable BBH merger rate under the no-lensing hypothesis by the strong lensing optical depth, which, as the Monte Carlo samples show, is effectively identical to the type II image optical depth.

We note that $\partial P_{\rm II}(\mu,z_s)/\partial \log_{10} \mu$ is independent from source redshift $z_s$ by construction, as Figure~\ref{P_dist} confirms. To explain this feature, we first note that the lens equation solution depends only on $\vec{\theta}/\theta_E$, where $\vec{\theta}=(x,y)$ and $\theta_E$ is defined in Eq.\eqref{eqn:thetaE}. Since the range of the possible lens angular positions, $\vec{\theta}$, is directly determined by $\theta_E$, the image solution (image count, magnification, etc.) and its distribution remain constant under the scaling. The only remaining redshift-dependent quantity is the galaxy velocity dispersion function. However, Eq.\eqref{eqn:psi_vz} shows that only the overall magnitude of $\Psi(\sigma_v,z_l)$ changes with redshift. Consequently, we expect a universal normalized magnification distribution for all redshifts. 

Finally, we fit $d P_{\rm{II}}(\mu)/d \log_{10} \mu$ by a log normal distribution with a mean of $-0.35$ and standard deviation of $0.57$, truncated at $\log_{10}\mu=-2,1$.

\subsection{GW Source Population}

We adopt GW source population models provided by \cite{Li2018,Cao}. In summary, we assume the merger rate of the primary black hole in a binary to be proportional to the formation rate of black holes and their progenitor stars. The merger rate is then calibrated to the observed BBH merger density in the local universe. We follow the prescription and the chosen astrophysical models in \cite{Li2018,Cao}, and we provide more details in Appendix~\ref{app:rmerg} for reference. The BBH merger population is calibrated to a local merger rate of 103 ${\rm{Gpc}}^{-3}\rm{yr}^{-1}$ based on LIGO detection data up until GW170104 \cite{GW170104}. With the second LIGO-Virgo Gravitational-Wave Transient Catalog (GWTC-2), the local BBH merger rate is more tightly constrained to be $23.9^{+14.3}_{-8.6}~{\rm{Gpc}}^{-3}\rm{yr}^{-1}$\cite{GW170104}. The data slightly favor that the merger rate increases with redshift, but remain statistically consistent with a non-evolving merger rate hypothesis \cite{GW170104}. This updated value suggests that our merger rate model could be an overestimate. However, the local rate difference is less than an order of magnitude, and the high-redshift merger rates are not constrained by LIGO data. As the local merger rate only acts as an overall scaling in the population model, our result can be easily scaled to reflect any differences. Consequently, our predictions serve as an adequate reference and can be easily adapted in light of new data and more accurate BBH population models.

\begin{figure}
    \centering
    \includegraphics[width=\columnwidth]{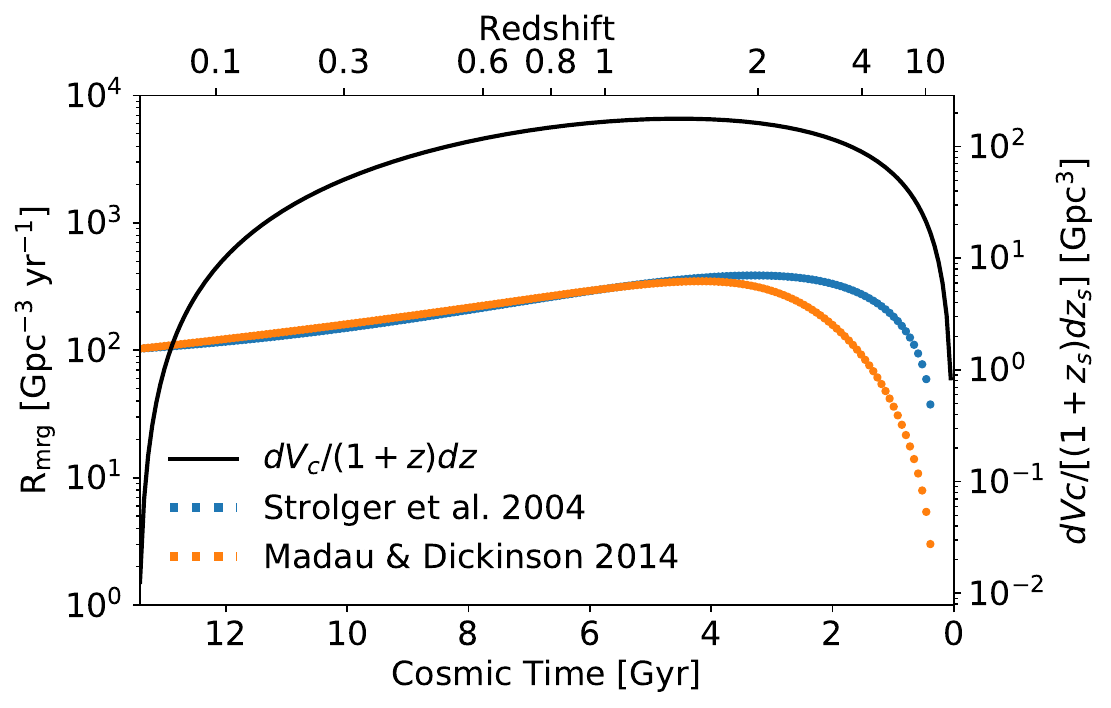}
    \caption{\small{BBH merger rate density, assuming two SFR models in \cite{Madau} and \cite{Strolger}. The blue over-arching trace plots the modified differential comoving volume and corresponds to values on the right axis. The merger rate density is directly analogous to Figure 1 in \cite{Li2018}.}}
    \label{Rmrg}
\end{figure}

Figure~\ref{Rmrg} replicates Figure 1 in \cite{Li2018} and shows the predicted BBH merger rate density under the two Star Formation Rate (SFR) models in \cite{Madau} and \cite{Strolger}, respectively. Due to the intrinsic uncertainty in these analytical SFR models, we choose one, the more optimistic SFR in \cite{Strolger}, for the following population estimates.

The total BBH merger rate per source redshift and the detectable merger rate are plotted in Figure~\ref{dN_red}. Since the strong lensing optical depth is in general smaller than 0.1\% at the redshift with the most GW sources, we will neglect the magnification effect when calculating the total detectable GW events. To keep the detectable population estimate general, we assume the detectors to always be online. To incorporate the detector duty cycle, the detectable and distinguishable populations simply scale proportionally with the fraction of detector online time, since the type II waveforms can be identified by their own waveforms, and the duty cycle does not disproportionally affect particular GW image types. We estimate a total of $2.17\times10^5$ BBH mergers per year up to $z_s=23$. The detectable total merger number is $2.17\times10^5$ for CE (99.96\%), $1.96\times10^5$ for ET (90.3\%) and $7.59\times10^4$ for LIGO Voyager (35.0\%). We note that the detection rate is not only affected by the detector sensitivity, but also by the redshift distribution of BBH mergers and the comoving volume. Even though ET has lower sensitivity than CE overall, it already covers the redshift range with peak GW source count ($z_s\sim2$). At large redshift with $z_s\gg 7$, BBH mergers happen far less frequently due to a lack of black hole formation and the decreasing comoving volume per redshift. Consequently, the detection rate of ET is only slightly lower than CE. In the case of LIGO Voyager, the lower sensitivity excludes many sources from $z_s\sim2$, leading to a larger loss in the detectable source fraction. 

\subsection{type II Image Rate}

In this section, we combine lensing statistics and GW source population models to study the rate of detectable and distinguishable type II images in third generation GW detectors.

\begin{figure*}[!htb]
    \centering
    \includegraphics[width=\linewidth]{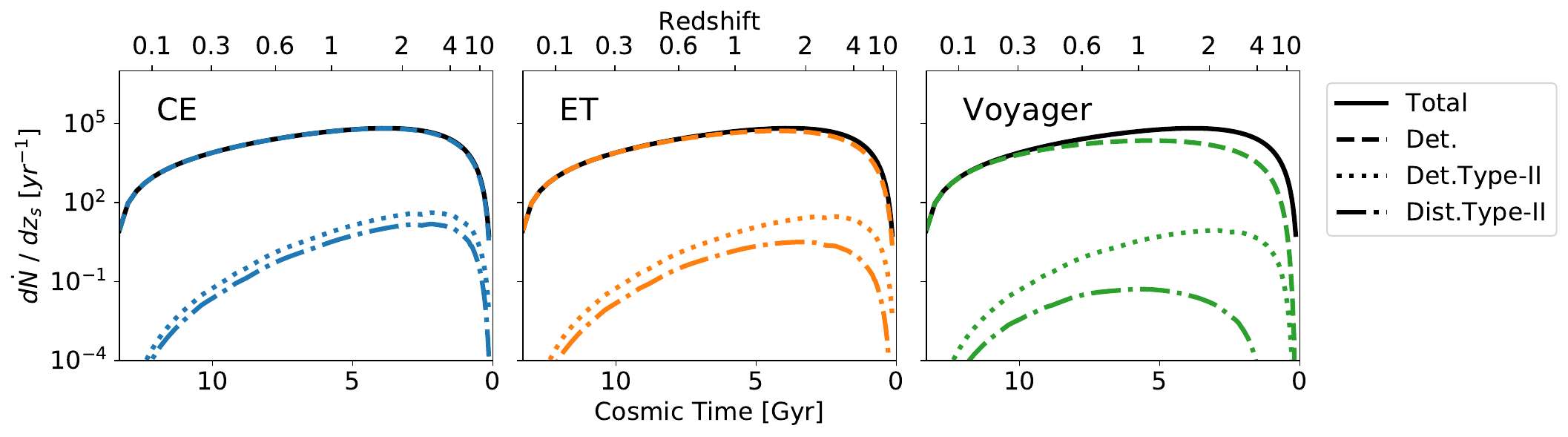}
    \caption{\small{Yearly detected population per unit reshift prediction as a function of redshift. The panels from left to right show the detection population for CE, ET and LIGO Voyager. In all panels, the solid black line denotes the total BBH merger rate. The dashed curves show the rate of detectable GW sources (i.e., $\rho>8$) when unlensed. The dotted curve shows the rate of GW sources with a detectable type II image. The dot-dash curve shows the event population with a distinguishable type II image. See text for total detection rates.}}
    \label{dN_red}
\end{figure*}

Figure~\ref{dN_red} shows the differential event rate as a function of redshift for three detectors. In each panel, four different populations are shown. The total rate of BBH mergers are plotted as solid black curves. The dashed curves show the rate of detectable GW events. The dotted curves show the rate of events with a detectable type II image as in Eq.\eqref{eqn:NdetII}. The dot-dash curves show the rate of GW events with at least one distinguishable type II image.

As expected, the rate of BBH mergers in all three categories decreases with the detector sensitivity, especially at high redshifts. For LIGO Voyager, in particular, the rate of expected GW sources with distinguishable type II images drops quickly with redshift, consistent with the trend of the distinguishable fraction in Figure~\ref{fig:M_int}. 

\begin{figure*}[!htb]
\centering
\includegraphics[width=0.85\linewidth]{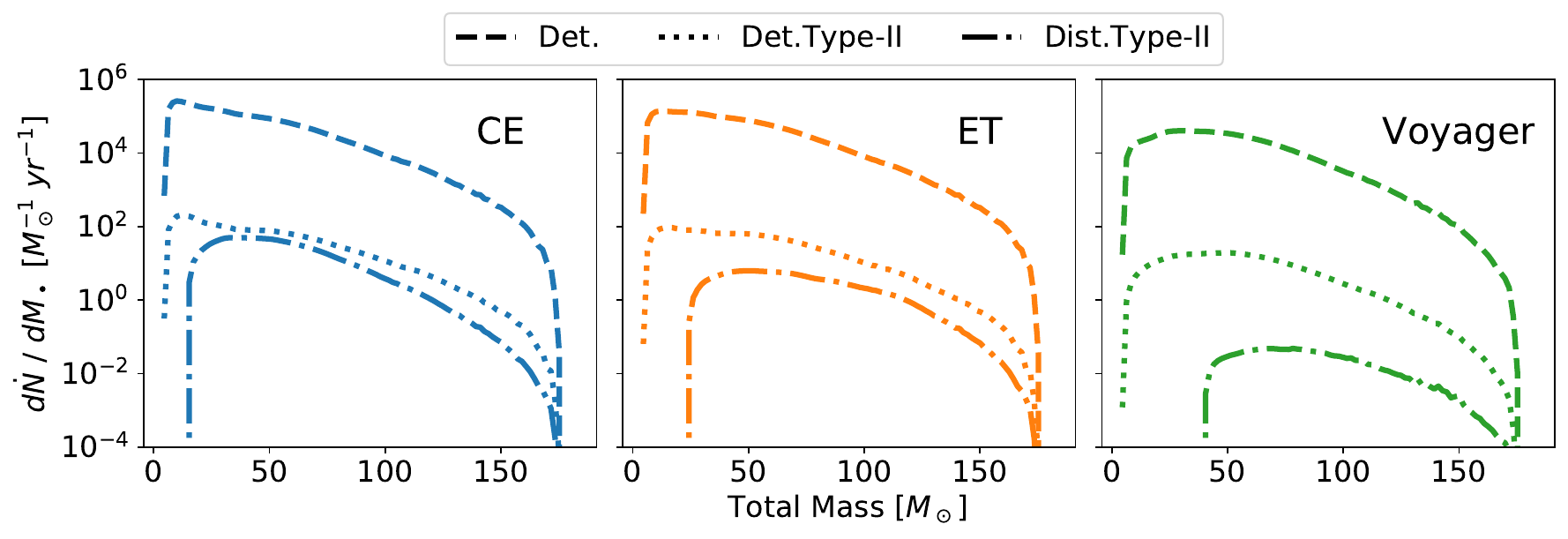}
\caption{\small{Detection rate as a function of BBH intrinsic mass. The panels from left to right shows the detection prediction for CE, ET and LIGO Voyager. The line styles are consistent with those in Figure~\ref{dN_red}.}}
\label{dN_mass}
\end{figure*}

Figure~\ref{dN_mass} plots the same population prediction binned by the total mass of the BBH, with consistent line styles as in Figure~\ref{dN_red}. For all detectors, the detection rate decreases with increasing total mass, consistent with the underlying initial mass function. The detection rate of events with distinguishable type II images shows a cutoff at small total mass, which is primarily due to two factors. When the waveform mismatch for low-mass BBHs is smaller than the imposed mismatch resolution (i.e., $\epsilon<0.001$), their type II images are considered indistinguishable from type I images. When the mismatch has just exceeded the resolution threshold, distinguishability requires very large SNRs, which may not be achievable depending on the detector sensitivity. Consequently, we observe a mass cutoff in all three detectors, which shifts to higher masses as the detector sensitivity decreases. 

Overall, we predict that CE will detect roughly 184.7 strongly lensed GW events per year, among which 172.2 have at least one detectable type II image. Among these strongly lensed GW sources, 56.9 per year have a type II image distinguishable via the log Bayes factor test. ET will be able to detect 157.1 strongly lensed events per year, and 118.2 of these have detectable type II images. However, due to reduced sensitivity, the number of sources with a distinguishable type II image drops to 8.6 per year. For LIGO Voyager, the yearly detection rate of GW events with detectable type II images is 27.4 per year out of the 38.4 strongly lensed events. The distinguishable type II image rate is 0.06 per year, which suggests that the possibility of observing a GW source with distinguishable type II images with LIGO Voyager is relatively slim. The detection rates are summarized in Table~\ref{tab:sum}.

\subsection{Discussion}

In this section, we discuss the implication of the predicted detection rates for GW sources with distinguishable type II images. We re-examine assumptions in our analysis and explore how relaxing these assumptions lead to more an optimistic detection prediction.

As Figure~\ref{dN_mass} shows, the yearly detection rates of GW sources with distinguishable type II images are 56.9, 8.6 and 0.06 for CE, ET and LIGO Voyager, respectively. In particular, in the case of CE, more than 30\% of all detectable strongly lensed sources will have distinguishable type II images. For such sources, detection of the type II image alone can confirm the existence of strongly lensed images, without pair-wise GW event inference on the strong lensing hypothesis. Once such images are identified, the inferred source parameter values can act as a prior during the subsequent and more elaborate catalog search for the other images. 

For ET and LIGO Voyager, the expected detection rate is smaller, thus the distinguishable type II images will not be as powerful for confirming the strong lensing hypothesis as in the case of CE. However, we emphasize that if several of our conservative constraints can be relaxed, distinguishable type II images can still contribute to the identification of strong lensing. 

The first condition we revisit is the waveform mismatch resolution. Throughout the analysis, we consistently adopt $\epsilon_{\rm{min}}=0.001$, which excludes the binaries at small inclinations, and the distinguishable fraction is ``saturated'' at roughly 70\% (see, e.g., Figure~\ref{fig:frac_z} and Figure~\ref{fig:eps_thresh}). As Figure~\ref{fig:eps_thresh} suggests, the waveform mismatch resolution significantly affects the fraction of distinguishable type II images. If we can expect a better waveform resolution from third-generation GW detectors, the distinguishable fraction should increase considerably; as Figure~\ref{fig:eps_thresh} shows, the distinguishable fraction roughly doubles as the mismatch resolution improves from $\mathcal{O}(1\%)$ to $\mathcal{O}(0.1\%)$ assuming CE sensitivity. For CE and ET, this increase results in many more detectable sources at small redshifts ($z_s\sim1,2$), where the BBH population also peaks. This requirement has two implications for third-generation GW detector performance and data analysis process. As is discussed briefly in Section~\ref{section:mismatch}, the error in detector calibration should be much smaller, such that the waveform mismatch is not obscured by systematic uncertainties. In terms of the data analysis process, the density of the matching template bank should be such that the waveform difference is large compared with the template spacing. If such conditions are not satisfied, the high SNR detection offered by the third-generation detectors cannot be taken full advantage of to maximize the scientific output. 

\begin{figure}
    \includegraphics[width=0.8\columnwidth]{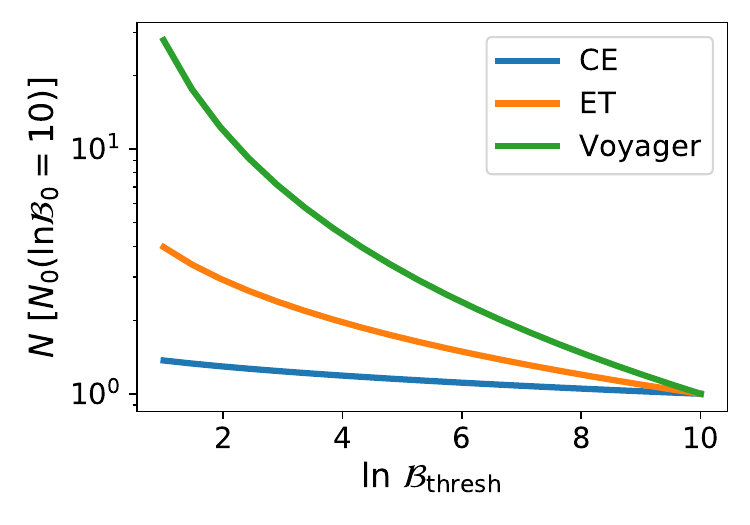}
    \caption{\small{The number of GW sources with distinguishable type II images for different log Bayes factor threshold values, expressed as a fraction of the distinguishable number with the threshold value $\ln\mathcal{B}_{\rm{thresh}}=10$.}}
    \label{fig:dn_ratio}
\end{figure}

\begin{figure*}[!htb]
    \centering
    \includegraphics[width=\linewidth]{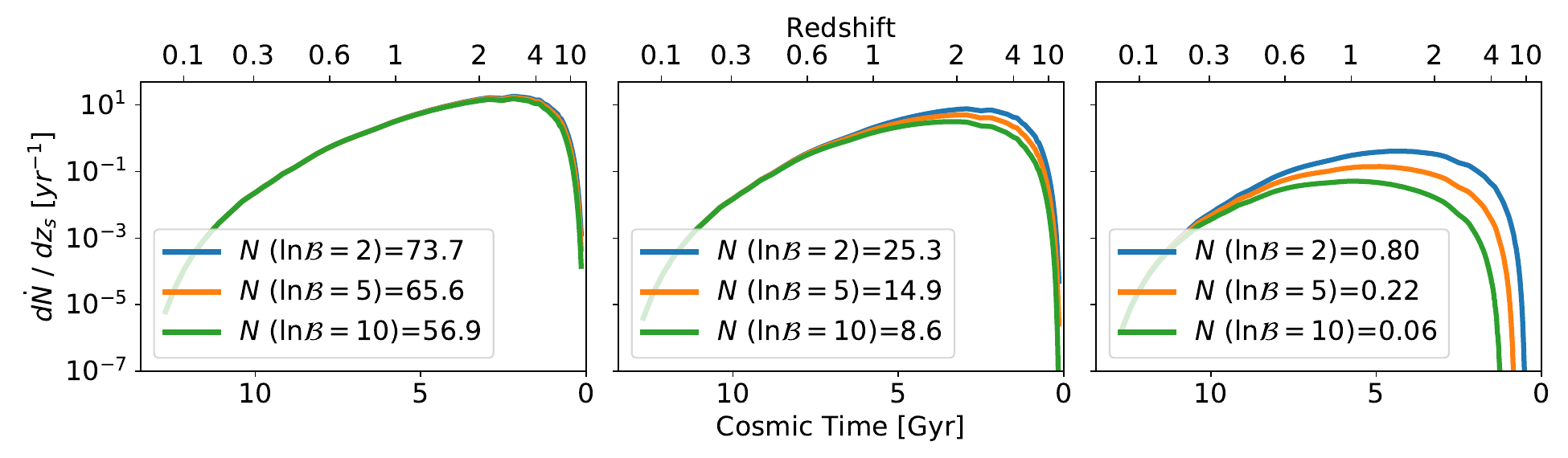}
    \caption{\small{The expected number of GW sources with distinguishable type II images with $\ln\mathcal{B}_{\rm{thresh}}=2,~5,$ and 10. The expected yearly detection count for each threshold value is shown in the legend. The panels from left to right correspond to CE, ET and LIGO Voyager. As expected, the detection number increase is most significant at large redshift for all three detectors.}}
    \label{fig:dnchangel}
\end{figure*}

We have also taken a conservative estimate by setting the threshold log Bayes factor to be 10. Even for $\ln\mathcal{B}_{\rm{thresh}}=5$, the type II image hypothesis is more than 100 times more likely than the type I image hypothesis, and an even smaller threshold value may be sufficient for realistic data analysis. Figure~\ref{fig:dn_ratio} shows the increase in the number of events with distinguishable type II images with a lower $\ln\mathcal{B}_{\rm{thresh}}$, normalized to the number when $\ln\mathcal{B}_{\rm{thresh}}=10$. We observe that the increase is the most dramatic for LIGO Voyager, as a lower threshold extends the sensitive range to higher redshift $(z_s\sim2)$, where the GW source population peaks. For CE and ET, the increase is more modest, as they already detect most sources at $z_s\sim 2$ with high SNR. The extended range is then expected to add relatively fewer GW sources in comparison. Figure~\ref{fig:dnchangel} shows the redshift distribution of the GW sources with distinguishable type II images with $\ln\mathcal{B}_{\rm{thresh}}=2,~5,$ and 10. As expected, the distinguishable rate increase is more significant at high redshift, and the effect is the strongest for LIGO Voyager; at $\ln\mathcal{B}_{\rm{thresh}}=2$, 42.8\% of all strongly lensed GW sources in CE are accompanied by at least one distinguishable type II image and 21.4\% for ET. For LIGO Voyager, the distinguishable number is still small, but at $\sim1/$yr, it is more promising that such an event will appear in the LIGO Voyager catalog with a few years of observing run. The predicted detection rates are summarized in Table~\ref{tab:sum}. 

In addition, we have so far considered the single-detector scenario, and we estimate the advantage of a detector network via the simplifying assumption that both GW polarizations can be independently detected, i.e., the time-domain waveform for calculating the overlap and $\rho$ is complex. If ET implements a triangular design, the detector itself is sufficient to capture the polarization content \cite{ET2011}. For LIGO Voyager and CE, the polarization content can be obtained if a concurrent detector network exists. In the upcoming decades, more GW observatories across the globe will start to observe, such as the expansion of the LIGO network to include IndiGO\footnote{\texttt{http://www.gw-indigo.org/tiki-index.php}} \cite{IndIGO}. This global network offers increased detector-networks $\rho$ and an increased detection spatial resolution. On the other hand, the uncertainty in the polarization content from a realistic detector-network model may be partially degenerate with the Hilbert transform signal, thus subtracting away from the type I/II waveform difference and their distinguishability. A thorough investigation on realistic detector network effect is deferred to future studies.

\begin{table}[]
\renewcommand{\arraystretch}{1.3}
\begin{tabular}{|cc|c|c|c|}
\hline
 &   & CE & ET & Voyager          \\ \hline 
\multicolumn{2}{|c|}{Det.}                                         & $2.17\times10^5$ & $1.96\times10^5$ & $7.59\times10^4$ \\ \hline
\multicolumn{2}{|c|}{Det. SL}                                  & 184.7            & 157.1            & 38.4             \\ \hline
\multicolumn{2}{|c|}{Det. type II}                                  & 172.2            & 118.2            & 27.4             \\ \hline
\multicolumn{1}{|c|}{\multirow{3}{*}{Dist. type II}} & $\ln\mathcal{B}\geq 10$ & 56.9 (33.1\%) & 8.6 (7.3\%) & 0.06 (0.22\%)  \\
\cline{2-5} 
\multicolumn{1}{|c|}{} &$\ln\mathcal{B}\geq5$  & 65.6 (38.1\%)& 14.9 (12.6\%)& 0.22 (0.81\%)\\
\cline{2-5} 
\multicolumn{1}{|c|}{}& $\ln\mathcal{B}\geq2$  & 73.7 (42.8\%)& 25.3 (21.4\%)& 0.80 (2.93\%)\\ \hline
\end{tabular}
\caption{\small Predicted yearly detection rates. The columns show the detectable BBH merger rates, the rates for strongly lensed (SL) BBH mergers, the rates for BBH mergers with detectable type II images and the rates for BBH mergers with distinguishable type II images. The distinguishable event rates are given with $\ln\mathcal{B} \geq 10,5,2$. The fraction of events with detectable type II images that are also distinguishable is shown in the parenthesis.}
\label{tab:sum}
\end{table}

\section{Conclusions}
\label{section:conclusion} 
In this paper, we study an intrinsic waveform signature of type II images of strongly lensed GW sources. For CE, ET and LIGO Voyager, we compute the best-match overlap between type I/II waveforms. We then calculate the required threshold orbit inclination to establish the type II waveform hypothesis by a favoring log Bayes factor of 10. The fraction of GW sources with distinguishable type II images is computed from the threshold inclination accordingly. For all three detectors, we find that significant fractions of type II images (e.g., $50-70\%$) of sufficiently high SNR GW events can be identified. In other words, if such a type II image is detected with reasonable SNR, it can likely be distinguished from regular type I images and used as the tell-tale evidence of strongly lensed events. 

We also assess the effects of the type II signature in the context of the current LIGO data analysis process. We apply the targeted sub-threshold search method described in \cite{Li:2019osa} on an example high-mass-ratio compact binary coalescence event GW190814 \cite{GW190814}. We generate a reduced template bank based on injection run results using simulated type I lensed injections of the target event. The resulting reduced bank is used, then, to search for the same set of simulated lensed injections in two different searches, in which they are injected as type I images (original waveforms) and type II images (Hilbert transform of the same waveforms) respectively.

Our preliminary result shows that there is a slight increase in the number of injections missed when they are treated as type II images. This hints at the possibility that the current search scheme may suffer from sensitivity loss without considering type II images. However, we remark that the current results in this study are only preliminary and will require further studies. 

We then incorporate GW source population model and lensing probabilities to predict the expected number of GW sources with distinguishable type II images in CE, ET and LIGO Voyager respectively. For these three detectors, we predict the yearly detection rates are 56.9, 8.6 and 0.06 with a conservative threshold at $\ln\mathcal{B}_{\rm{thresh}}=10$. A relaxed log Bayes factor threshold boosts the expected detection rates, especially for LIGO Voyager; at $\ln\mathcal{B}_{\rm{thresh}}=2$, the yearly detection rate for LIGO Voyager approaches 1$/$yr.

Such distinguishable type II images are ``short-cuts'' for identifying strongly lensed events, as they guarantee the existence of at least one other lensed image. They also improve the computational efficiency of searching for the companion images, as the estimated parameters, such as the redshifted mass, mass ratio and sky location, can inform a more comprehensive catalog search. As illustrated, this method will be most powerful with the unprecedented sensitivity offered by third-generation GW detectors. 

Our work can be extended and refined in several directions. We can relax the constraints on GW source range by including spin and orbit eccentricity, as is studied in Ref.~\cite{Esquiaga2020}. On one hand, the Hilbert transform of GWs from such sources may have a larger mismatch from the original waveform, favoring type II image distinguishability. On the other hand, the Hilbert transform may be partially degenerate with a parameter bias with the additional degrees of freedom. The effect of these competing factors warrants careful treatment. 

We may consider realistic detector networks instead of assuming complete knowledge on both GW polarizations, which adds to the underlying waveform uncertainties. Similar to the hypothesized effect of binary spin and orbital eccentricity, uncertainty in the polarization may be partially degenerate with the Hilbert transform signature. However, a detector network yields larger signal SNR, which should promote the distinguishability of type II images.  

It is also important to refine the lens modeling. While theoretical works on various lens types and their respective image characteristics abound, to our knowledge, the effect of the model choice on predictions for realistic detection has yet to be systematically investigated. Therefore, such a follow-up study is essential for understanding the uncertainty and robustness of this strongly-lensing detection forecast.

In conclusion, this study shows that the intrinsic waveform characteristics of type II images can be a powerful supplemental tool for hunting strongly lensed events in the catalog of third-generation GW detectors, when tens of such events may be identified.

\acknowledgements
We thank the anonymous referee for helpful suggestions that improve this manuscript. YW would like to thank the David and Ellen Lee Distinguished Fellowship for support during this research.  Research of YW and YC are supported by the Simons Foundation (Grant No. 568762), and the National Science Foundation, through Grants PHY-2011961, PHY-2011968, and PHY-1836809. RKLL would also like to thank the Croucher Foundation for support during this research. The computations presented here were conducted on the Caltech High Performance Cluster partially supported by a grant from the Gordon and Betty Moore Foundation. This paper carries LIGO Document Number LIGO-P2100002.

\bibliographystyle{apsrev4-2}
\bibliography{ref}

\providecommand{\noopsort}[1]{}\providecommand{\singleletter}[1]{#1}%
\begin{thebibliography}{78}%
\makeatletter
\providecommand \@ifxundefined [1]{%
 \@ifx{#1\undefined}
}%
\providecommand \@ifnum [1]{%
 \ifnum #1\expandafter \@firstoftwo
 \else \expandafter \@secondoftwo
 \fi
}%
\providecommand \@ifx [1]{%
 \ifx #1\expandafter \@firstoftwo
 \else \expandafter \@secondoftwo
 \fi
}%
\providecommand \natexlab [1]{#1}%
\providecommand \enquote  [1]{``#1''}%
\providecommand \bibnamefont  [1]{#1}%
\providecommand \bibfnamefont [1]{#1}%
\providecommand \citenamefont [1]{#1}%
\providecommand \href@noop [0]{\@secondoftwo}%
\providecommand \href [0]{\begingroup \@sanitize@url \@href}%
\providecommand \@href[1]{\@@startlink{#1}\@@href}%
\providecommand \@@href[1]{\endgroup#1\@@endlink}%
\providecommand \@sanitize@url [0]{\catcode `\\12\catcode `\$12\catcode
  `\&12\catcode `\#12\catcode `\^12\catcode `\_12\catcode `\%12\relax}%
\providecommand \@@startlink[1]{}%
\providecommand \@@endlink[0]{}%
\providecommand \url  [0]{\begingroup\@sanitize@url \@url }%
\providecommand \@url [1]{\endgroup\@href {#1}{\urlprefix }}%
\providecommand \urlprefix  [0]{URL }%
\providecommand \Eprint [0]{\href }%
\providecommand \doibase [0]{https://doi.org/}%
\providecommand \selectlanguage [0]{\@gobble}%
\providecommand \bibinfo  [0]{\@secondoftwo}%
\providecommand \bibfield  [0]{\@secondoftwo}%
\providecommand \translation [1]{[#1]}%
\providecommand \BibitemOpen [0]{}%
\providecommand \bibitemStop [0]{}%
\providecommand \bibitemNoStop [0]{.\EOS\space}%
\providecommand \EOS [0]{\spacefactor3000\relax}%
\providecommand \BibitemShut  [1]{\csname bibitem#1\endcsname}%
\let\auto@bib@innerbib\@empty
\bibitem [{\citenamefont {{Abbott}}\ \emph
  {et~al.}(2020{\natexlab{a}})\citenamefont {{Abbott}}, \citenamefont
  {{Abbott}}, \citenamefont {{Abraham}}, \citenamefont {{Acernese}},
  \citenamefont {{Ackley}}, \citenamefont {{Adams}}, \citenamefont
  {{Adhikari}}, \citenamefont {{Adya}}, \citenamefont {{Affeldt}},
  \citenamefont {{Agathos}}, \citenamefont {{Agatsuma}}, \citenamefont
  {{Aggarwal}}, \citenamefont {{Aguiar}}, \citenamefont {{Aich}}, \citenamefont
  {{Aiello}}, \citenamefont {{Ain}}, \citenamefont {{Ajith}}, \citenamefont
  {{Akcay}}, \citenamefont {{Allen}}, \citenamefont {others}, \citenamefont
  {{LIGO Scientific Collaboration}},\ and\ \citenamefont {{Virgo
  Collaboration}}}]{GW1908142020}%
  \BibitemOpen
  \bibfield  {author} {\bibinfo {author} {\bibfnamefont {R.}~\bibnamefont
  {{Abbott}}}, \bibinfo {author} {\bibfnamefont {T.~D.}\ \bibnamefont
  {{Abbott}}}, \bibinfo {author} {\bibfnamefont {S.}~\bibnamefont {{Abraham}}},
  \bibinfo {author} {\bibfnamefont {F.}~\bibnamefont {{Acernese}}}, \bibinfo
  {author} {\bibfnamefont {K.}~\bibnamefont {{Ackley}}}, \bibinfo {author}
  {\bibfnamefont {C.}~\bibnamefont {{Adams}}}, \bibinfo {author} {\bibfnamefont
  {R.~X.}\ \bibnamefont {{Adhikari}}}, \bibinfo {author} {\bibfnamefont
  {V.~B.}\ \bibnamefont {{Adya}}}, \bibinfo {author} {\bibfnamefont
  {C.}~\bibnamefont {{Affeldt}}}, \bibinfo {author} {\bibfnamefont
  {M.}~\bibnamefont {{Agathos}}}, \bibinfo {author} {\bibfnamefont
  {K.}~\bibnamefont {{Agatsuma}}}, \bibinfo {author} {\bibfnamefont
  {N.}~\bibnamefont {{Aggarwal}}}, \bibinfo {author} {\bibfnamefont {O.~D.}\
  \bibnamefont {{Aguiar}}}, \bibinfo {author} {\bibfnamefont {A.}~\bibnamefont
  {{Aich}}}, \bibinfo {author} {\bibfnamefont {L.}~\bibnamefont {{Aiello}}},
  \bibinfo {author} {\bibfnamefont {A.}~\bibnamefont {{Ain}}}, \bibinfo
  {author} {\bibfnamefont {P.}~\bibnamefont {{Ajith}}}, \bibinfo {author}
  {\bibfnamefont {S.}~\bibnamefont {{Akcay}}}, \bibinfo {author} {\bibfnamefont
  {G.}~\bibnamefont {{Allen}}}, \bibinfo {author} {\bibnamefont {others}},
  \bibinfo {author} {\bibnamefont {{LIGO Scientific Collaboration}}},\ and\
  \bibinfo {author} {\bibnamefont {{Virgo Collaboration}}},\ }\href
  {https://doi.org/10.3847/2041-8213/ab960f} {\bibfield  {journal} {\bibinfo
  {journal} {\apjl}\ }\textbf {\bibinfo {volume} {896}},\ \bibinfo {eid} {L44}
  (\bibinfo {year} {2020}{\natexlab{a}})},\ \Eprint
  {https://arxiv.org/abs/2006.12611} {arXiv:2006.12611 [astro-ph.HE]}
  \BibitemShut {NoStop}%
\bibitem [{\citenamefont {{The LIGO Scientific Collaboration}}\ \emph
  {et~al.}(2020)\citenamefont {{The LIGO Scientific Collaboration}},
  \citenamefont {{the Virgo Collaboration}}, \citenamefont {{Abbott}},
  \citenamefont {{Abbott}}, \citenamefont {{Abraham}}, \citenamefont
  {{Acernese}}, \citenamefont {{Ackley}}, \citenamefont {{Adams}},
  \citenamefont {{Adhikari}}, \citenamefont {{Adya}}, \citenamefont
  {{Affeldt}}, \citenamefont {{Agathos}}, \citenamefont {{Agatsuma}},
  \citenamefont {{Aggarwal}}, \citenamefont {{Aguiar}}, \citenamefont {{Aich}},
  \citenamefont {{Aiello}}, \citenamefont {{Ain}}, \citenamefont {{Ajith}},
  \citenamefont {{Akcay}}, \citenamefont {{Allen}} \emph
  {et~al.}}]{GW1905212020}%
  \BibitemOpen
  \bibfield  {author} {\bibinfo {author} {\bibnamefont {{The LIGO Scientific
  Collaboration}}}, \bibinfo {author} {\bibnamefont {{the Virgo
  Collaboration}}}, \bibinfo {author} {\bibfnamefont {R.}~\bibnamefont
  {{Abbott}}}, \bibinfo {author} {\bibfnamefont {T.~D.}\ \bibnamefont
  {{Abbott}}}, \bibinfo {author} {\bibfnamefont {S.}~\bibnamefont {{Abraham}}},
  \bibinfo {author} {\bibfnamefont {F.}~\bibnamefont {{Acernese}}}, \bibinfo
  {author} {\bibfnamefont {K.}~\bibnamefont {{Ackley}}}, \bibinfo {author}
  {\bibfnamefont {C.}~\bibnamefont {{Adams}}}, \bibinfo {author} {\bibfnamefont
  {R.~X.}\ \bibnamefont {{Adhikari}}}, \bibinfo {author} {\bibfnamefont
  {V.~B.}\ \bibnamefont {{Adya}}}, \bibinfo {author} {\bibfnamefont
  {C.}~\bibnamefont {{Affeldt}}}, \bibinfo {author} {\bibfnamefont
  {M.}~\bibnamefont {{Agathos}}}, \bibinfo {author} {\bibfnamefont
  {K.}~\bibnamefont {{Agatsuma}}}, \bibinfo {author} {\bibfnamefont
  {N.}~\bibnamefont {{Aggarwal}}}, \bibinfo {author} {\bibfnamefont {O.~D.}\
  \bibnamefont {{Aguiar}}}, \bibinfo {author} {\bibfnamefont {A.}~\bibnamefont
  {{Aich}}}, \bibinfo {author} {\bibfnamefont {L.}~\bibnamefont {{Aiello}}},
  \bibinfo {author} {\bibfnamefont {A.}~\bibnamefont {{Ain}}}, \bibinfo
  {author} {\bibfnamefont {P.}~\bibnamefont {{Ajith}}}, \bibinfo {author}
  {\bibfnamefont {S.}~\bibnamefont {{Akcay}}}, \bibinfo {author} {\bibfnamefont
  {G.}~\bibnamefont {{Allen}}}, \emph {et~al.},\ }\href@noop {} {\bibfield
  {journal} {\bibinfo  {journal} {arXiv e-prints}\ ,\ \bibinfo {eid}
  {arXiv:2009.01190}} (\bibinfo {year} {2020})},\ \Eprint
  {https://arxiv.org/abs/2009.01190} {arXiv:2009.01190 [astro-ph.HE]}
  \BibitemShut {NoStop}%
\bibitem [{\citenamefont {{Lai}}\ \emph {et~al.}(2018)\citenamefont {{Lai}},
  \citenamefont {{Hannuksela}}, \citenamefont {{Herrera-Mart{\'\i}n}},
  \citenamefont {{Diego}}, \citenamefont {{Broadhurst}},\ and\ \citenamefont
  {{Li}}}]{Lai2018}%
  \BibitemOpen
  \bibfield  {author} {\bibinfo {author} {\bibfnamefont {K.-H.}\ \bibnamefont
  {{Lai}}}, \bibinfo {author} {\bibfnamefont {O.~A.}\ \bibnamefont
  {{Hannuksela}}}, \bibinfo {author} {\bibfnamefont {A.}~\bibnamefont
  {{Herrera-Mart{\'\i}n}}}, \bibinfo {author} {\bibfnamefont {J.~M.}\
  \bibnamefont {{Diego}}}, \bibinfo {author} {\bibfnamefont {T.}~\bibnamefont
  {{Broadhurst}}},\ and\ \bibinfo {author} {\bibfnamefont {T.~G.~F.}\
  \bibnamefont {{Li}}},\ }\href {https://doi.org/10.1103/PhysRevD.98.083005}
  {\bibfield  {journal} {\bibinfo  {journal} {\prd}\ }\textbf {\bibinfo
  {volume} {98}},\ \bibinfo {eid} {083005} (\bibinfo {year} {2018})},\ \Eprint
  {https://arxiv.org/abs/1801.07840} {arXiv:1801.07840 [gr-qc]} \BibitemShut
  {NoStop}%
\bibitem [{\citenamefont {{Meena}}\ and\ \citenamefont
  {{Bagla}}(2020)}]{Meena}%
  \BibitemOpen
  \bibfield  {author} {\bibinfo {author} {\bibfnamefont {A.~K.}\ \bibnamefont
  {{Meena}}}\ and\ \bibinfo {author} {\bibfnamefont {J.~S.}\ \bibnamefont
  {{Bagla}}},\ }\href {https://doi.org/10.1093/mnras/stz3509} {\bibfield
  {journal} {\bibinfo  {journal} {\mnras}\ }\textbf {\bibinfo {volume} {492}},\
  \bibinfo {pages} {1127} (\bibinfo {year} {2020})},\ \Eprint
  {https://arxiv.org/abs/1903.11809} {arXiv:1903.11809 [astro-ph.CO]}
  \BibitemShut {NoStop}%
\bibitem [{\citenamefont {{Pardo}}\ \emph {et~al.}(2018)\citenamefont
  {{Pardo}}, \citenamefont {{Fishbach}}, \citenamefont {{Holz}},\ and\
  \citenamefont {{Spergel}}}]{Pardo}%
  \BibitemOpen
  \bibfield  {author} {\bibinfo {author} {\bibfnamefont {K.}~\bibnamefont
  {{Pardo}}}, \bibinfo {author} {\bibfnamefont {M.}~\bibnamefont {{Fishbach}}},
  \bibinfo {author} {\bibfnamefont {D.~E.}\ \bibnamefont {{Holz}}},\ and\
  \bibinfo {author} {\bibfnamefont {D.~N.}\ \bibnamefont {{Spergel}}},\ }\href
  {https://doi.org/10.1088/1475-7516/2018/07/048} {\bibfield  {journal}
  {\bibinfo  {journal} {\jcap}\ }\textbf {\bibinfo {volume} {2018}},\ \bibinfo
  {eid} {048} (\bibinfo {year} {2018})},\ \Eprint
  {https://arxiv.org/abs/1801.08160} {arXiv:1801.08160 [gr-qc]} \BibitemShut
  {NoStop}%
\bibitem [{\citenamefont {{Abbott}}\ \emph
  {et~al.}(2019{\natexlab{a}})\citenamefont {{Abbott}}, \citenamefont
  {{Abbott}}, \citenamefont {{Abbott}}, \citenamefont {{Abraham}},
  \citenamefont {{Acernese}}, \citenamefont {{Ackley}}, \citenamefont
  {{Adams}}, \citenamefont {{Adhikari}}, \citenamefont {{Adya}}, \citenamefont
  {{Affeldt}}, \citenamefont {{Agathos}}, \citenamefont {{Agatsuma}},
  \citenamefont {others}, \citenamefont {{LIGO Scientific Collaboration}},\
  and\ \citenamefont {{Virgo Collaboration}}}]{GWTC1TestGR}%
  \BibitemOpen
  \bibfield  {author} {\bibinfo {author} {\bibfnamefont {B.~P.}\ \bibnamefont
  {{Abbott}}}, \bibinfo {author} {\bibfnamefont {R.}~\bibnamefont {{Abbott}}},
  \bibinfo {author} {\bibfnamefont {T.~D.}\ \bibnamefont {{Abbott}}}, \bibinfo
  {author} {\bibfnamefont {S.}~\bibnamefont {{Abraham}}}, \bibinfo {author}
  {\bibfnamefont {F.}~\bibnamefont {{Acernese}}}, \bibinfo {author}
  {\bibfnamefont {K.}~\bibnamefont {{Ackley}}}, \bibinfo {author}
  {\bibfnamefont {C.}~\bibnamefont {{Adams}}}, \bibinfo {author} {\bibfnamefont
  {R.~X.}\ \bibnamefont {{Adhikari}}}, \bibinfo {author} {\bibfnamefont
  {V.~B.}\ \bibnamefont {{Adya}}}, \bibinfo {author} {\bibfnamefont
  {C.}~\bibnamefont {{Affeldt}}}, \bibinfo {author} {\bibfnamefont
  {M.}~\bibnamefont {{Agathos}}}, \bibinfo {author} {\bibfnamefont
  {K.}~\bibnamefont {{Agatsuma}}}, \bibinfo {author} {\bibnamefont {others}},
  \bibinfo {author} {\bibnamefont {{LIGO Scientific Collaboration}}},\ and\
  \bibinfo {author} {\bibnamefont {{Virgo Collaboration}}},\ }\href
  {https://doi.org/10.1103/PhysRevD.100.104036} {\bibfield  {journal} {\bibinfo
   {journal} {\prd}\ }\textbf {\bibinfo {volume} {100}},\ \bibinfo {eid}
  {104036} (\bibinfo {year} {2019}{\natexlab{a}})},\ \Eprint
  {https://arxiv.org/abs/1903.04467} {arXiv:1903.04467 [gr-qc]} \BibitemShut
  {NoStop}%
\bibitem [{\citenamefont {{Perkins}}\ and\ \citenamefont
  {{Yunes}}(2019)}]{Perkins}%
  \BibitemOpen
  \bibfield  {author} {\bibinfo {author} {\bibfnamefont {S.}~\bibnamefont
  {{Perkins}}}\ and\ \bibinfo {author} {\bibfnamefont {N.}~\bibnamefont
  {{Yunes}}},\ }\href {https://doi.org/10.1088/1361-6382/aafce6} {\bibfield
  {journal} {\bibinfo  {journal} {Classical and Quantum Gravity}\ }\textbf
  {\bibinfo {volume} {36}},\ \bibinfo {eid} {055013} (\bibinfo {year}
  {2019})},\ \Eprint {https://arxiv.org/abs/1811.02533} {arXiv:1811.02533
  [gr-qc]} \BibitemShut {NoStop}%
\bibitem [{\citenamefont {{Vijaykumar}}\ \emph {et~al.}(2020)\citenamefont
  {{Vijaykumar}}, \citenamefont {{Saketh}}, \citenamefont {{Kumar}},
  \citenamefont {{Ajith}},\ and\ \citenamefont {{Choudhury}}}]{Vijaykumar}%
  \BibitemOpen
  \bibfield  {author} {\bibinfo {author} {\bibfnamefont {A.}~\bibnamefont
  {{Vijaykumar}}}, \bibinfo {author} {\bibfnamefont {M.~V.~S.}\ \bibnamefont
  {{Saketh}}}, \bibinfo {author} {\bibfnamefont {S.}~\bibnamefont {{Kumar}}},
  \bibinfo {author} {\bibfnamefont {P.}~\bibnamefont {{Ajith}}},\ and\ \bibinfo
  {author} {\bibfnamefont {T.~R.}\ \bibnamefont {{Choudhury}}},\ }\href@noop {}
  {\bibfield  {journal} {\bibinfo  {journal} {arXiv e-prints}\ ,\ \bibinfo
  {eid} {arXiv:2005.01111}} (\bibinfo {year} {2020})},\ \Eprint
  {https://arxiv.org/abs/2005.01111} {arXiv:2005.01111 [astro-ph.CO]}
  \BibitemShut {NoStop}%
\bibitem [{\citenamefont {{Biesiada}}\ \emph {et~al.}(2014)\citenamefont
  {{Biesiada}}, \citenamefont {{Ding}}, \citenamefont {{Pi{\'o}rkowska}},\ and\
  \citenamefont {{Zhu}}}]{Biesiada2014}%
  \BibitemOpen
  \bibfield  {author} {\bibinfo {author} {\bibfnamefont {M.}~\bibnamefont
  {{Biesiada}}}, \bibinfo {author} {\bibfnamefont {X.}~\bibnamefont {{Ding}}},
  \bibinfo {author} {\bibfnamefont {A.}~\bibnamefont {{Pi{\'o}rkowska}}},\ and\
  \bibinfo {author} {\bibfnamefont {Z.-H.}\ \bibnamefont {{Zhu}}},\ }\href
  {https://doi.org/10.1088/1475-7516/2014/10/080} {\bibfield  {journal}
  {\bibinfo  {journal} {\jcap}\ }\textbf {\bibinfo {volume} {2014}},\ \bibinfo
  {eid} {080} (\bibinfo {year} {2014})},\ \Eprint
  {https://arxiv.org/abs/1409.8360} {arXiv:1409.8360 [astro-ph.HE]}
  \BibitemShut {NoStop}%
\bibitem [{\citenamefont {{Li}}\ \emph {et~al.}(2018)\citenamefont {{Li}},
  \citenamefont {{Mao}}, \citenamefont {{Zhao}},\ and\ \citenamefont
  {{Lu}}}]{Li2018}%
  \BibitemOpen
  \bibfield  {author} {\bibinfo {author} {\bibfnamefont {S.-S.}\ \bibnamefont
  {{Li}}}, \bibinfo {author} {\bibfnamefont {S.}~\bibnamefont {{Mao}}},
  \bibinfo {author} {\bibfnamefont {Y.}~\bibnamefont {{Zhao}}},\ and\ \bibinfo
  {author} {\bibfnamefont {Y.}~\bibnamefont {{Lu}}},\ }\href
  {https://doi.org/10.1093/mnras/sty411} {\bibfield  {journal} {\bibinfo
  {journal} {\mnras}\ }\textbf {\bibinfo {volume} {476}},\ \bibinfo {pages}
  {2220} (\bibinfo {year} {2018})},\ \Eprint {https://arxiv.org/abs/1802.05089}
  {arXiv:1802.05089 [astro-ph.CO]} \BibitemShut {NoStop}%
\bibitem [{\citenamefont {{Oguri}}(2018)}]{Oguri2018}%
  \BibitemOpen
  \bibfield  {author} {\bibinfo {author} {\bibfnamefont {M.}~\bibnamefont
  {{Oguri}}},\ }\href {https://doi.org/10.1093/mnras/sty2145} {\bibfield
  {journal} {\bibinfo  {journal} {\mnras}\ }\textbf {\bibinfo {volume} {480}},\
  \bibinfo {pages} {3842} (\bibinfo {year} {2018})},\ \Eprint
  {https://arxiv.org/abs/1807.02584} {arXiv:1807.02584 [astro-ph.CO]}
  \BibitemShut {NoStop}%
\bibitem [{\citenamefont {{Yang}}\ \emph {et~al.}(2019)\citenamefont {{Yang}},
  \citenamefont {{Ding}}, \citenamefont {{Biesiada}}, \citenamefont {{Liao}},\
  and\ \citenamefont {{Zhu}}}]{Yang}%
  \BibitemOpen
  \bibfield  {author} {\bibinfo {author} {\bibfnamefont {L.}~\bibnamefont
  {{Yang}}}, \bibinfo {author} {\bibfnamefont {X.}~\bibnamefont {{Ding}}},
  \bibinfo {author} {\bibfnamefont {M.}~\bibnamefont {{Biesiada}}}, \bibinfo
  {author} {\bibfnamefont {K.}~\bibnamefont {{Liao}}},\ and\ \bibinfo {author}
  {\bibfnamefont {Z.-H.}\ \bibnamefont {{Zhu}}},\ }\href
  {https://doi.org/10.3847/1538-4357/ab095c} {\bibfield  {journal} {\bibinfo
  {journal} {\apj}\ }\textbf {\bibinfo {volume} {874}},\ \bibinfo {eid} {139}
  (\bibinfo {year} {2019})},\ \Eprint {https://arxiv.org/abs/1903.11079}
  {arXiv:1903.11079 [astro-ph.GA]} \BibitemShut {NoStop}%
\bibitem [{\citenamefont {{Ding}}\ \emph {et~al.}(2015)\citenamefont {{Ding}},
  \citenamefont {{Biesiada}},\ and\ \citenamefont {{Zhu}}}]{Ding2015}%
  \BibitemOpen
  \bibfield  {author} {\bibinfo {author} {\bibfnamefont {X.}~\bibnamefont
  {{Ding}}}, \bibinfo {author} {\bibfnamefont {M.}~\bibnamefont {{Biesiada}}},\
  and\ \bibinfo {author} {\bibfnamefont {Z.-H.}\ \bibnamefont {{Zhu}}},\ }\href
  {https://doi.org/10.1088/1475-7516/2015/12/006} {\bibfield  {journal}
  {\bibinfo  {journal} {\jcap}\ }\textbf {\bibinfo {volume} {2015}},\ \bibinfo
  {eid} {006} (\bibinfo {year} {2015})},\ \Eprint
  {https://arxiv.org/abs/1508.05000} {arXiv:1508.05000 [astro-ph.HE]}
  \BibitemShut {NoStop}%
\bibitem [{\citenamefont {Liao}\ \emph {et~al.}(2017)\citenamefont {Liao},
  \citenamefont {Fan}, \citenamefont {Ding}, \citenamefont {Biesiada},\ and\
  \citenamefont {Zhu}}]{liao2017precision}%
  \BibitemOpen
  \bibfield  {author} {\bibinfo {author} {\bibfnamefont {K.}~\bibnamefont
  {Liao}}, \bibinfo {author} {\bibfnamefont {X.-L.}\ \bibnamefont {Fan}},
  \bibinfo {author} {\bibfnamefont {X.}~\bibnamefont {Ding}}, \bibinfo {author}
  {\bibfnamefont {M.}~\bibnamefont {Biesiada}},\ and\ \bibinfo {author}
  {\bibfnamefont {Z.-H.}\ \bibnamefont {Zhu}},\ }\href@noop {} {\bibfield
  {journal} {\bibinfo  {journal} {Nature Communications}\ }\textbf {\bibinfo
  {volume} {8}},\ \bibinfo {pages} {1} (\bibinfo {year} {2017})}\BibitemShut
  {NoStop}%
\bibitem [{\citenamefont {{Dai}}\ and\ \citenamefont
  {{Venumadhav}}(2017)}]{Dai2017}%
  \BibitemOpen
  \bibfield  {author} {\bibinfo {author} {\bibfnamefont {L.}~\bibnamefont
  {{Dai}}}\ and\ \bibinfo {author} {\bibfnamefont {T.}~\bibnamefont
  {{Venumadhav}}},\ }\href@noop {} {\bibfield  {journal} {\bibinfo  {journal}
  {arXiv e-prints}\ ,\ \bibinfo {eid} {arXiv:1702.04724}} (\bibinfo {year}
  {2017})},\ \Eprint {https://arxiv.org/abs/1702.04724} {arXiv:1702.04724
  [gr-qc]} \BibitemShut {NoStop}%
\bibitem [{\citenamefont {{Ng}}\ \emph {et~al.}(2018)\citenamefont {{Ng}},
  \citenamefont {{Wong}}, \citenamefont {{Broadhurst}},\ and\ \citenamefont
  {{Li}}}]{Ng2018}%
  \BibitemOpen
  \bibfield  {author} {\bibinfo {author} {\bibfnamefont {K.~K.~Y.}\
  \bibnamefont {{Ng}}}, \bibinfo {author} {\bibfnamefont {K.~W.~K.}\
  \bibnamefont {{Wong}}}, \bibinfo {author} {\bibfnamefont {T.}~\bibnamefont
  {{Broadhurst}}},\ and\ \bibinfo {author} {\bibfnamefont {T.~G.~F.}\
  \bibnamefont {{Li}}},\ }\href {https://doi.org/10.1103/PhysRevD.97.023012}
  {\bibfield  {journal} {\bibinfo  {journal} {\prd}\ }\textbf {\bibinfo
  {volume} {97}},\ \bibinfo {eid} {023012} (\bibinfo {year} {2018})},\ \Eprint
  {https://arxiv.org/abs/1703.06319} {arXiv:1703.06319 [astro-ph.CO]}
  \BibitemShut {NoStop}%
\bibitem [{\citenamefont {Fan}\ \emph {et~al.}(2017)\citenamefont {Fan},
  \citenamefont {Liao}, \citenamefont {Biesiada}, \citenamefont
  {Pi{\'o}rkowska-Kurpas},\ and\ \citenamefont {Zhu}}]{fan2017speed}%
  \BibitemOpen
  \bibfield  {author} {\bibinfo {author} {\bibfnamefont {X.-L.}\ \bibnamefont
  {Fan}}, \bibinfo {author} {\bibfnamefont {K.}~\bibnamefont {Liao}}, \bibinfo
  {author} {\bibfnamefont {M.}~\bibnamefont {Biesiada}}, \bibinfo {author}
  {\bibfnamefont {A.}~\bibnamefont {Pi{\'o}rkowska-Kurpas}},\ and\ \bibinfo
  {author} {\bibfnamefont {Z.-H.}\ \bibnamefont {Zhu}},\ }\href@noop {}
  {\bibfield  {journal} {\bibinfo  {journal} {Physical Review Letters}\
  }\textbf {\bibinfo {volume} {118}},\ \bibinfo {pages} {091102} (\bibinfo
  {year} {2017})}\BibitemShut {NoStop}%
\bibitem [{\citenamefont {{Collett}}\ and\ \citenamefont
  {{Bacon}}(2017)}]{Collett}%
  \BibitemOpen
  \bibfield  {author} {\bibinfo {author} {\bibfnamefont {T.~E.}\ \bibnamefont
  {{Collett}}}\ and\ \bibinfo {author} {\bibfnamefont {D.}~\bibnamefont
  {{Bacon}}},\ }\href {https://doi.org/10.1103/PhysRevLett.118.091101}
  {\bibfield  {journal} {\bibinfo  {journal} {\prl}\ }\textbf {\bibinfo
  {volume} {118}},\ \bibinfo {eid} {091101} (\bibinfo {year} {2017})},\ \Eprint
  {https://arxiv.org/abs/1602.05882} {arXiv:1602.05882 [astro-ph.HE]}
  \BibitemShut {NoStop}%
\bibitem [{\citenamefont {{Mukherjee}}\ \emph
  {et~al.}(2020{\natexlab{a}})\citenamefont {{Mukherjee}}, \citenamefont
  {{Wandelt}},\ and\ \citenamefont {{Silk}}}]{Mukherjee_MM}%
  \BibitemOpen
  \bibfield  {author} {\bibinfo {author} {\bibfnamefont {S.}~\bibnamefont
  {{Mukherjee}}}, \bibinfo {author} {\bibfnamefont {B.~D.}\ \bibnamefont
  {{Wandelt}}},\ and\ \bibinfo {author} {\bibfnamefont {J.}~\bibnamefont
  {{Silk}}},\ }\href {https://doi.org/10.1103/PhysRevD.101.103509} {\bibfield
  {journal} {\bibinfo  {journal} {\prd}\ }\textbf {\bibinfo {volume} {101}},\
  \bibinfo {eid} {103509} (\bibinfo {year} {2020}{\natexlab{a}})},\ \Eprint
  {https://arxiv.org/abs/1908.08950} {arXiv:1908.08950 [astro-ph.CO]}
  \BibitemShut {NoStop}%
\bibitem [{\citenamefont {{Mukherjee}}\ \emph
  {et~al.}(2020{\natexlab{b}})\citenamefont {{Mukherjee}}, \citenamefont
  {{Wandelt}},\ and\ \citenamefont {{Silk}}}]{Mukherjee_Grav}%
  \BibitemOpen
  \bibfield  {author} {\bibinfo {author} {\bibfnamefont {S.}~\bibnamefont
  {{Mukherjee}}}, \bibinfo {author} {\bibfnamefont {B.~D.}\ \bibnamefont
  {{Wandelt}}},\ and\ \bibinfo {author} {\bibfnamefont {J.}~\bibnamefont
  {{Silk}}},\ }\href {https://doi.org/10.1093/mnras/staa827} {\bibfield
  {journal} {\bibinfo  {journal} {\mnras}\ }\textbf {\bibinfo {volume} {494}},\
  \bibinfo {pages} {1956} (\bibinfo {year} {2020}{\natexlab{b}})},\ \Eprint
  {https://arxiv.org/abs/1908.08951} {arXiv:1908.08951 [astro-ph.CO]}
  \BibitemShut {NoStop}%
\bibitem [{\citenamefont {{Dai}}\ \emph {et~al.}(2017)\citenamefont {{Dai}},
  \citenamefont {{Venumadhav}},\ and\ \citenamefont
  {{Sigurdson}}}]{DaiPop2017}%
  \BibitemOpen
  \bibfield  {author} {\bibinfo {author} {\bibfnamefont {L.}~\bibnamefont
  {{Dai}}}, \bibinfo {author} {\bibfnamefont {T.}~\bibnamefont
  {{Venumadhav}}},\ and\ \bibinfo {author} {\bibfnamefont {K.}~\bibnamefont
  {{Sigurdson}}},\ }\href {https://doi.org/10.1103/PhysRevD.95.044011}
  {\bibfield  {journal} {\bibinfo  {journal} {\prd}\ }\textbf {\bibinfo
  {volume} {95}},\ \bibinfo {eid} {044011} (\bibinfo {year} {2017})},\ \Eprint
  {https://arxiv.org/abs/1605.09398} {arXiv:1605.09398 [astro-ph.CO]}
  \BibitemShut {NoStop}%
\bibitem [{\citenamefont {{Hannuksela}}\ \emph {et~al.}(2019)\citenamefont
  {{Hannuksela}}, \citenamefont {{Haris}}, \citenamefont {{Ng}}, \citenamefont
  {{Kumar}}, \citenamefont {{Mehta}}, \citenamefont {{Keitel}}, \citenamefont
  {{Li}},\ and\ \citenamefont {{Ajith}}}]{Hannuksela2019}%
  \BibitemOpen
  \bibfield  {author} {\bibinfo {author} {\bibfnamefont {O.~A.}\ \bibnamefont
  {{Hannuksela}}}, \bibinfo {author} {\bibfnamefont {K.}~\bibnamefont
  {{Haris}}}, \bibinfo {author} {\bibfnamefont {K.~K.~Y.}\ \bibnamefont
  {{Ng}}}, \bibinfo {author} {\bibfnamefont {S.}~\bibnamefont {{Kumar}}},
  \bibinfo {author} {\bibfnamefont {A.~K.}\ \bibnamefont {{Mehta}}}, \bibinfo
  {author} {\bibfnamefont {D.}~\bibnamefont {{Keitel}}}, \bibinfo {author}
  {\bibfnamefont {T.~G.~F.}\ \bibnamefont {{Li}}},\ and\ \bibinfo {author}
  {\bibfnamefont {P.}~\bibnamefont {{Ajith}}},\ }\href
  {https://doi.org/10.3847/2041-8213/ab0c0f} {\bibfield  {journal} {\bibinfo
  {journal} {\apjl}\ }\textbf {\bibinfo {volume} {874}},\ \bibinfo {eid} {L2}
  (\bibinfo {year} {2019})},\ \Eprint {https://arxiv.org/abs/1901.02674}
  {arXiv:1901.02674 [gr-qc]} \BibitemShut {NoStop}%
\bibitem [{\citenamefont {{Dai}}\ \emph {et~al.}(2020)\citenamefont {{Dai}},
  \citenamefont {{Zackay}}, \citenamefont {{Venumadhav}}, \citenamefont
  {{Roulet}},\ and\ \citenamefont {{Zaldarriaga}}}]{Dai2020}%
  \BibitemOpen
  \bibfield  {author} {\bibinfo {author} {\bibfnamefont {L.}~\bibnamefont
  {{Dai}}}, \bibinfo {author} {\bibfnamefont {B.}~\bibnamefont {{Zackay}}},
  \bibinfo {author} {\bibfnamefont {T.}~\bibnamefont {{Venumadhav}}}, \bibinfo
  {author} {\bibfnamefont {J.}~\bibnamefont {{Roulet}}},\ and\ \bibinfo
  {author} {\bibfnamefont {M.}~\bibnamefont {{Zaldarriaga}}},\ }\href@noop {}
  {\bibfield  {journal} {\bibinfo  {journal} {arXiv e-prints}\ ,\ \bibinfo
  {eid} {arXiv:2007.12709}} (\bibinfo {year} {2020})},\ \Eprint
  {https://arxiv.org/abs/2007.12709} {arXiv:2007.12709 [astro-ph.HE]}
  \BibitemShut {NoStop}%
\bibitem [{\citenamefont {{Broadhurst}}\ \emph {et~al.}(2019)\citenamefont
  {{Broadhurst}}, \citenamefont {{Diego}},\ and\ \citenamefont
  {{Smoot}}}]{Broadhurst}%
  \BibitemOpen
  \bibfield  {author} {\bibinfo {author} {\bibfnamefont {T.}~\bibnamefont
  {{Broadhurst}}}, \bibinfo {author} {\bibfnamefont {J.~M.}\ \bibnamefont
  {{Diego}}},\ and\ \bibinfo {author} {\bibfnamefont {I.}~\bibnamefont
  {{Smoot}}, \bibfnamefont {George~F.}},\ }\href@noop {} {\bibfield  {journal}
  {\bibinfo  {journal} {arXiv e-prints}\ ,\ \bibinfo {eid} {arXiv:1901.03190}}
  (\bibinfo {year} {2019})},\ \Eprint {https://arxiv.org/abs/1901.03190}
  {arXiv:1901.03190 [astro-ph.CO]} \BibitemShut {NoStop}%
\bibitem [{\citenamefont {{Broadhurst}}\ \emph {et~al.}(2018)\citenamefont
  {{Broadhurst}}, \citenamefont {{Diego}},\ and\ \citenamefont
  {{Smoot}}}]{Broadhurst_pop}%
  \BibitemOpen
  \bibfield  {author} {\bibinfo {author} {\bibfnamefont {T.}~\bibnamefont
  {{Broadhurst}}}, \bibinfo {author} {\bibfnamefont {J.~M.}\ \bibnamefont
  {{Diego}}},\ and\ \bibinfo {author} {\bibfnamefont {I.}~\bibnamefont
  {{Smoot}}, \bibfnamefont {George}},\ }\href@noop {} {\bibfield  {journal}
  {\bibinfo  {journal} {arXiv e-prints}\ ,\ \bibinfo {eid} {arXiv:1802.05273}}
  (\bibinfo {year} {2018})},\ \Eprint {https://arxiv.org/abs/1802.05273}
  {arXiv:1802.05273 [astro-ph.CO]} \BibitemShut {NoStop}%
\bibitem [{\citenamefont {{Nakamura}}(1998)}]{Nakamura1998}%
  \BibitemOpen
  \bibfield  {author} {\bibinfo {author} {\bibfnamefont {T.~T.}\ \bibnamefont
  {{Nakamura}}},\ }\href {https://doi.org/10.1103/PhysRevLett.80.1138}
  {\bibfield  {journal} {\bibinfo  {journal} {\prl}\ }\textbf {\bibinfo
  {volume} {80}},\ \bibinfo {pages} {1138} (\bibinfo {year}
  {1998})}\BibitemShut {NoStop}%
\bibitem [{\citenamefont {{Takahashi}}\ and\ \citenamefont
  {{Nakamura}}(2003)}]{Takahashi2003}%
  \BibitemOpen
  \bibfield  {author} {\bibinfo {author} {\bibfnamefont {R.}~\bibnamefont
  {{Takahashi}}}\ and\ \bibinfo {author} {\bibfnamefont {T.}~\bibnamefont
  {{Nakamura}}},\ }\href {https://doi.org/10.1086/377430} {\bibfield  {journal}
  {\bibinfo  {journal} {\apj}\ }\textbf {\bibinfo {volume} {595}},\ \bibinfo
  {pages} {1039} (\bibinfo {year} {2003})},\ \Eprint
  {https://arxiv.org/abs/astro-ph/0305055} {arXiv:astro-ph/0305055 [astro-ph]}
  \BibitemShut {NoStop}%
\bibitem [{\citenamefont {{Diego}}\ \emph {et~al.}(2019)\citenamefont
  {{Diego}}, \citenamefont {{Hannuksela}}, \citenamefont {{Kelly}},
  \citenamefont {{Pagano}}, \citenamefont {{Broadhurst}}, \citenamefont
  {{Kim}}, \citenamefont {{Li}},\ and\ \citenamefont {{Smoot}}}]{Diego}%
  \BibitemOpen
  \bibfield  {author} {\bibinfo {author} {\bibfnamefont {J.~M.}\ \bibnamefont
  {{Diego}}}, \bibinfo {author} {\bibfnamefont {O.~A.}\ \bibnamefont
  {{Hannuksela}}}, \bibinfo {author} {\bibfnamefont {P.~L.}\ \bibnamefont
  {{Kelly}}}, \bibinfo {author} {\bibfnamefont {G.}~\bibnamefont {{Pagano}}},
  \bibinfo {author} {\bibfnamefont {T.}~\bibnamefont {{Broadhurst}}}, \bibinfo
  {author} {\bibfnamefont {K.}~\bibnamefont {{Kim}}}, \bibinfo {author}
  {\bibfnamefont {T.~G.~F.}\ \bibnamefont {{Li}}},\ and\ \bibinfo {author}
  {\bibfnamefont {G.~F.}\ \bibnamefont {{Smoot}}},\ }\href
  {https://doi.org/10.1051/0004-6361/201935490} {\bibfield  {journal} {\bibinfo
   {journal} {\aap}\ }\textbf {\bibinfo {volume} {627}},\ \bibinfo {eid} {A130}
  (\bibinfo {year} {2019})},\ \Eprint {https://arxiv.org/abs/1903.04513}
  {arXiv:1903.04513 [astro-ph.CO]} \BibitemShut {NoStop}%
\bibitem [{\citenamefont {{Cheung}}\ \emph {et~al.}(2020)\citenamefont
  {{Cheung}}, \citenamefont {{Gais}}, \citenamefont {{Hannuksela}},\ and\
  \citenamefont {{Li}}}]{Cheung}%
  \BibitemOpen
  \bibfield  {author} {\bibinfo {author} {\bibfnamefont {M.~H.~Y.}\
  \bibnamefont {{Cheung}}}, \bibinfo {author} {\bibfnamefont {J.}~\bibnamefont
  {{Gais}}}, \bibinfo {author} {\bibfnamefont {O.~A.}\ \bibnamefont
  {{Hannuksela}}},\ and\ \bibinfo {author} {\bibfnamefont {T.~G.~F.}\
  \bibnamefont {{Li}}},\ }\href@noop {} {\bibfield  {journal} {\bibinfo
  {journal} {arXiv e-prints}\ ,\ \bibinfo {eid} {arXiv:2012.07800}} (\bibinfo
  {year} {2020})},\ \Eprint {https://arxiv.org/abs/2012.07800}
  {arXiv:2012.07800 [astro-ph.HE]} \BibitemShut {NoStop}%
\bibitem [{\citenamefont {{Schneider}}\ \emph {et~al.}(1992)\citenamefont
  {{Schneider}}, \citenamefont {{Ehlers}},\ and\ \citenamefont
  {{Falco}}}]{Schneider1992}%
  \BibitemOpen
  \bibfield  {author} {\bibinfo {author} {\bibfnamefont {P.}~\bibnamefont
  {{Schneider}}}, \bibinfo {author} {\bibfnamefont {J.}~\bibnamefont
  {{Ehlers}}},\ and\ \bibinfo {author} {\bibfnamefont {E.~E.}\ \bibnamefont
  {{Falco}}},\ }\href {https://doi.org/10.1007/978-3-662-03758-4} {\emph
  {\bibinfo {title} {{Gravitational Lenses}}}}\ (\bibinfo {year}
  {1992})\BibitemShut {NoStop}%
\bibitem [{\citenamefont {{Mar{\'\i}a Ezquiaga}}\ \emph
  {et~al.}(2020)\citenamefont {{Mar{\'\i}a Ezquiaga}}, \citenamefont {{Holz}},
  \citenamefont {{Hu}}, \citenamefont {{Lagos}},\ and\ \citenamefont
  {{Wald}}}]{Esquiaga2020}%
  \BibitemOpen
  \bibfield  {author} {\bibinfo {author} {\bibfnamefont {J.}~\bibnamefont
  {{Mar{\'\i}a Ezquiaga}}}, \bibinfo {author} {\bibfnamefont {D.~E.}\
  \bibnamefont {{Holz}}}, \bibinfo {author} {\bibfnamefont {W.}~\bibnamefont
  {{Hu}}}, \bibinfo {author} {\bibfnamefont {M.}~\bibnamefont {{Lagos}}},\ and\
  \bibinfo {author} {\bibfnamefont {R.~M.}\ \bibnamefont {{Wald}}},\
  }\href@noop {} {\bibfield  {journal} {\bibinfo  {journal} {arXiv e-prints}\
  ,\ \bibinfo {eid} {arXiv:2008.12814}} (\bibinfo {year} {2020})},\ \Eprint
  {https://arxiv.org/abs/2008.12814} {arXiv:2008.12814 [gr-qc]} \BibitemShut
  {NoStop}%
\bibitem [{\citenamefont {{Adhikari}}\ \emph {et~al.}(2020)\citenamefont
  {{Adhikari}}, \citenamefont {{Arai}}, \citenamefont {{Brooks}}, \citenamefont
  {{Wipf}}, \citenamefont {{Aguiar}}, \citenamefont {{Altin}}, \citenamefont
  {{Barr}}, \citenamefont {{Barsotti}}, \citenamefont {{Bassiri}},
  \citenamefont {{Bell}} \emph {et~al.}}]{Voyager2020}%
  \BibitemOpen
  \bibfield  {author} {\bibinfo {author} {\bibfnamefont {R.~X.}\ \bibnamefont
  {{Adhikari}}}, \bibinfo {author} {\bibfnamefont {K.}~\bibnamefont {{Arai}}},
  \bibinfo {author} {\bibfnamefont {A.~F.}\ \bibnamefont {{Brooks}}}, \bibinfo
  {author} {\bibfnamefont {C.}~\bibnamefont {{Wipf}}}, \bibinfo {author}
  {\bibfnamefont {O.}~\bibnamefont {{Aguiar}}}, \bibinfo {author}
  {\bibfnamefont {P.}~\bibnamefont {{Altin}}}, \bibinfo {author} {\bibfnamefont
  {B.}~\bibnamefont {{Barr}}}, \bibinfo {author} {\bibfnamefont
  {L.}~\bibnamefont {{Barsotti}}}, \bibinfo {author} {\bibfnamefont
  {R.}~\bibnamefont {{Bassiri}}}, \bibinfo {author} {\bibfnamefont
  {A.}~\bibnamefont {{Bell}}}, \emph {et~al.},\ }\href
  {https://doi.org/10.1088/1361-6382/ab9143} {\bibfield  {journal} {\bibinfo
  {journal} {Classical and Quantum Gravity}\ }\textbf {\bibinfo {volume}
  {37}},\ \bibinfo {eid} {165003} (\bibinfo {year} {2020})},\ \Eprint
  {https://arxiv.org/abs/2001.11173} {arXiv:2001.11173 [astro-ph.IM]}
  \BibitemShut {NoStop}%
\bibitem [{\citenamefont {{Sathyaprakash}}\ \emph {et~al.}(2011)\citenamefont
  {{Sathyaprakash}}, \citenamefont {{Abernathy}}, \citenamefont {{Acernese}},
  \citenamefont {{Amaro-Seoane}}, \citenamefont {{Andersson}}, \citenamefont
  {{Arun}}, \citenamefont {{Barone}}, \citenamefont {{Barr}}, \citenamefont
  {{Barsuglia}}, \citenamefont {{Beker}}, \citenamefont {{Beveridge}},
  \citenamefont {{Birindelli}}, \citenamefont {{Bose}}, \citenamefont {{Bosi}},
  \citenamefont {{Braccini}}, \citenamefont {{Bradaschia}}, \citenamefont
  {{Bulik}} \emph {et~al.}}]{ET2011}%
  \BibitemOpen
  \bibfield  {author} {\bibinfo {author} {\bibfnamefont {B.}~\bibnamefont
  {{Sathyaprakash}}}, \bibinfo {author} {\bibfnamefont {M.}~\bibnamefont
  {{Abernathy}}}, \bibinfo {author} {\bibfnamefont {F.}~\bibnamefont
  {{Acernese}}}, \bibinfo {author} {\bibfnamefont {P.}~\bibnamefont
  {{Amaro-Seoane}}}, \bibinfo {author} {\bibfnamefont {N.}~\bibnamefont
  {{Andersson}}}, \bibinfo {author} {\bibfnamefont {K.}~\bibnamefont {{Arun}}},
  \bibinfo {author} {\bibfnamefont {F.}~\bibnamefont {{Barone}}}, \bibinfo
  {author} {\bibfnamefont {B.}~\bibnamefont {{Barr}}}, \bibinfo {author}
  {\bibfnamefont {M.}~\bibnamefont {{Barsuglia}}}, \bibinfo {author}
  {\bibfnamefont {M.}~\bibnamefont {{Beker}}}, \bibinfo {author} {\bibfnamefont
  {N.}~\bibnamefont {{Beveridge}}}, \bibinfo {author} {\bibfnamefont
  {S.}~\bibnamefont {{Birindelli}}}, \bibinfo {author} {\bibfnamefont
  {S.}~\bibnamefont {{Bose}}}, \bibinfo {author} {\bibfnamefont
  {L.}~\bibnamefont {{Bosi}}}, \bibinfo {author} {\bibfnamefont
  {S.}~\bibnamefont {{Braccini}}}, \bibinfo {author} {\bibfnamefont
  {C.}~\bibnamefont {{Bradaschia}}}, \bibinfo {author} {\bibfnamefont
  {T.}~\bibnamefont {{Bulik}}}, \emph {et~al.},\ }\href@noop {} {\bibfield
  {journal} {\bibinfo  {journal} {arXiv e-prints}\ ,\ \bibinfo {eid}
  {arXiv:1108.1423}} (\bibinfo {year} {2011})},\ \Eprint
  {https://arxiv.org/abs/1108.1423} {arXiv:1108.1423 [gr-qc]} \BibitemShut
  {NoStop}%
\bibitem [{\citenamefont {{Reitze}}\ \emph {et~al.}(2019)\citenamefont
  {{Reitze}}, \citenamefont {{Adhikari}}, \citenamefont {{Ballmer}},
  \citenamefont {{Barish}}, \citenamefont {{Barsotti}}, \citenamefont
  {{Billingsley}}, \citenamefont {{Brown}}, \citenamefont {{Chen}} \emph
  {et~al.}}]{CE2019}%
  \BibitemOpen
  \bibfield  {author} {\bibinfo {author} {\bibfnamefont {D.}~\bibnamefont
  {{Reitze}}}, \bibinfo {author} {\bibfnamefont {R.~X.}\ \bibnamefont
  {{Adhikari}}}, \bibinfo {author} {\bibfnamefont {S.}~\bibnamefont
  {{Ballmer}}}, \bibinfo {author} {\bibfnamefont {B.}~\bibnamefont {{Barish}}},
  \bibinfo {author} {\bibfnamefont {L.}~\bibnamefont {{Barsotti}}}, \bibinfo
  {author} {\bibfnamefont {G.}~\bibnamefont {{Billingsley}}}, \bibinfo {author}
  {\bibfnamefont {D.~A.}\ \bibnamefont {{Brown}}}, \bibinfo {author}
  {\bibfnamefont {Y.}~\bibnamefont {{Chen}}}, \emph {et~al.},\ }in\ \href@noop
  {} {\emph {\bibinfo {booktitle} {Bulletin of the American Astronomical
  Society}}},\ Vol.~\bibinfo {volume} {51}\ (\bibinfo {year} {2019})\
  p.~\bibinfo {pages} {35},\ \Eprint {https://arxiv.org/abs/1907.04833}
  {arXiv:1907.04833 [astro-ph.IM]} \BibitemShut {NoStop}%
\bibitem [{\citenamefont {{Nakamura}}\ and\ \citenamefont
  {{Deguchi}}(1999)}]{NakamuraDeguchi1999}%
  \BibitemOpen
  \bibfield  {author} {\bibinfo {author} {\bibfnamefont {T.~T.}\ \bibnamefont
  {{Nakamura}}}\ and\ \bibinfo {author} {\bibfnamefont {S.}~\bibnamefont
  {{Deguchi}}},\ }\href {https://doi.org/10.1143/PTPS.133.137} {\bibfield
  {journal} {\bibinfo  {journal} {Progress of Theoretical Physics Supplement}\
  }\textbf {\bibinfo {volume} {133}},\ \bibinfo {pages} {137} (\bibinfo {year}
  {1999})}\BibitemShut {NoStop}%
\bibitem [{\citenamefont {{Favata}}(2010)}]{Favata}%
  \BibitemOpen
  \bibfield  {author} {\bibinfo {author} {\bibfnamefont {M.}~\bibnamefont
  {{Favata}}},\ }\href {https://doi.org/10.1088/0264-9381/27/8/084036}
  {\bibfield  {journal} {\bibinfo  {journal} {Classical and Quantum Gravity}\
  }\textbf {\bibinfo {volume} {27}},\ \bibinfo {eid} {084036} (\bibinfo {year}
  {2010})},\ \Eprint {https://arxiv.org/abs/1003.3486} {arXiv:1003.3486
  [gr-qc]} \BibitemShut {NoStop}%
\bibitem [{\citenamefont {{Favata}}(2009)}]{Favata2009}%
  \BibitemOpen
  \bibfield  {author} {\bibinfo {author} {\bibfnamefont {M.}~\bibnamefont
  {{Favata}}},\ }\href {https://doi.org/10.1103/PhysRevD.80.024002} {\bibfield
  {journal} {\bibinfo  {journal} {\prd}\ }\textbf {\bibinfo {volume} {80}},\
  \bibinfo {eid} {024002} (\bibinfo {year} {2009})},\ \Eprint
  {https://arxiv.org/abs/0812.0069} {arXiv:0812.0069 [gr-qc]} \BibitemShut
  {NoStop}%
\bibitem [{\citenamefont {{H{\"u}bner}}\ \emph {et~al.}(2020)\citenamefont
  {{H{\"u}bner}}, \citenamefont {{Talbot}}, \citenamefont {{Lasky}},\ and\
  \citenamefont {{Thrane}}}]{Hubner}%
  \BibitemOpen
  \bibfield  {author} {\bibinfo {author} {\bibfnamefont {M.}~\bibnamefont
  {{H{\"u}bner}}}, \bibinfo {author} {\bibfnamefont {C.}~\bibnamefont
  {{Talbot}}}, \bibinfo {author} {\bibfnamefont {P.~D.}\ \bibnamefont
  {{Lasky}}},\ and\ \bibinfo {author} {\bibfnamefont {E.}~\bibnamefont
  {{Thrane}}},\ }\href {https://doi.org/10.1103/PhysRevD.101.023011} {\bibfield
   {journal} {\bibinfo  {journal} {\prd}\ }\textbf {\bibinfo {volume} {101}},\
  \bibinfo {eid} {023011} (\bibinfo {year} {2020})},\ \Eprint
  {https://arxiv.org/abs/1911.12496} {arXiv:1911.12496 [astro-ph.HE]}
  \BibitemShut {NoStop}%
\bibitem [{\citenamefont {{Zaldarriaga}}\ \emph {et~al.}(2018)\citenamefont
  {{Zaldarriaga}}, \citenamefont {{Kushnir}},\ and\ \citenamefont
  {{Kollmeier}}}]{Zaldarriaga}%
  \BibitemOpen
  \bibfield  {author} {\bibinfo {author} {\bibfnamefont {M.}~\bibnamefont
  {{Zaldarriaga}}}, \bibinfo {author} {\bibfnamefont {D.}~\bibnamefont
  {{Kushnir}}},\ and\ \bibinfo {author} {\bibfnamefont {J.~A.}\ \bibnamefont
  {{Kollmeier}}},\ }\href {https://doi.org/10.1093/mnras/stx2577} {\bibfield
  {journal} {\bibinfo  {journal} {\mnras}\ }\textbf {\bibinfo {volume} {473}},\
  \bibinfo {pages} {4174} (\bibinfo {year} {2018})},\ \Eprint
  {https://arxiv.org/abs/1702.00885} {arXiv:1702.00885 [astro-ph.HE]}
  \BibitemShut {NoStop}%
\bibitem [{\citenamefont {{Abbott}}\ \emph
  {et~al.}(2019{\natexlab{b}})\citenamefont {{Abbott}}, \citenamefont
  {{Abbott}}, \citenamefont {{Abbott}}, \citenamefont {{Abraham}},
  \citenamefont {{Acernese}}, \citenamefont {{Ackley}}, \citenamefont
  {{Adams}}, \citenamefont {{Adams}}, \citenamefont {{Adhikari}}, \citenamefont
  {{Adya}}, \citenamefont {{Affeldt}}, \citenamefont {{Agathos}}, \citenamefont
  {{Agatsuma}}, \citenamefont {{Aggarwal}}, \citenamefont {{Aguiar}},
  \citenamefont {{Aiello}}, \citenamefont {{Ain}}, \citenamefont {{Ajith}},
  \citenamefont {{Allen}}, \citenamefont {{Allocca}}, \citenamefont {others},
  \citenamefont {{LIGO Scientific Collaboration}},\ and\ \citenamefont {{Virgo
  Collaboration}}}]{search_ecc}%
  \BibitemOpen
  \bibfield  {author} {\bibinfo {author} {\bibfnamefont {B.~P.}\ \bibnamefont
  {{Abbott}}}, \bibinfo {author} {\bibfnamefont {R.}~\bibnamefont {{Abbott}}},
  \bibinfo {author} {\bibfnamefont {T.~D.}\ \bibnamefont {{Abbott}}}, \bibinfo
  {author} {\bibfnamefont {S.}~\bibnamefont {{Abraham}}}, \bibinfo {author}
  {\bibfnamefont {F.}~\bibnamefont {{Acernese}}}, \bibinfo {author}
  {\bibfnamefont {K.}~\bibnamefont {{Ackley}}}, \bibinfo {author}
  {\bibfnamefont {A.}~\bibnamefont {{Adams}}}, \bibinfo {author} {\bibfnamefont
  {C.}~\bibnamefont {{Adams}}}, \bibinfo {author} {\bibfnamefont {R.~X.}\
  \bibnamefont {{Adhikari}}}, \bibinfo {author} {\bibfnamefont {V.~B.}\
  \bibnamefont {{Adya}}}, \bibinfo {author} {\bibfnamefont {C.}~\bibnamefont
  {{Affeldt}}}, \bibinfo {author} {\bibfnamefont {M.}~\bibnamefont
  {{Agathos}}}, \bibinfo {author} {\bibfnamefont {K.}~\bibnamefont
  {{Agatsuma}}}, \bibinfo {author} {\bibfnamefont {N.}~\bibnamefont
  {{Aggarwal}}}, \bibinfo {author} {\bibfnamefont {O.~D.}\ \bibnamefont
  {{Aguiar}}}, \bibinfo {author} {\bibfnamefont {L.}~\bibnamefont {{Aiello}}},
  \bibinfo {author} {\bibfnamefont {A.}~\bibnamefont {{Ain}}}, \bibinfo
  {author} {\bibfnamefont {P.}~\bibnamefont {{Ajith}}}, \bibinfo {author}
  {\bibfnamefont {G.}~\bibnamefont {{Allen}}}, \bibinfo {author} {\bibfnamefont
  {A.}~\bibnamefont {{Allocca}}}, \bibinfo {author} {\bibnamefont {others}},
  \bibinfo {author} {\bibnamefont {{LIGO Scientific Collaboration}}},\ and\
  \bibinfo {author} {\bibnamefont {{Virgo Collaboration}}},\ }\href
  {https://doi.org/10.1103/PhysRevD.100.064064} {\bibfield  {journal} {\bibinfo
   {journal} {\prd}\ }\textbf {\bibinfo {volume} {100}},\ \bibinfo {eid}
  {064064} (\bibinfo {year} {2019}{\natexlab{b}})},\ \Eprint
  {https://arxiv.org/abs/1907.09384} {arXiv:1907.09384 [astro-ph.HE]}
  \BibitemShut {NoStop}%
\bibitem [{\citenamefont {{Varma}}\ \emph {et~al.}(2019)\citenamefont
  {{Varma}}, \citenamefont {{Field}}, \citenamefont {{Scheel}}, \citenamefont
  {{Blackman}}, \citenamefont {{Gerosa}}, \citenamefont {{Stein}},
  \citenamefont {{Kidder}},\ and\ \citenamefont {{Pfeiffer}}}]{sur2019}%
  \BibitemOpen
  \bibfield  {author} {\bibinfo {author} {\bibfnamefont {V.}~\bibnamefont
  {{Varma}}}, \bibinfo {author} {\bibfnamefont {S.~E.}\ \bibnamefont
  {{Field}}}, \bibinfo {author} {\bibfnamefont {M.~A.}\ \bibnamefont
  {{Scheel}}}, \bibinfo {author} {\bibfnamefont {J.}~\bibnamefont
  {{Blackman}}}, \bibinfo {author} {\bibfnamefont {D.}~\bibnamefont
  {{Gerosa}}}, \bibinfo {author} {\bibfnamefont {L.~C.}\ \bibnamefont
  {{Stein}}}, \bibinfo {author} {\bibfnamefont {L.~E.}\ \bibnamefont
  {{Kidder}}},\ and\ \bibinfo {author} {\bibfnamefont {H.~P.}\ \bibnamefont
  {{Pfeiffer}}},\ }\href {https://doi.org/10.1103/PhysRevResearch.1.033015}
  {\bibfield  {journal} {\bibinfo  {journal} {Physical Review Research}\
  }\textbf {\bibinfo {volume} {1}},\ \bibinfo {eid} {033015} (\bibinfo {year}
  {2019})},\ \Eprint {https://arxiv.org/abs/1905.09300} {arXiv:1905.09300
  [gr-qc]} \BibitemShut {NoStop}%
\bibitem [{\citenamefont {{Field}}\ \emph {et~al.}(2018)\citenamefont
  {{Field}}, \citenamefont {{Galley}},\ and\ \citenamefont
  {{Blackman}}}]{gwsur2018}%
  \BibitemOpen
  \bibfield  {author} {\bibinfo {author} {\bibfnamefont {S.}~\bibnamefont
  {{Field}}}, \bibinfo {author} {\bibfnamefont {C.}~\bibnamefont {{Galley}}},\
  and\ \bibinfo {author} {\bibfnamefont {J.}~\bibnamefont {{Blackman}}},\ }in\
  \href@noop {} {\emph {\bibinfo {booktitle} {APS April Meeting Abstracts}}},\
  \bibinfo {series} {APS Meeting Abstracts}, Vol.\ \bibinfo {volume} {2018}\
  (\bibinfo {year} {2018})\ p.\ \bibinfo {pages} {G14.005}\BibitemShut
  {NoStop}%
\bibitem [{\citenamefont {{van der Walt}}\ \emph {et~al.}(2011)\citenamefont
  {{van der Walt}}, \citenamefont {{Colbert}},\ and\ \citenamefont
  {{Varoquaux}}}]{numpy}%
  \BibitemOpen
  \bibfield  {author} {\bibinfo {author} {\bibfnamefont {S.}~\bibnamefont {{van
  der Walt}}}, \bibinfo {author} {\bibfnamefont {S.~C.}\ \bibnamefont
  {{Colbert}}},\ and\ \bibinfo {author} {\bibfnamefont {G.}~\bibnamefont
  {{Varoquaux}}},\ }\href {https://doi.org/10.1109/MCSE.2011.37} {\bibfield
  {journal} {\bibinfo  {journal} {Computing in Science and Engineering}\
  }\textbf {\bibinfo {volume} {13}},\ \bibinfo {pages} {22} (\bibinfo {year}
  {2011})},\ \Eprint {https://arxiv.org/abs/1102.1523} {arXiv:1102.1523
  [cs.MS]} \BibitemShut {NoStop}%
\bibitem [{\citenamefont {{Virtanen}}\ \emph {et~al.}(2020)\citenamefont
  {{Virtanen}}, \citenamefont {{Gommers}}, \citenamefont {{Oliphant}},
  \citenamefont {{Haberland}}, \citenamefont {{Reddy}}, \citenamefont
  {{Cournapeau}}, \citenamefont {{Burovski}}, \citenamefont {{Peterson}},
  \citenamefont {{Weckesser}}, \citenamefont {{Bright}}, \citenamefont {{van
  der Walt}} \emph {et~al.}}]{scipy}%
  \BibitemOpen
  \bibfield  {author} {\bibinfo {author} {\bibfnamefont {P.}~\bibnamefont
  {{Virtanen}}}, \bibinfo {author} {\bibfnamefont {R.}~\bibnamefont
  {{Gommers}}}, \bibinfo {author} {\bibfnamefont {T.~E.}\ \bibnamefont
  {{Oliphant}}}, \bibinfo {author} {\bibfnamefont {M.}~\bibnamefont
  {{Haberland}}}, \bibinfo {author} {\bibfnamefont {T.}~\bibnamefont
  {{Reddy}}}, \bibinfo {author} {\bibfnamefont {D.}~\bibnamefont
  {{Cournapeau}}}, \bibinfo {author} {\bibfnamefont {E.}~\bibnamefont
  {{Burovski}}}, \bibinfo {author} {\bibfnamefont {P.}~\bibnamefont
  {{Peterson}}}, \bibinfo {author} {\bibfnamefont {W.}~\bibnamefont
  {{Weckesser}}}, \bibinfo {author} {\bibfnamefont {J.}~\bibnamefont
  {{Bright}}}, \bibinfo {author} {\bibfnamefont {S.~J.}\ \bibnamefont {{van der
  Walt}}}, \emph {et~al.},\ }\href
  {https://doi.org/https://doi.org/10.1038/s41592-019-0686-2} {\bibfield
  {journal} {\bibinfo  {journal} {Nature Methods}\ }\textbf {\bibinfo {volume}
  {17}},\ \bibinfo {pages} {261} (\bibinfo {year} {2020})}\BibitemShut
  {NoStop}%
\bibitem [{\citenamefont {{Moore}}\ \emph {et~al.}(2018)\citenamefont
  {{Moore}}, \citenamefont {{Robson}}, \citenamefont {{Loutrel}},\ and\
  \citenamefont {{Yunes}}}]{Moore2018}%
  \BibitemOpen
  \bibfield  {author} {\bibinfo {author} {\bibfnamefont {B.}~\bibnamefont
  {{Moore}}}, \bibinfo {author} {\bibfnamefont {T.}~\bibnamefont {{Robson}}},
  \bibinfo {author} {\bibfnamefont {N.}~\bibnamefont {{Loutrel}}},\ and\
  \bibinfo {author} {\bibfnamefont {N.}~\bibnamefont {{Yunes}}},\ }\href
  {https://doi.org/10.1088/1361-6382/aaea00} {\bibfield  {journal} {\bibinfo
  {journal} {Classical and Quantum Gravity}\ }\textbf {\bibinfo {volume}
  {35}},\ \bibinfo {eid} {235006} (\bibinfo {year} {2018})},\ \Eprint
  {https://arxiv.org/abs/1807.07163} {arXiv:1807.07163 [gr-qc]} \BibitemShut
  {NoStop}%
\bibitem [{\citenamefont {{Evans}}\ \emph {et~al.}(2018)\citenamefont
  {{Evans}}, \citenamefont {{Sturani}}, \citenamefont {{Vitale}},\ and\
  \citenamefont {{Hall}}}]{VoyagerPSD}%
  \BibitemOpen
  \bibfield  {author} {\bibinfo {author} {\bibfnamefont {M.}~\bibnamefont
  {{Evans}}}, \bibinfo {author} {\bibfnamefont {R.}~\bibnamefont {{Sturani}}},
  \bibinfo {author} {\bibfnamefont {S.}~\bibnamefont {{Vitale}}},\ and\
  \bibinfo {author} {\bibfnamefont {E.}~\bibnamefont {{Hall}}} (\bibinfo
  {collaboration} {LIGO Scientific Collaboration}),\ }\href
  {https://dcc.ligo.org/LIGO-T1500293-v11/public} {\emph {\bibinfo {title}
  {{Unofficial sensitivity curves (ASD) for aLIGO, Kagra, Virgo, Voyager,
  Cosmic Explorer and ET}}}},\ \bibinfo {type} {Tech. Rep.}\ \bibinfo {number}
  {LIGO-T1500293-v11}\ (\bibinfo {year} {2018})\BibitemShut {NoStop}%
\bibitem [{\citenamefont {{Allen}}(2005)}]{Allen2005}%
  \BibitemOpen
  \bibfield  {author} {\bibinfo {author} {\bibfnamefont {B.}~\bibnamefont
  {{Allen}}},\ }\href {https://doi.org/10.1103/PhysRevD.71.062001} {\bibfield
  {journal} {\bibinfo  {journal} {\prd}\ }\textbf {\bibinfo {volume} {71}},\
  \bibinfo {eid} {062001} (\bibinfo {year} {2005})},\ \Eprint
  {https://arxiv.org/abs/gr-qc/0405045} {arXiv:gr-qc/0405045 [gr-qc]}
  \BibitemShut {NoStop}%
\bibitem [{\citenamefont {{Li}}\ \emph {et~al.}(2019)\citenamefont {{Li}},
  \citenamefont {{Lo}}, \citenamefont {{Sachdev}}, \citenamefont {{Chan}},
  \citenamefont {{Lin}}, \citenamefont {{Li}},\ and\ \citenamefont
  {{Weinstein}}}]{Li:2019osa}%
  \BibitemOpen
  \bibfield  {author} {\bibinfo {author} {\bibfnamefont {A.~K.~Y.}\
  \bibnamefont {{Li}}}, \bibinfo {author} {\bibfnamefont {R.~K.~L.}\
  \bibnamefont {{Lo}}}, \bibinfo {author} {\bibfnamefont {S.}~\bibnamefont
  {{Sachdev}}}, \bibinfo {author} {\bibfnamefont {C.~L.}\ \bibnamefont
  {{Chan}}}, \bibinfo {author} {\bibfnamefont {E.~T.}\ \bibnamefont {{Lin}}},
  \bibinfo {author} {\bibfnamefont {T.~G.~F.}\ \bibnamefont {{Li}}},\ and\
  \bibinfo {author} {\bibfnamefont {A.~J.}\ \bibnamefont {{Weinstein}}},\
  }\href@noop {} {\bibfield  {journal} {\bibinfo  {journal} {arXiv e-prints}\
  ,\ \bibinfo {eid} {arXiv:1904.06020}} (\bibinfo {year} {2019})},\ \Eprint
  {https://arxiv.org/abs/1904.06020} {arXiv:1904.06020 [gr-qc]} \BibitemShut
  {NoStop}%
\bibitem [{\citenamefont {{McIsaac}}\ \emph {et~al.}(2020)\citenamefont
  {{McIsaac}}, \citenamefont {{Keitel}}, \citenamefont {{Collett}},
  \citenamefont {{Harry}}, \citenamefont {{Mozzon}}, \citenamefont {{Edy}},\
  and\ \citenamefont {{Bacon}}}]{McIsaac:2019use}%
  \BibitemOpen
  \bibfield  {author} {\bibinfo {author} {\bibfnamefont {C.}~\bibnamefont
  {{McIsaac}}}, \bibinfo {author} {\bibfnamefont {D.}~\bibnamefont {{Keitel}}},
  \bibinfo {author} {\bibfnamefont {T.}~\bibnamefont {{Collett}}}, \bibinfo
  {author} {\bibfnamefont {I.}~\bibnamefont {{Harry}}}, \bibinfo {author}
  {\bibfnamefont {S.}~\bibnamefont {{Mozzon}}}, \bibinfo {author}
  {\bibfnamefont {O.}~\bibnamefont {{Edy}}},\ and\ \bibinfo {author}
  {\bibfnamefont {D.}~\bibnamefont {{Bacon}}},\ }\href
  {https://doi.org/10.1103/PhysRevD.102.084031} {\bibfield  {journal} {\bibinfo
   {journal} {\prd}\ }\textbf {\bibinfo {volume} {102}},\ \bibinfo {eid}
  {084031} (\bibinfo {year} {2020})},\ \Eprint
  {https://arxiv.org/abs/1912.05389} {arXiv:1912.05389 [gr-qc]} \BibitemShut
  {NoStop}%
\bibitem [{\citenamefont {{Abbott}}\ \emph
  {et~al.}(2020{\natexlab{b}})\citenamefont {{Abbott}}, \citenamefont
  {{Abbott}}, \citenamefont {{Abraham}}, \citenamefont {{Acernese}},
  \citenamefont {{Ackley}}, \citenamefont {{Adams}}, \citenamefont
  {{Adhikari}}, \citenamefont {{Adya}}, \citenamefont {{Affeldt}},
  \citenamefont {{Agathos}}, \citenamefont {{Agatsuma}}, \citenamefont
  {others}, \citenamefont {{LIGO Scientific Collaboration}},\ and\
  \citenamefont {{Virgo Collaboration}}}]{GW190814}%
  \BibitemOpen
  \bibfield  {author} {\bibinfo {author} {\bibfnamefont {R.}~\bibnamefont
  {{Abbott}}}, \bibinfo {author} {\bibfnamefont {T.~D.}\ \bibnamefont
  {{Abbott}}}, \bibinfo {author} {\bibfnamefont {S.}~\bibnamefont {{Abraham}}},
  \bibinfo {author} {\bibfnamefont {F.}~\bibnamefont {{Acernese}}}, \bibinfo
  {author} {\bibfnamefont {K.}~\bibnamefont {{Ackley}}}, \bibinfo {author}
  {\bibfnamefont {C.}~\bibnamefont {{Adams}}}, \bibinfo {author} {\bibfnamefont
  {R.~X.}\ \bibnamefont {{Adhikari}}}, \bibinfo {author} {\bibfnamefont
  {V.~B.}\ \bibnamefont {{Adya}}}, \bibinfo {author} {\bibfnamefont
  {C.}~\bibnamefont {{Affeldt}}}, \bibinfo {author} {\bibfnamefont
  {M.}~\bibnamefont {{Agathos}}}, \bibinfo {author} {\bibfnamefont
  {K.}~\bibnamefont {{Agatsuma}}}, \bibinfo {author} {\bibnamefont {others}},
  \bibinfo {author} {\bibnamefont {{LIGO Scientific Collaboration}}},\ and\
  \bibinfo {author} {\bibnamefont {{Virgo Collaboration}}},\ }\href
  {https://doi.org/10.3847/2041-8213/ab960f} {\bibfield  {journal} {\bibinfo
  {journal} {\apjl}\ }\textbf {\bibinfo {volume} {896}},\ \bibinfo {eid} {L44}
  (\bibinfo {year} {2020}{\natexlab{b}})},\ \Eprint
  {https://arxiv.org/abs/2006.12611} {arXiv:2006.12611 [astro-ph.HE]}
  \BibitemShut {NoStop}%
\bibitem [{\citenamefont {{Pratten}}\ \emph {et~al.}(2020)\citenamefont
  {{Pratten}}, \citenamefont {{Garc{\'\i}a-Quir{\'o}s}}, \citenamefont
  {{Colleoni}}, \citenamefont {{Ramos-Buades}}, \citenamefont {{Estell{\'e}s}},
  \citenamefont {{Mateu-Lucena}}, \citenamefont {{Jaume}}, \citenamefont
  {{Haney}}, \citenamefont {{Keitel}}, \citenamefont {{Thompson}},\ and\
  \citenamefont {{Husa}}}]{Pratten2020}%
  \BibitemOpen
  \bibfield  {author} {\bibinfo {author} {\bibfnamefont {G.}~\bibnamefont
  {{Pratten}}}, \bibinfo {author} {\bibfnamefont {C.}~\bibnamefont
  {{Garc{\'\i}a-Quir{\'o}s}}}, \bibinfo {author} {\bibfnamefont
  {M.}~\bibnamefont {{Colleoni}}}, \bibinfo {author} {\bibfnamefont
  {A.}~\bibnamefont {{Ramos-Buades}}}, \bibinfo {author} {\bibfnamefont
  {H.}~\bibnamefont {{Estell{\'e}s}}}, \bibinfo {author} {\bibfnamefont
  {M.}~\bibnamefont {{Mateu-Lucena}}}, \bibinfo {author} {\bibfnamefont
  {R.}~\bibnamefont {{Jaume}}}, \bibinfo {author} {\bibfnamefont
  {M.}~\bibnamefont {{Haney}}}, \bibinfo {author} {\bibfnamefont
  {D.}~\bibnamefont {{Keitel}}}, \bibinfo {author} {\bibfnamefont {J.~E.}\
  \bibnamefont {{Thompson}}},\ and\ \bibinfo {author} {\bibfnamefont
  {S.}~\bibnamefont {{Husa}}},\ }\href@noop {} {\bibfield  {journal} {\bibinfo
  {journal} {arXiv e-prints}\ ,\ \bibinfo {eid} {arXiv:2004.06503}} (\bibinfo
  {year} {2020})},\ \Eprint {https://arxiv.org/abs/2004.06503}
  {arXiv:2004.06503 [gr-qc]} \BibitemShut {NoStop}%
\bibitem [{\citenamefont {{Messick}}\ \emph {et~al.}(2017)\citenamefont
  {{Messick}}, \citenamefont {{Blackburn}}, \citenamefont {{Brady}},
  \citenamefont {{Brockill}}, \citenamefont {{Cannon}}, \citenamefont
  {{Cariou}}, \citenamefont {{Caudill}}, \citenamefont {{Chamberlin}},
  \citenamefont {{Creighton}}, \citenamefont {{Everett}}, \citenamefont
  {{Hanna}}, \citenamefont {{Keppel}}, \citenamefont {{Lang}}, \citenamefont
  {{Li}}, \citenamefont {{Meacher}}, \citenamefont {{Nielsen}}, \citenamefont
  {{Pankow}}, \citenamefont {{Privitera}}, \citenamefont {{Qi}}, \citenamefont
  {{Sachdev}}, \citenamefont {{Sadeghian}}, \citenamefont {{Singer}},
  \citenamefont {{Thomas}}, \citenamefont {{Wade}}, \citenamefont {{Wade}},
  \citenamefont {{Weinstein}},\ and\ \citenamefont {{Wiesner}}}]{gstlal2017}%
  \BibitemOpen
  \bibfield  {author} {\bibinfo {author} {\bibfnamefont {C.}~\bibnamefont
  {{Messick}}}, \bibinfo {author} {\bibfnamefont {K.}~\bibnamefont
  {{Blackburn}}}, \bibinfo {author} {\bibfnamefont {P.}~\bibnamefont
  {{Brady}}}, \bibinfo {author} {\bibfnamefont {P.}~\bibnamefont {{Brockill}}},
  \bibinfo {author} {\bibfnamefont {K.}~\bibnamefont {{Cannon}}}, \bibinfo
  {author} {\bibfnamefont {R.}~\bibnamefont {{Cariou}}}, \bibinfo {author}
  {\bibfnamefont {S.}~\bibnamefont {{Caudill}}}, \bibinfo {author}
  {\bibfnamefont {S.~J.}\ \bibnamefont {{Chamberlin}}}, \bibinfo {author}
  {\bibfnamefont {J.~D.~E.}\ \bibnamefont {{Creighton}}}, \bibinfo {author}
  {\bibfnamefont {R.}~\bibnamefont {{Everett}}}, \bibinfo {author}
  {\bibfnamefont {C.}~\bibnamefont {{Hanna}}}, \bibinfo {author} {\bibfnamefont
  {D.}~\bibnamefont {{Keppel}}}, \bibinfo {author} {\bibfnamefont {R.~N.}\
  \bibnamefont {{Lang}}}, \bibinfo {author} {\bibfnamefont {T.~G.~F.}\
  \bibnamefont {{Li}}}, \bibinfo {author} {\bibfnamefont {D.}~\bibnamefont
  {{Meacher}}}, \bibinfo {author} {\bibfnamefont {A.}~\bibnamefont
  {{Nielsen}}}, \bibinfo {author} {\bibfnamefont {C.}~\bibnamefont {{Pankow}}},
  \bibinfo {author} {\bibfnamefont {S.}~\bibnamefont {{Privitera}}}, \bibinfo
  {author} {\bibfnamefont {H.}~\bibnamefont {{Qi}}}, \bibinfo {author}
  {\bibfnamefont {S.}~\bibnamefont {{Sachdev}}}, \bibinfo {author}
  {\bibfnamefont {L.}~\bibnamefont {{Sadeghian}}}, \bibinfo {author}
  {\bibfnamefont {L.}~\bibnamefont {{Singer}}}, \bibinfo {author}
  {\bibfnamefont {E.~G.}\ \bibnamefont {{Thomas}}}, \bibinfo {author}
  {\bibfnamefont {L.}~\bibnamefont {{Wade}}}, \bibinfo {author} {\bibfnamefont
  {M.}~\bibnamefont {{Wade}}}, \bibinfo {author} {\bibfnamefont
  {A.}~\bibnamefont {{Weinstein}}},\ and\ \bibinfo {author} {\bibfnamefont
  {K.}~\bibnamefont {{Wiesner}}},\ }\href
  {https://doi.org/10.1103/PhysRevD.95.042001} {\bibfield  {journal} {\bibinfo
  {journal} {\prd}\ }\textbf {\bibinfo {volume} {95}},\ \bibinfo {eid} {042001}
  (\bibinfo {year} {2017})},\ \Eprint {https://arxiv.org/abs/1604.04324}
  {arXiv:1604.04324 [astro-ph.IM]} \BibitemShut {NoStop}%
\bibitem [{\citenamefont {Abbott}\ \emph {et~al.}(2019)\citenamefont {Abbott}
  \emph {et~al.}}]{GWTC1:2018mvr}%
  \BibitemOpen
  \bibfield  {author} {\bibinfo {author} {\bibfnamefont {B.~P.}\ \bibnamefont
  {Abbott}} \emph {et~al.} (\bibinfo {collaboration} {LIGO Scientific,
  Virgo}),\ }\href {https://doi.org/10.1103/PhysRevX.9.031040} {\bibfield
  {journal} {\bibinfo  {journal} {Phys. Rev. X}\ }\textbf {\bibinfo {volume}
  {9}},\ \bibinfo {pages} {031040} (\bibinfo {year} {2019})},\ \Eprint
  {https://arxiv.org/abs/1811.12907} {arXiv:1811.12907 [astro-ph.HE]}
  \BibitemShut {NoStop}%
\bibitem [{\citenamefont {Cornish}\ \emph {et~al.}(2011)\citenamefont
  {Cornish}, \citenamefont {Sampson}, \citenamefont {Yunes},\ and\
  \citenamefont {Pretorius}}]{Cornish:2011ys}%
  \BibitemOpen
  \bibfield  {author} {\bibinfo {author} {\bibfnamefont {N.}~\bibnamefont
  {Cornish}}, \bibinfo {author} {\bibfnamefont {L.}~\bibnamefont {Sampson}},
  \bibinfo {author} {\bibfnamefont {N.}~\bibnamefont {Yunes}},\ and\ \bibinfo
  {author} {\bibfnamefont {F.}~\bibnamefont {Pretorius}},\ }\href
  {https://doi.org/10.1103/PhysRevD.84.062003} {\bibfield  {journal} {\bibinfo
  {journal} {Phys. Rev. D}\ }\textbf {\bibinfo {volume} {84}},\ \bibinfo
  {pages} {062003} (\bibinfo {year} {2011})},\ \Eprint
  {https://arxiv.org/abs/1105.2088} {arXiv:1105.2088 [gr-qc]} \BibitemShut
  {NoStop}%
\bibitem [{\citenamefont {Vallisneri}(2012)}]{Vallisneri:2012qq}%
  \BibitemOpen
  \bibfield  {author} {\bibinfo {author} {\bibfnamefont {M.}~\bibnamefont
  {Vallisneri}},\ }\href {https://doi.org/10.1103/PhysRevD.86.082001}
  {\bibfield  {journal} {\bibinfo  {journal} {Phys. Rev. D}\ }\textbf {\bibinfo
  {volume} {86}},\ \bibinfo {pages} {082001} (\bibinfo {year} {2012})},\
  \Eprint {https://arxiv.org/abs/1207.4759} {arXiv:1207.4759 [gr-qc]}
  \BibitemShut {NoStop}%
\bibitem [{\citenamefont {Del~Pozzo}\ \emph {et~al.}(2014)\citenamefont
  {Del~Pozzo}, \citenamefont {Grover}, \citenamefont {Mandel},\ and\
  \citenamefont {Vecchio}}]{DelPozzo:2014cla}%
  \BibitemOpen
  \bibfield  {author} {\bibinfo {author} {\bibfnamefont {W.}~\bibnamefont
  {Del~Pozzo}}, \bibinfo {author} {\bibfnamefont {K.}~\bibnamefont {Grover}},
  \bibinfo {author} {\bibfnamefont {I.}~\bibnamefont {Mandel}},\ and\ \bibinfo
  {author} {\bibfnamefont {A.}~\bibnamefont {Vecchio}},\ }\href
  {https://doi.org/10.1088/0264-9381/31/20/205006} {\bibfield  {journal}
  {\bibinfo  {journal} {Class. Quant. Grav.}\ }\textbf {\bibinfo {volume}
  {31}},\ \bibinfo {pages} {205006} (\bibinfo {year} {2014})},\ \Eprint
  {https://arxiv.org/abs/1408.2356} {arXiv:1408.2356 [gr-qc]} \BibitemShut
  {NoStop}%
\bibitem [{\citenamefont {Ashton}\ \emph {et~al.}(2019)\citenamefont {Ashton}
  \emph {et~al.}}]{Ashton:2018jfp}%
  \BibitemOpen
  \bibfield  {author} {\bibinfo {author} {\bibfnamefont {G.}~\bibnamefont
  {Ashton}} \emph {et~al.},\ }\href {https://doi.org/10.3847/1538-4365/ab06fc}
  {\bibfield  {journal} {\bibinfo  {journal} {Astrophys. J. Suppl.}\ }\textbf
  {\bibinfo {volume} {241}},\ \bibinfo {pages} {27} (\bibinfo {year} {2019})},\
  \Eprint {https://arxiv.org/abs/1811.02042} {arXiv:1811.02042 [astro-ph.IM]}
  \BibitemShut {NoStop}%
\bibitem [{\citenamefont {{Speagle}}(2020)}]{2020MNRAS.493.3132S}%
  \BibitemOpen
  \bibfield  {author} {\bibinfo {author} {\bibfnamefont {J.~S.}\ \bibnamefont
  {{Speagle}}},\ }\href {https://doi.org/10.1093/mnras/staa278} {\bibfield
  {journal} {\bibinfo  {journal} {\mnras}\ }\textbf {\bibinfo {volume} {493}},\
  \bibinfo {pages} {3132} (\bibinfo {year} {2020})},\ \Eprint
  {https://arxiv.org/abs/1904.02180} {arXiv:1904.02180 [astro-ph.IM]}
  \BibitemShut {NoStop}%
\bibitem [{\citenamefont {Garc\'\i{}a-Quir\'os}\ \emph
  {et~al.}(2020)\citenamefont {Garc\'\i{}a-Quir\'os}, \citenamefont {Colleoni},
  \citenamefont {Husa}, \citenamefont {Estell\'es}, \citenamefont {Pratten},
  \citenamefont {Ramos-Buades}, \citenamefont {Mateu-Lucena},\ and\
  \citenamefont {Jaume}}]{Garcia-Quiros:2020qpx}%
  \BibitemOpen
  \bibfield  {author} {\bibinfo {author} {\bibfnamefont {C.}~\bibnamefont
  {Garc\'\i{}a-Quir\'os}}, \bibinfo {author} {\bibfnamefont {M.}~\bibnamefont
  {Colleoni}}, \bibinfo {author} {\bibfnamefont {S.}~\bibnamefont {Husa}},
  \bibinfo {author} {\bibfnamefont {H.}~\bibnamefont {Estell\'es}}, \bibinfo
  {author} {\bibfnamefont {G.}~\bibnamefont {Pratten}}, \bibinfo {author}
  {\bibfnamefont {A.}~\bibnamefont {Ramos-Buades}}, \bibinfo {author}
  {\bibfnamefont {M.}~\bibnamefont {Mateu-Lucena}},\ and\ \bibinfo {author}
  {\bibfnamefont {R.}~\bibnamefont {Jaume}},\ }\href
  {https://doi.org/10.1103/PhysRevD.102.064002} {\bibfield  {journal} {\bibinfo
   {journal} {Phys. Rev. D}\ }\textbf {\bibinfo {volume} {102}},\ \bibinfo
  {pages} {064002} (\bibinfo {year} {2020})},\ \Eprint
  {https://arxiv.org/abs/2001.10914} {arXiv:2001.10914 [gr-qc]} \BibitemShut
  {NoStop}%
\bibitem [{\citenamefont {{Cutler}}\ and\ \citenamefont
  {{Flanagan}}(1994)}]{Cutler1994}%
  \BibitemOpen
  \bibfield  {author} {\bibinfo {author} {\bibfnamefont {C.}~\bibnamefont
  {{Cutler}}}\ and\ \bibinfo {author} {\bibfnamefont {{\'E}.~E.}\ \bibnamefont
  {{Flanagan}}},\ }\href {https://doi.org/10.1103/PhysRevD.49.2658} {\bibfield
  {journal} {\bibinfo  {journal} {\prd}\ }\textbf {\bibinfo {volume} {49}},\
  \bibinfo {pages} {2658} (\bibinfo {year} {1994})},\ \Eprint
  {https://arxiv.org/abs/gr-qc/9402014} {arXiv:gr-qc/9402014 [gr-qc]}
  \BibitemShut {NoStop}%
\bibitem [{\citenamefont {{Sesana}}\ \emph {et~al.}(2004)\citenamefont
  {{Sesana}}, \citenamefont {{Haardt}}, \citenamefont {{Madau}},\ and\
  \citenamefont {{Volonteri}}}]{Sesana2004}%
  \BibitemOpen
  \bibfield  {author} {\bibinfo {author} {\bibfnamefont {A.}~\bibnamefont
  {{Sesana}}}, \bibinfo {author} {\bibfnamefont {F.}~\bibnamefont {{Haardt}}},
  \bibinfo {author} {\bibfnamefont {P.}~\bibnamefont {{Madau}}},\ and\ \bibinfo
  {author} {\bibfnamefont {M.}~\bibnamefont {{Volonteri}}},\ }\href
  {https://doi.org/10.1086/422185} {\bibfield  {journal} {\bibinfo  {journal}
  {\apj}\ }\textbf {\bibinfo {volume} {611}},\ \bibinfo {pages} {623} (\bibinfo
  {year} {2004})},\ \Eprint {https://arxiv.org/abs/astro-ph/0401543}
  {arXiv:astro-ph/0401543 [astro-ph]} \BibitemShut {NoStop}%
\bibitem [{\citenamefont {{Kalaghatgi}}\ \emph {et~al.}(2020)\citenamefont
  {{Kalaghatgi}}, \citenamefont {{Hannam}},\ and\ \citenamefont
  {{Raymond}}}]{PhenomHM}%
  \BibitemOpen
  \bibfield  {author} {\bibinfo {author} {\bibfnamefont {C.}~\bibnamefont
  {{Kalaghatgi}}}, \bibinfo {author} {\bibfnamefont {M.}~\bibnamefont
  {{Hannam}}},\ and\ \bibinfo {author} {\bibfnamefont {V.}~\bibnamefont
  {{Raymond}}},\ }\href {https://doi.org/10.1103/PhysRevD.101.103004}
  {\bibfield  {journal} {\bibinfo  {journal} {\prd}\ }\textbf {\bibinfo
  {volume} {101}},\ \bibinfo {eid} {103004} (\bibinfo {year} {2020})},\ \Eprint
  {https://arxiv.org/abs/1909.10010} {arXiv:1909.10010 [gr-qc]} \BibitemShut
  {NoStop}%
\bibitem [{\citenamefont {{Nitz}}\ \emph {et~al.}(2020)\citenamefont {{Nitz}},
  \citenamefont {{Harry}}, \citenamefont {{Brown}}, \citenamefont {{Biwer}},
  \citenamefont {{Willis}}, \citenamefont {{Dal Canton}}, \citenamefont
  {{Capano}}, \citenamefont {{Pekowsky}}, \citenamefont {{Dent}}, \citenamefont
  {{Williamson}}, \citenamefont {{Davies}}, \citenamefont {{De}}, \citenamefont
  {{Cabero}}, \citenamefont {{Machenschalk}}, \citenamefont {{Kumar}},
  \citenamefont {{Reyes}}, \citenamefont {{Macleod}}, \citenamefont
  {{Dfinstad}}, \citenamefont {{Pannarale}}, \citenamefont {{Massinger}},
  \citenamefont {{Kumar}}, \citenamefont {{T{\'a}pai}}, \citenamefont
  {{Singer}}, \citenamefont {{Khan}}, \citenamefont {{Fairhurst}},
  \citenamefont {{Nielsen}}, \citenamefont {{Singh}},\ and\ \citenamefont
  {{Shasvath}}}]{pycbc}%
  \BibitemOpen
  \bibfield  {author} {\bibinfo {author} {\bibfnamefont {A.}~\bibnamefont
  {{Nitz}}}, \bibinfo {author} {\bibfnamefont {I.}~\bibnamefont {{Harry}}},
  \bibinfo {author} {\bibfnamefont {D.}~\bibnamefont {{Brown}}}, \bibinfo
  {author} {\bibfnamefont {C.~M.}\ \bibnamefont {{Biwer}}}, \bibinfo {author}
  {\bibfnamefont {J.}~\bibnamefont {{Willis}}}, \bibinfo {author}
  {\bibfnamefont {T.}~\bibnamefont {{Dal Canton}}}, \bibinfo {author}
  {\bibfnamefont {C.}~\bibnamefont {{Capano}}}, \bibinfo {author}
  {\bibfnamefont {L.}~\bibnamefont {{Pekowsky}}}, \bibinfo {author}
  {\bibfnamefont {T.}~\bibnamefont {{Dent}}}, \bibinfo {author} {\bibfnamefont
  {A.~R.}\ \bibnamefont {{Williamson}}}, \bibinfo {author} {\bibfnamefont
  {G.~S.}\ \bibnamefont {{Davies}}}, \bibinfo {author} {\bibfnamefont
  {S.}~\bibnamefont {{De}}}, \bibinfo {author} {\bibfnamefont {M.}~\bibnamefont
  {{Cabero}}}, \bibinfo {author} {\bibfnamefont {B.}~\bibnamefont
  {{Machenschalk}}}, \bibinfo {author} {\bibfnamefont {P.}~\bibnamefont
  {{Kumar}}}, \bibinfo {author} {\bibfnamefont {S.}~\bibnamefont {{Reyes}}},
  \bibinfo {author} {\bibfnamefont {D.}~\bibnamefont {{Macleod}}}, \bibinfo
  {author} {\bibnamefont {{Dfinstad}}}, \bibinfo {author} {\bibfnamefont
  {F.}~\bibnamefont {{Pannarale}}}, \bibinfo {author} {\bibfnamefont
  {T.}~\bibnamefont {{Massinger}}}, \bibinfo {author} {\bibfnamefont
  {S.}~\bibnamefont {{Kumar}}}, \bibinfo {author} {\bibfnamefont
  {M.}~\bibnamefont {{T{\'a}pai}}}, \bibinfo {author} {\bibfnamefont
  {L.}~\bibnamefont {{Singer}}}, \bibinfo {author} {\bibfnamefont
  {S.}~\bibnamefont {{Khan}}}, \bibinfo {author} {\bibfnamefont
  {S.}~\bibnamefont {{Fairhurst}}}, \bibinfo {author} {\bibfnamefont
  {A.}~\bibnamefont {{Nielsen}}}, \bibinfo {author} {\bibfnamefont
  {S.}~\bibnamefont {{Singh}}},\ and\ \bibinfo {author} {\bibnamefont
  {{Shasvath}}},\ }\href {https://doi.org/10.5281/zenodo.596388} {\bibinfo
  {title} {{gwastro/pycbc: PyCBC release v1.16.11}}} (\bibinfo {year}
  {2020})\BibitemShut {NoStop}%
\bibitem [{\citenamefont {{Robertson}}\ \emph {et~al.}(2020)\citenamefont
  {{Robertson}}, \citenamefont {{Smith}}, \citenamefont {{Massey}},
  \citenamefont {{Eke}}, \citenamefont {{Jauzac}}, \citenamefont {{Bianconi}},\
  and\ \citenamefont {{Ryczanowski}}}]{Robertson}%
  \BibitemOpen
  \bibfield  {author} {\bibinfo {author} {\bibfnamefont {A.}~\bibnamefont
  {{Robertson}}}, \bibinfo {author} {\bibfnamefont {G.~P.}\ \bibnamefont
  {{Smith}}}, \bibinfo {author} {\bibfnamefont {R.}~\bibnamefont {{Massey}}},
  \bibinfo {author} {\bibfnamefont {V.}~\bibnamefont {{Eke}}}, \bibinfo
  {author} {\bibfnamefont {M.}~\bibnamefont {{Jauzac}}}, \bibinfo {author}
  {\bibfnamefont {M.}~\bibnamefont {{Bianconi}}},\ and\ \bibinfo {author}
  {\bibfnamefont {D.}~\bibnamefont {{Ryczanowski}}},\ }\href
  {https://doi.org/10.1093/mnras/staa1429} {\bibfield  {journal} {\bibinfo
  {journal} {\mnras}\ }\textbf {\bibinfo {volume} {495}},\ \bibinfo {pages}
  {3727} (\bibinfo {year} {2020})},\ \Eprint {https://arxiv.org/abs/2002.01479}
  {arXiv:2002.01479 [astro-ph.CO]} \BibitemShut {NoStop}%
\bibitem [{\citenamefont {{Finkelstein}}(2016)}]{Finkelstein}%
  \BibitemOpen
  \bibfield  {author} {\bibinfo {author} {\bibfnamefont {S.~L.}\ \bibnamefont
  {{Finkelstein}}},\ }\href {https://doi.org/10.1017/pasa.2016.26} {\bibfield
  {journal} {\bibinfo  {journal} {\pasa}\ }\textbf {\bibinfo {volume} {33}},\
  \bibinfo {eid} {e037} (\bibinfo {year} {2016})},\ \Eprint
  {https://arxiv.org/abs/1511.05558} {arXiv:1511.05558 [astro-ph.GA]}
  \BibitemShut {NoStop}%
\bibitem [{\citenamefont {{Choi}}\ \emph {et~al.}(2007)\citenamefont {{Choi}},
  \citenamefont {{Park}},\ and\ \citenamefont {{Vogeley}}}]{Choi}%
  \BibitemOpen
  \bibfield  {author} {\bibinfo {author} {\bibfnamefont {Y.-Y.}\ \bibnamefont
  {{Choi}}}, \bibinfo {author} {\bibfnamefont {C.}~\bibnamefont {{Park}}},\
  and\ \bibinfo {author} {\bibfnamefont {M.~S.}\ \bibnamefont {{Vogeley}}},\
  }\href {https://doi.org/10.1086/511060} {\bibfield  {journal} {\bibinfo
  {journal} {\apj}\ }\textbf {\bibinfo {volume} {658}},\ \bibinfo {pages} {884}
  (\bibinfo {year} {2007})},\ \Eprint {https://arxiv.org/abs/astro-ph/0611607}
  {arXiv:astro-ph/0611607 [astro-ph]} \BibitemShut {NoStop}%
\bibitem [{\citenamefont {{Torrey}}\ \emph {et~al.}(2015)\citenamefont
  {{Torrey}}, \citenamefont {{Wellons}}, \citenamefont {{Machado}},
  \citenamefont {{Griffen}}, \citenamefont {{Nelson}}, \citenamefont
  {{Rodriguez-Gomez}}, \citenamefont {{McKinnon}}, \citenamefont {{Pillepich}},
  \citenamefont {{Ma}}, \citenamefont {{Vogelsberger}}, \citenamefont
  {{Springel}},\ and\ \citenamefont {{Hernquist}}}]{Torrey}%
  \BibitemOpen
  \bibfield  {author} {\bibinfo {author} {\bibfnamefont {P.}~\bibnamefont
  {{Torrey}}}, \bibinfo {author} {\bibfnamefont {S.}~\bibnamefont {{Wellons}}},
  \bibinfo {author} {\bibfnamefont {F.}~\bibnamefont {{Machado}}}, \bibinfo
  {author} {\bibfnamefont {B.}~\bibnamefont {{Griffen}}}, \bibinfo {author}
  {\bibfnamefont {D.}~\bibnamefont {{Nelson}}}, \bibinfo {author}
  {\bibfnamefont {V.}~\bibnamefont {{Rodriguez-Gomez}}}, \bibinfo {author}
  {\bibfnamefont {R.}~\bibnamefont {{McKinnon}}}, \bibinfo {author}
  {\bibfnamefont {A.}~\bibnamefont {{Pillepich}}}, \bibinfo {author}
  {\bibfnamefont {C.-P.}\ \bibnamefont {{Ma}}}, \bibinfo {author}
  {\bibfnamefont {M.}~\bibnamefont {{Vogelsberger}}}, \bibinfo {author}
  {\bibfnamefont {V.}~\bibnamefont {{Springel}}},\ and\ \bibinfo {author}
  {\bibfnamefont {L.}~\bibnamefont {{Hernquist}}},\ }\href
  {https://doi.org/10.1093/mnras/stv1986} {\bibfield  {journal} {\bibinfo
  {journal} {\mnras}\ }\textbf {\bibinfo {volume} {454}},\ \bibinfo {pages}
  {2770} (\bibinfo {year} {2015})},\ \Eprint {https://arxiv.org/abs/1507.01942}
  {arXiv:1507.01942 [astro-ph.GA]} \BibitemShut {NoStop}%
\bibitem [{\citenamefont {{Birrer}}\ and\ \citenamefont
  {{Amara}}(2018)}]{lenstronomy}%
  \BibitemOpen
  \bibfield  {author} {\bibinfo {author} {\bibfnamefont {S.}~\bibnamefont
  {{Birrer}}}\ and\ \bibinfo {author} {\bibfnamefont {A.}~\bibnamefont
  {{Amara}}},\ }\href {https://doi.org/10.1016/j.dark.2018.11.002} {\bibfield
  {journal} {\bibinfo  {journal} {Physics of the Dark Universe}\ }\textbf
  {\bibinfo {volume} {22}},\ \bibinfo {pages} {189} (\bibinfo {year} {2018})},\
  \Eprint {https://arxiv.org/abs/1803.09746} {arXiv:1803.09746 [astro-ph.CO]}
  \BibitemShut {NoStop}%
\bibitem [{\citenamefont {{Hilbert}}\ \emph {et~al.}(2007)\citenamefont
  {{Hilbert}}, \citenamefont {{White}}, \citenamefont {{Hartlap}},\ and\
  \citenamefont {{Schneider}}}]{Hilbert2007}%
  \BibitemOpen
  \bibfield  {author} {\bibinfo {author} {\bibfnamefont {S.}~\bibnamefont
  {{Hilbert}}}, \bibinfo {author} {\bibfnamefont {S.~D.~M.}\ \bibnamefont
  {{White}}}, \bibinfo {author} {\bibfnamefont {J.}~\bibnamefont {{Hartlap}}},\
  and\ \bibinfo {author} {\bibfnamefont {P.}~\bibnamefont {{Schneider}}},\
  }\href {https://doi.org/10.1111/j.1365-2966.2007.12391.x} {\bibfield
  {journal} {\bibinfo  {journal} {\mnras}\ }\textbf {\bibinfo {volume} {382}},\
  \bibinfo {pages} {121} (\bibinfo {year} {2007})},\ \Eprint
  {https://arxiv.org/abs/astro-ph/0703803} {arXiv:astro-ph/0703803 [astro-ph]}
  \BibitemShut {NoStop}%
\bibitem [{\citenamefont {{Hilbert}}\ \emph {et~al.}(2008)\citenamefont
  {{Hilbert}}, \citenamefont {{White}}, \citenamefont {{Hartlap}},\ and\
  \citenamefont {{Schneider}}}]{Hilbert2008}%
  \BibitemOpen
  \bibfield  {author} {\bibinfo {author} {\bibfnamefont {S.}~\bibnamefont
  {{Hilbert}}}, \bibinfo {author} {\bibfnamefont {S.~D.~M.}\ \bibnamefont
  {{White}}}, \bibinfo {author} {\bibfnamefont {J.}~\bibnamefont {{Hartlap}}},\
  and\ \bibinfo {author} {\bibfnamefont {P.}~\bibnamefont {{Schneider}}},\
  }\href {https://doi.org/10.1111/j.1365-2966.2008.13190.x} {\bibfield
  {journal} {\bibinfo  {journal} {\mnras}\ }\textbf {\bibinfo {volume} {386}},\
  \bibinfo {pages} {1845} (\bibinfo {year} {2008})},\ \Eprint
  {https://arxiv.org/abs/0712.1593} {arXiv:0712.1593 [astro-ph]} \BibitemShut
  {NoStop}%
\bibitem [{\citenamefont {{Cao}}\ \emph {et~al.}(2018)\citenamefont {{Cao}},
  \citenamefont {{Lu}},\ and\ \citenamefont {{Zhao}}}]{Cao}%
  \BibitemOpen
  \bibfield  {author} {\bibinfo {author} {\bibfnamefont {L.}~\bibnamefont
  {{Cao}}}, \bibinfo {author} {\bibfnamefont {Y.}~\bibnamefont {{Lu}}},\ and\
  \bibinfo {author} {\bibfnamefont {Y.}~\bibnamefont {{Zhao}}},\ }\href
  {https://doi.org/10.1093/mnras/stx3087} {\bibfield  {journal} {\bibinfo
  {journal} {\mnras}\ }\textbf {\bibinfo {volume} {474}},\ \bibinfo {pages}
  {4997} (\bibinfo {year} {2018})},\ \Eprint {https://arxiv.org/abs/1711.09190}
  {arXiv:1711.09190 [astro-ph.GA]} \BibitemShut {NoStop}%
\bibitem [{\citenamefont {{Abbott}}\ \emph {et~al.}(2017)\citenamefont
  {{Abbott}}, \citenamefont {{Abbott}}, \citenamefont {{Abbott}}, \citenamefont
  {{Acernese}}, \citenamefont {{Ackley}}, \citenamefont {{Adams}},
  \citenamefont {{Adams}}, \citenamefont {{Addesso}}, \citenamefont
  {{Adhikari}}, \citenamefont {{Adya}}, \citenamefont {{Affeldt}},
  \citenamefont {{Afrough}}, \citenamefont {{Agarwal}}, \citenamefont
  {{Agathos}}, \citenamefont {{Agatsuma}}, \citenamefont {{Aggarwal}},
  \citenamefont {{Aguiar}}, \citenamefont {{Aiello}}, \citenamefont {{Ain}},
  \citenamefont {{Ajith}}, \citenamefont {others}, \citenamefont {{LIGO
  Scientific}},\ and\ \citenamefont {{Virgo Collaboration}}}]{GW170104}%
  \BibitemOpen
  \bibfield  {author} {\bibinfo {author} {\bibfnamefont {B.~P.}\ \bibnamefont
  {{Abbott}}}, \bibinfo {author} {\bibfnamefont {R.}~\bibnamefont {{Abbott}}},
  \bibinfo {author} {\bibfnamefont {T.~D.}\ \bibnamefont {{Abbott}}}, \bibinfo
  {author} {\bibfnamefont {F.}~\bibnamefont {{Acernese}}}, \bibinfo {author}
  {\bibfnamefont {K.}~\bibnamefont {{Ackley}}}, \bibinfo {author}
  {\bibfnamefont {C.}~\bibnamefont {{Adams}}}, \bibinfo {author} {\bibfnamefont
  {T.}~\bibnamefont {{Adams}}}, \bibinfo {author} {\bibfnamefont
  {P.}~\bibnamefont {{Addesso}}}, \bibinfo {author} {\bibfnamefont {R.~X.}\
  \bibnamefont {{Adhikari}}}, \bibinfo {author} {\bibfnamefont {V.~B.}\
  \bibnamefont {{Adya}}}, \bibinfo {author} {\bibfnamefont {C.}~\bibnamefont
  {{Affeldt}}}, \bibinfo {author} {\bibfnamefont {M.}~\bibnamefont
  {{Afrough}}}, \bibinfo {author} {\bibfnamefont {B.}~\bibnamefont
  {{Agarwal}}}, \bibinfo {author} {\bibfnamefont {M.}~\bibnamefont
  {{Agathos}}}, \bibinfo {author} {\bibfnamefont {K.}~\bibnamefont
  {{Agatsuma}}}, \bibinfo {author} {\bibfnamefont {N.}~\bibnamefont
  {{Aggarwal}}}, \bibinfo {author} {\bibfnamefont {O.~D.}\ \bibnamefont
  {{Aguiar}}}, \bibinfo {author} {\bibfnamefont {L.}~\bibnamefont {{Aiello}}},
  \bibinfo {author} {\bibfnamefont {A.}~\bibnamefont {{Ain}}}, \bibinfo
  {author} {\bibfnamefont {P.}~\bibnamefont {{Ajith}}}, \bibinfo {author}
  {\bibnamefont {others}}, \bibinfo {author} {\bibnamefont {{LIGO
  Scientific}}},\ and\ \bibinfo {author} {\bibnamefont {{Virgo
  Collaboration}}},\ }\href {https://doi.org/10.1103/PhysRevLett.118.221101}
  {\bibfield  {journal} {\bibinfo  {journal} {\prl}\ }\textbf {\bibinfo
  {volume} {118}},\ \bibinfo {eid} {221101} (\bibinfo {year} {2017})},\ \Eprint
  {https://arxiv.org/abs/1706.01812} {arXiv:1706.01812 [gr-qc]} \BibitemShut
  {NoStop}%
\bibitem [{\citenamefont {{Madau}}\ and\ \citenamefont
  {{Dickinson}}(2014)}]{Madau}%
  \BibitemOpen
  \bibfield  {author} {\bibinfo {author} {\bibfnamefont {P.}~\bibnamefont
  {{Madau}}}\ and\ \bibinfo {author} {\bibfnamefont {M.}~\bibnamefont
  {{Dickinson}}},\ }\href {https://doi.org/10.1146/annurev-astro-081811-125615}
  {\bibfield  {journal} {\bibinfo  {journal} {\araa}\ }\textbf {\bibinfo
  {volume} {52}},\ \bibinfo {pages} {415} (\bibinfo {year} {2014})},\ \Eprint
  {https://arxiv.org/abs/1403.0007} {arXiv:1403.0007 [astro-ph.CO]}
  \BibitemShut {NoStop}%
\bibitem [{\citenamefont {{Strolger}}\ \emph {et~al.}(2004)\citenamefont
  {{Strolger}}, \citenamefont {{Riess}}, \citenamefont {{Dahlen}},
  \citenamefont {{Livio}}, \citenamefont {{Panagia}}, \citenamefont
  {{Challis}}, \citenamefont {{Tonry}}, \citenamefont {{Filippenko}},
  \citenamefont {{Chornock}}, \citenamefont {{Ferguson}}, \citenamefont
  {{Koekemoer}}, \citenamefont {{Mobasher}}, \citenamefont {{Dickinson}} \emph
  {et~al.}}]{Strolger}%
  \BibitemOpen
  \bibfield  {author} {\bibinfo {author} {\bibfnamefont {L.-G.}\ \bibnamefont
  {{Strolger}}}, \bibinfo {author} {\bibfnamefont {A.~G.}\ \bibnamefont
  {{Riess}}}, \bibinfo {author} {\bibfnamefont {T.}~\bibnamefont {{Dahlen}}},
  \bibinfo {author} {\bibfnamefont {M.}~\bibnamefont {{Livio}}}, \bibinfo
  {author} {\bibfnamefont {N.}~\bibnamefont {{Panagia}}}, \bibinfo {author}
  {\bibfnamefont {P.}~\bibnamefont {{Challis}}}, \bibinfo {author}
  {\bibfnamefont {J.~L.}\ \bibnamefont {{Tonry}}}, \bibinfo {author}
  {\bibfnamefont {A.~V.}\ \bibnamefont {{Filippenko}}}, \bibinfo {author}
  {\bibfnamefont {R.}~\bibnamefont {{Chornock}}}, \bibinfo {author}
  {\bibfnamefont {H.}~\bibnamefont {{Ferguson}}}, \bibinfo {author}
  {\bibfnamefont {A.}~\bibnamefont {{Koekemoer}}}, \bibinfo {author}
  {\bibfnamefont {B.}~\bibnamefont {{Mobasher}}}, \bibinfo {author}
  {\bibfnamefont {M.}~\bibnamefont {{Dickinson}}}, \emph {et~al.},\ }\href
  {https://doi.org/10.1086/422901} {\bibfield  {journal} {\bibinfo  {journal}
  {\apj}\ }\textbf {\bibinfo {volume} {613}},\ \bibinfo {pages} {200} (\bibinfo
  {year} {2004})},\ \Eprint {https://arxiv.org/abs/astro-ph/0406546}
  {arXiv:astro-ph/0406546 [astro-ph]} \BibitemShut {NoStop}%
\bibitem [{\citenamefont {{Iyer}}\ \emph {et~al.}(2011)\citenamefont {{Iyer}},
  \citenamefont {{Souradeep}}, \citenamefont {{Unnikrishnan}}, \citenamefont
  {{Dhurandhar}}, \citenamefont {{Raja}},\ and\ \citenamefont
  {{Sengupta}}}]{IndIGO}%
  \BibitemOpen
  \bibfield  {author} {\bibinfo {author} {\bibfnamefont {B.}~\bibnamefont
  {{Iyer}}}, \bibinfo {author} {\bibfnamefont {T.}~\bibnamefont {{Souradeep}}},
  \bibinfo {author} {\bibfnamefont {C.}~\bibnamefont {{Unnikrishnan}}},
  \bibinfo {author} {\bibfnamefont {S.}~\bibnamefont {{Dhurandhar}}}, \bibinfo
  {author} {\bibfnamefont {S.}~\bibnamefont {{Raja}}},\ and\ \bibinfo {author}
  {\bibfnamefont {A.}~\bibnamefont {{Sengupta}}} (\bibinfo {collaboration}
  {LIGO Scientific Collaboration}),\ }\href
  {https://dcc.ligo.org/cgi-bin/DocDB/ShowDocument?docid=75988} {\emph
  {\bibinfo {title} {{LIGO-India, Proposal of the Consortium for Indian
  Initiative in Gravitational-wave Observations (IndIGO),[Public]}}}},\
  \bibinfo {type} {Tech. Rep.}\ \bibinfo {number} {LIGO-M1100296-v2}\ (\bibinfo
  {year} {2011})\BibitemShut {NoStop}%
\bibitem [{\citenamefont {{Spera}}\ \emph {et~al.}(2015)\citenamefont
  {{Spera}}, \citenamefont {{Mapelli}},\ and\ \citenamefont
  {{Bressan}}}]{spera}%
  \BibitemOpen
  \bibfield  {author} {\bibinfo {author} {\bibfnamefont {M.}~\bibnamefont
  {{Spera}}}, \bibinfo {author} {\bibfnamefont {M.}~\bibnamefont {{Mapelli}}},\
  and\ \bibinfo {author} {\bibfnamefont {A.}~\bibnamefont {{Bressan}}},\ }\href
  {https://doi.org/10.1093/mnras/stv1161} {\bibfield  {journal} {\bibinfo
  {journal} {\mnras}\ }\textbf {\bibinfo {volume} {451}},\ \bibinfo {pages}
  {4086} (\bibinfo {year} {2015})},\ \Eprint {https://arxiv.org/abs/1505.05201}
  {arXiv:1505.05201 [astro-ph.SR]} \BibitemShut {NoStop}%
\bibitem [{\citenamefont {{Belczynski}}\ \emph {et~al.}(2016)\citenamefont
  {{Belczynski}}, \citenamefont {{Holz}}, \citenamefont {{Bulik}},\ and\
  \citenamefont {{O'Shaughnessy}}}]{BelczynskiHolz}%
  \BibitemOpen
  \bibfield  {author} {\bibinfo {author} {\bibfnamefont {K.}~\bibnamefont
  {{Belczynski}}}, \bibinfo {author} {\bibfnamefont {D.~E.}\ \bibnamefont
  {{Holz}}}, \bibinfo {author} {\bibfnamefont {T.}~\bibnamefont {{Bulik}}},\
  and\ \bibinfo {author} {\bibfnamefont {R.}~\bibnamefont {{O'Shaughnessy}}},\
  }\href {https://doi.org/10.1038/nature18322} {\bibfield  {journal} {\bibinfo
  {journal} {\nat}\ }\textbf {\bibinfo {volume} {534}},\ \bibinfo {pages} {512}
  (\bibinfo {year} {2016})},\ \Eprint {https://arxiv.org/abs/1602.04531}
  {arXiv:1602.04531 [astro-ph.HE]} \BibitemShut {NoStop}%
\bibitem [{\citenamefont {{Chabrier}}(2003)}]{Chabrier}%
  \BibitemOpen
  \bibfield  {author} {\bibinfo {author} {\bibfnamefont {G.}~\bibnamefont
  {{Chabrier}}},\ }\href {https://doi.org/10.1086/376392} {\bibfield  {journal}
  {\bibinfo  {journal} {\pasp}\ }\textbf {\bibinfo {volume} {115}},\ \bibinfo
  {pages} {763} (\bibinfo {year} {2003})},\ \Eprint
  {https://arxiv.org/abs/astro-ph/0304382} {arXiv:astro-ph/0304382 [astro-ph]}
  \BibitemShut {NoStop}%
\end{thebibliography}%

\appendix
\section{Binary Black Hole Merger Rate}
\label{app:rmerg}
In this appendix, we elaborate on the astrophysical models adopted to calculate the merger rate of binary black holes. In summary, we compute the BBH merger rate from population models on black hole progenitor stars and calibrate to the observed rate in the local universe.

Adapted from Eq.~(B1) and (B2) in \cite{Li2018}, the birth rate of individual black holes with mass $m_\bullet$ at redshift $z_s$, $R_{\rm{birth}}(m_\bullet,z_s)$, is given by

\begin{widetext}
\begin{equation}
    R_{\rm{mrg}}(m_\bullet,z_s) = \int dm_\star~dt_d~dZ~ \phi(m_\star )\dot{\psi}\left(t\left(z_s\right)-t_d\right)P(t_d)P\left(Z,t\left(z_s\right)-t_d\right)\delta\left[m_\star -g^{-1}(m_\bullet,Z)\right]\;,
\end{equation}
\end{widetext}
where $m_\star$ is the mass of the progenitor star, $Z$ is stellar metallicity and $t(z_s)$ is the cosmic time as a function of redshift. $g^{-1}(m_\bullet,Z)$ gives the stellar mass $m_\star$ with metallicity $Z$ that leaves a black hole remnant with mass $m_\bullet$. The expression of remnant black hole mass as a function of stellar mass and metallicity is given in \cite{spera}, with $20~M_\odot<m_\star<105~M_\odot$ and $-5<\log_{10}Z<-1.7$. $P(Z,t_z)$ is the redshift-dependent distribution of metallicity. The mean log metallicity at any redshift is given in \cite{BelczynskiHolz}. At each redshift, the metallicity follows a log normal distribution \cite{BelczynskiHolz}. 

$t_d$ is the time delay between black hole formation and its merger with another black hole. $P(t_d)$ is the distribution of time delay, and we adopt the form $P(t_d)\propto t_d^{-1}$, truncated at $t_d = 50~{\rm{Myr}}$ and the Hubble time \cite{Li2018}. Note that we ignore the time delay between the formation of a star and the formation of its remnant. Since stellar evolution is on the order of Myr, which is negligibly small compare to the evolution time scale of galaxies and hence that of black holes, we can neglect it for model simplicity without incurring large errors. 

The quantity $\phi(m_\star)$ is the initial mass function that describes the stellar mass distribution, which we assume to remain constant across redshift. Specifically, we adopt the Chabrier initial mass function \cite{Chabrier} for $m_\star >1~M_\odot$, where $\phi(m_\star)\propto m_\star^{-2.3}$. The quantity $\dot{\psi}(t)$ is the Star Formation Rate (SFR) including all $m_\star$ at cosmic time $t$. We adopt the analytic SFR expression in \cite{Strolger}. We calibrate the merger rate at $z=0$ to be 103 ${\rm{Gpc}}^{-3}\rm{yr}^{-1}$, which is the expected local black hole merger rate given LIGO detection data up until GW170104\cite{GW170104,Cao}. See text for the effects of an updated local merger rate based on GWTC-2.

We note that $R_{\rm{mrg}}(m_\bullet,z_s)$ is the rates for black hole binary at $z_s$ whose \textit{primary} black hole, i.e. the heavier one, is $m_\bullet$. We then assign a mass ratio value according to the distribution $P(q)\propto q$, with $q$ truncated at 1.2 and 3.2. We can then directly convert the rates into $R_{\rm{mrg}}(M_\bullet,q,z_s)$ where $M_\bullet$ is the total binary mass. 

\end{document}